\documentclass[a4paper,12pt]{article}
\pdfoutput=1
\usepackage{graphicx, rotating,slashed,ifpdf}
\usepackage[dvipsnames]{xcolor}

\ifx\pdfoutput\undefined
\usepackage[dvips,bookmarks=false]{hyperref}	% This is for arXiv.org
\else
\usepackage{hyperref}	% This is for pdftex
\fi
\hypersetup{colorlinks,bookmarksopen,bookmarksnumbered,citecolor=verdes,
linkcolor=blus,pdfstartview=FitH,urlcolor=BrickRed}
\def\hhref#1{\href{http://arxiv.org/abs/#1}{#1}} % in bibliography
\usepackage{amsmath}
\usepackage{slashed}

\newcommand{\MS}{\overline{\mbox{\sc ms}}}

\newcommand{\Mtexp}{173.34}
\newcommand{\Mterr}{0.76}
\newcommand{\Mhexp}{125.15}
\newcommand{\Mherr}{0.24}
\newcommand{\MWexp}{80.384}
\newcommand{\MWerr}{0.014}
\newcommand{\Mhexperr}{\Mhexp\pm\Mherr~\GeV}
\newcommand{\Mtdiff}{ \bigg(\frac{M_t}{\GeV}-\Mtexp\bigg)}
\newcommand{\asdiff}{\, \frac{\alpha_3(M_Z)-0.1184}{0.0007} }
\newcommand{\Mhdiff}{\bigg(\frac{M_h}{\GeV}-\Mhexp\bigg)}
\newcommand{\MWdiff}{\frac{M_W - \MWexp\GeV}{\MWerr\GeV}  }
\newcommand{\beq}{\begin{equation}}
\newcommand{\eeq}{\end{equation}}
\newcommand{\fig}[1]{~\ref{fig:#1}}
\newcommand{\be}{\begin{equation}}
\newcommand{\ee}{\end{equation}}

\newcount\Mac  \Mac=1  % devo mettere Mac=1 se sto lavorando sul file Mac
\newcommand{\ifMac}[2]{\ifnum\Mac=1 #1 \else #2 \fi}
\def\putps(#1,#2)(#3,#4)#5#6{\ifnum\Mac=1 \put(#1,#2){\special{picture #5}}
\else  \put(#3,#4){\includegraphics{#6}} \fi}

\renewcommand{\Re}{\hbox{Re}\,}
\newcommand{\riga}[1]{\noalign{\hbox{\parbox{\textwidth}{#1}}}\nonumber}

\newcommand{\One}{\hbox{1\kern-.24em I}}

\newcommand{\GeV}{\,{\rm GeV}}

\newcommand{\eq}[1]{~{\rm(\ref{eq:#1})}}

\newcommand{\mub}{\bar\mu}
\newcommand{\Ord}{{\cal O}}

\def\eqg#1{eq.~(\ref{#1})}
\newcommand\mpl{M_{\rm Pl}}
\usepackage{cite}

\newcommand{\lascia}[1]{}
\makeatletter
\def\art{\@ifnextchar[{\eart}{\oart}}
\def\eart[#1]#2#3#4#5#6{{\rm #2}, {#3 #4} {\rm (#6) #5} [arXiv:{\hhref{#1}}]}
\def\hepart[#1]#2{{\rm #2, arXiv:\hhref{#1}}}
\newcommand{\oart}[5]{{\rm #1}, {#2 #3} {\rm (#5) #4}}

\newcounter{alphaequation}[equation]
\def\thealphaequation{\theequation\hbox to
0.6em{\hfil\alph{alphaequation}\hfil}}
\def\eqnsystem#1{
\def\@eqnnum{{\rm (\thealphaequation)}}
\def\@@eqncr{\let\@tempa\relax \ifcase\@eqcnt \def\@tempa{& & &} \or
  \def\@tempa{& &}\or \def\@tempa{&}\fi\@tempa
  \if@eqnsw\@eqnnum\refstepcounter{alphaequation}\fi
\global\@eqnswtrue\global\@eqcnt=0\cr}
\refstepcounter{equation} \let\@currentlabel\theequation \def\@tempb{#1}
\ifx\@tempb\empty\else\label{#1}\fi
\refstepcounter{alphaequation}
\let\@currentlabel\thealphaequation
\global\@eqnswtrue\global\@eqcnt=0 \tabskip\@centering\let\\=\@eqncr
$$\halign to \displaywidth\bgroup \@eqnsel\hskip\@centering
$\displaystyle\tabskip\z@{##}$&\global\@eqcnt\@ne
\hskip2\arraycolsep\hfil${##}$\hfil& \global\@eqcnt\tw@\hskip2\arraycolsep
$\displaystyle\tabskip\z@{##}$\hfil
\tabskip\@centering&\llap{##}\tabskip\z@\cr}

\def\endeqnsystem{\@@eqncr\egroup$$\global\@ignoretrue} \makeatother

\newcommand{\os}{{\rm OS}}

\def\Ord{{\cal O}}
\def\Lag{{\cal L}}
\def\SU{{\rm SU}}

\def\circa#1{\,\raise.3ex\hbox{$#1$\kern-.75em\lower1ex\hbox{$\sim$}}\,}
\def\ssW{{\scriptscriptstyle W}}

\def\ss0{{\scriptscriptstyle 0}}
\def\1l{{\scriptscriptstyle (1)}}
\def\2l{{\scriptscriptstyle (2)}}

\usepackage{multicol}
\usepackage{color}
\definecolor{rosso}{cmyk}{0,1,1,0.4}
\definecolor{rossos}{cmyk}{0,1,1,1}
\definecolor{rossoc}{cmyk}{0,1,1,0.2}
\definecolor{blu}{cmyk}{1,1,0,0.3}
\definecolor{blus}{cmyk}{1,1,0,0.6}
\definecolor{bluc}{cmyk}{1,1,0,0.1}
\definecolor{verde}{cmyk}{0.92,0,0.59,0.25}
\definecolor{verdec}{cmyk}{0.92,0,0.59,0.15}
\definecolor{verdes}{cmyk}{0.92,0,0.59,0.4}
\definecolor{grigio}{cmyk}{0,0,0,0.07}
\definecolor{rosa}{cmyk}{0,0.1,0.1,0.02}
\definecolor{rosino}{cmyk}{0,0.05,0.05,0.02}
\definecolor{rosas}{cmyk}{0,0.3,0.25,0.05}
\definecolor{celeste}{cmyk}{0.1,0,0,0.02}
\definecolor{giallino}{cmyk}{0,0,0.4,0.02}
\definecolor{rosso}{cmyk}{0,1,1,0.4}
\definecolor{rossos}{cmyk}{0,1,1,0.55}
\definecolor{rossoc}{cmyk}{0,1,1,0.2}
\definecolor{blu}{cmyk}{1,1,0,0.3}
\definecolor{bluc}{cmyk}{1,1,0,0.1}
\definecolor{blucc}{cmyk}{0.7,0.5,0,0}
\definecolor{viola}{cmyk}{0,1,0,0.6}
\definecolor{viola2}{cmyk}{0,1,0.2,0.6}
\definecolor{verde}{cmyk}{0.92,0,0.59,0.25}
\definecolor{verdec}{cmyk}{0.92,0,0.59,0.15}
\definecolor{verdes}{cmyk}{0.92,0,0.59,0.4}
\definecolor{verdino}{cmyk}{0.12,0,0.09,0.05}
\definecolor{giallo}{cmyk}{0,0,1,0}
\definecolor{gialloverde}{cmyk}{0.44,0,0.74,0}

\font\tenrsfs=rsfs10 at 12pt

\font\sevenrsfs=rsfs7
\font\fiversfs=rsfs5
\newfam\rsfsfam
\textfont\rsfsfam=\tenrsfs
\scriptfont\rsfsfam=\sevenrsfs
\scriptscriptfont\rsfsfam=\fiversfs
\def\mathscr#1{{\fam\rsfsfam\relax#1}}
\def\Lag{\mathscr{L}}

\def\order#1{{\cal O}(#1)}

\def\beq{\begin{equation}}
\def\eeq{\end{equation}}
\def\bea{\begin{eqnarray}}
\def\eea{\end{eqnarray}}
\def\bac{\begin{array} {c}}
\def\ea{\end{array}}

\def\dm2{\delta m^2}
\def\dv2{\delta v^2}

\def\nn{\nonumber}

\def\psl{\hbox{\hbox{${p}$}}\kern-1.9mm{\hbox{${/}$}}}

\begin{document}
CERN-PH-TH-2013-166\hfill 
FTUAM-13-20\hfill
IFT-UAM/CSIC-13-082\hfill
RM3-TH/13-9
\color{black}
\vspace{1cm}
\begin{center}
{\Huge\bf\color{black}\color{BrickRed}Investigating the near-criticality\\of the Higgs boson}\\
\bigskip\color{black}\vspace{0.6cm}{
{\large\bf Dario~Buttazzo$^{a,b}$, Giuseppe~Degrassi$^{c}$, 
Pier Paolo~Giardino$^{a,d}$,\\ Gian F.~Giudice$^{a}$, Filippo~Sala$^{b,e}$, Alberto~Salvio$^{b,f}$, Alessandro~Strumia$^{d}$}
} \\[7mm]
{\it (a) CERN, Theory Division, Geneva, Switzerland}\\[1mm]
{\it (b) Scuola Normale Superiore and INFN, sezione di  Pisa, Italy}\\[1mm]
{\it (c) Dipartimento di Matematica e Fisica, Universit{\`a} di Roma Tre and \\
 INFN sezione di Roma Tre, Italy}\\
% INFN, sezione di Roma Tre, Italy}\\[1mm]
{\it (d) Dipartimento di Fisica, Universit{\`a} di Pisa and INFN, sezione di Pisa, Italy}\\[1mm]
{\it (e) Theoretical Physics Group, Lawrence Berkeley National Laboratory, Berkeley, CA, USA
}\\[1mm]
{\it  (f) Departamento de F\'isica Te\'orica, Universidad Aut\'onoma de Madrid\\ and Instituto de F\'isica Te\'orica IFT-UAM/CSIC,  Madrid, Spain}\\[1mm]
\end{center}
\bigskip
\bigskip
\bigskip
\vspace{1cm}

\centerline{\large\bf\color{blus} Abstract}

\begin{quote}\large
We extract from data  the parameters of the Higgs potential, the top Yukawa coupling
and the electroweak gauge couplings
with full 2-loop NNLO precision, and we extrapolate the SM parameters
up to large energies with full 3-loop NNLO RGE precision. Then we study the phase
diagram of the Standard Model in terms of high-energy parameters, finding that the
measured Higgs mass roughly corresponds to the minimum values of the Higgs quartic and
top Yukawa and the maximum value of the gauge couplings allowed by vacuum metastability.
We discuss various theoretical interpretations of the near-criticality of the Higgs mass.

\end{quote}
\thispagestyle{empty}
\newpage

\tableofcontents

\section{Introduction}

The discovery of the Higgs boson~\cite{Higgsexp1,Higgsexp2} was
expected to be the herald of new physics soon to be found at the TeV
scale. So far, however, no signal of new physics nor any clear
deviation from the SM Higgs properties have been detected at the
LHC. Moreover, the Higgs mass has not provided unambiguous indications
for new physics. The measured value $M_h=\Mhexperr$~\cite{HiggsMass} is a bit high for
supersymmetry and a bit low for composite models, making theoretical
interpretations rather uncomfortable. Neither option is unequivocally
favoured, although neither option is excluded. On the other hand,
$M_h=\Mhexperr$ lies well within the parameter window in which the SM can
be extrapolated all the way up to the Planck mass $\mpl $, with no
problem of consistency other than remaining in the dark about
naturalness. Remarkably, in the context of the SM the measured value
of $M_h$ is special because it corresponds to a near-critical
situation in which the Higgs vacuum does not reside in the
configuration of minimal energy, but in a metastable state close to a
phase transition~\cite{DDEEGIS} (for earlier considerations see~\cite{Krive:1976sg,Krasnikov:1978pu,Maiani:1977cg,Politzer:1978ic,Hung:1979dn,Cabibbo:1979ay,Linde:1979ny,Lindner:1985uk,Lindner:1988ww,Sher:1988mj,Arnold:1989cb,Arnold:1991cv,Sher:1993mf,Altarelli:1994rb,Casas:1994qy,Espinosa:1995se,Casas:1996aq,Schrempp:1996fb,Hambye:1996wb,IRS,Espinosa:2007qp,Ellis:2009tp,Hall};
for related studies see~\cite{Holthausen:2011aa,EliasMiro:2011aa,Chen:2012faa,Lebedev:2012zw,EliasMiro:2012ay,Rodejohann:2012px,Bezrukov:2012sa,Datta:2012db,Alekhin:2012py,Chakrabortty:2012np,Anchordoqui:2012fq,Masina:2012tz,Chun:2012jw,Chung:2012vg,Chao:2012mx,Lebedev:2012sy,Nielsen:2012pu,Kobakhidze:2013tn,Tang:2013bz,Klinkhamer:2013sos,He:2013tla,Chun:2013soa,Jegerlehner:2013cta,Antipin:2013sga,Branchina}).

We believe that near-criticality of the SM vacuum is the most
important message we have learnt so far from experimental data on the
Higgs boson. Near-criticality gives us a unique opportunity to obtain
information about physics taking place at energy scales well beyond
the reach of any collider experiment. Its consequences are so
intriguing and potentially so revolutionary that they deserve accurate
calculations and dedicated studies. In this paper we continue our
programme of investigating the status and implications of
near-criticality. We make advancements on both sides: on the
computational side, we improve the calculation of the large-field
extrapolation of the Higgs potential and of the critical value of
$M_h$ for absolute stability; on the interpretation side, we explore
the significance of near-criticality in terms of high-energy SM
parameters.

The main new calculations presented in this paper are the results for
the $\MS$ quartic Higgs coupling $\lambda(\mub)$, for the
top Yukawa coupling $y_t(\mub)$, for the electroweak gauge couplings 
at NNLO precision (two loops) in terms
of physical observables: the pole masses of the Higgs ($M_h$), of the
top ($M_t$), of the $Z$ ($M_Z$), of the $W$ ($M_W$), the
$\MS$ strong coupling $\alpha_3(M_Z)$, and the Fermi
constant $G_\mu$.  We improve on the study in ref.~\cite{DDEEGIS}
where 2-loop threshold corrections to $\lambda(\mub)$ had been computed
in the limit of vanishing weak gauge couplings, and 2-loop electroweak threshold
corrections to $y_t(\mub)$ had been neglected. As a byproduct of our
two-loop calculation of  $\lambda(\mub)$ we also obtain the 
$\MS$ quadratic Higgs coupling $m^2(\mub)$ at the
 NNLO level.

Recently, many authors have contributed towards the completion of the
calculation of the renormalisation-group (RG) evolution
($\beta$-functions and thresholds) of the sizeable SM couplings at
NNLO precision. We summarise the present status of these calculations
in table~\ref{tab:status}. Our new calculation of threshold
corrections, together with the results collected in
table~\ref{tab:status}, allows us to refine the determination of the
critical value of $M_h$ that ensures absolute vacuum stability within
the Standard Model (SM) up to the Planck scale.  Furthermore, our
precision extrapolation of the SM to high energy scales is relevant
for testing any new physics scenario able of making predictions, such
as unification of gauge couplings constants, or high-scale supersymmetric models
that restrict or predict the quartic Higgs coupling.

The paper is organised as follows.  In section~\ref{sec:strategy} we
outline the general strategy for the two-loop computations and
describe our new results.  In section~\ref{sec:inputs} we present numerical
results for the $\MS$ couplings at the weak scale.
The implications of these results for Planck scale physics are discussed
in sections~\ref{sec:Planck}--\ref{sec:more}.  The results are summarised in 
the conclusions. We complemented the paper with several appendices where
we collect all the known results on the RG equations and the 
threshold corrections that we used in our computation.

\begin{table}
\begin{center}
\begin{tabular}{|c|cccc|}
\multicolumn{5}{c}{\color{BrickRed}\bf Renormalisation Group Equations}\\ \hline
& LO & NLO & NNLO & NNNLO\\ 
 & 1 loop & 2 loop & 3 loop & 4 loop\\   
 \hline
 $g_{3}$ & full~\hbox {\cite{Gross:1973id,Politzer:1973fx}} & $\Ord(\alpha_3^2)$~\hbox{\cite{Caswell:1974gg,Jones:1974mm}} & $\Ord(\alpha_3^3)$~\hbox{\cite{Tarasov:1980au,Larin:1993tp}} & $\Ord(\alpha_3^4)$~\hbox{\cite{vanRitbergen:1997va,Czakon:2004bu}}\\[-1mm]
  &  & $\Ord(\alpha_3\alpha_{1,2})$~\hbox{\cite{Jones:1981we}} & $\Ord(\alpha_3^2\alpha_t)$~\hbox{\cite{Steinhauser:1998cm}} & \\[-1mm]
 &  & full~\hbox{\cite{Machacek:1983tz}} & full~\hbox{\cite{RGE3-MSS,Mihaila:2012pz}} & \\[2mm]
$g_{1,2}$ & full~\hbox {\cite{Gross:1973id,Politzer:1973fx}} & full~\hbox{\cite{Machacek:1983tz}} & full~\hbox{\cite{RGE3-MSS,Mihaila:2012pz}} & --- \\[2mm]
$y_t$ & full~\hbox{\cite{Cheng:1973nv}} & $\Ord(\alpha_t^2,\alpha_3\alpha_t)$~\hbox{\cite{Fischler:1982du}} & full~\hbox{\cite{Chetyrkin:2012rz,Bednyakov:2012en}} & --- \\[-1mm]
 &  &  full~\hbox{\cite{Machacek:1983fi}}& & \\[2mm]
$\lambda, m^2$ & full~\hbox{\cite{Cheng:1973nv}} & full~\hbox{\cite{Machacek:1984zw,Luo:2002ey}} &full~\hbox{\cite{Chetyrkin:2013wya,Bednyakov:2013eba}} & --- \\
\hline 
%\vspace{4mm}
%
\multicolumn{5}{c}{}\\ 
\multicolumn{5}{c}{\color{BrickRed}\bf Threshold corrections at the weak scale$^{\phantom{X^X}}$}\\ \hline
 &LO & NLO & NNLO & NNNLO\\
 & 0 loop & 1 loop & 2 loop & 3 loop\\   
 \hline
%$g_3$ & & full~\hbox{\cite{}}  & full~\hbox{\cite{Mihaila:2012pz}} & --- 
%full~\cite{Mihaila:2012pz} 
%\\ %$\Ord(\alpha_3^3)$ only?? \\
%\hline
$g_{2}$ &$ 2M_W/V$
& full~\hbox{\cite{Sirlin:1980nh,Marciano:1980pb}} & full~[This work]    & --- %full~\cite{Mihaila:2012pz} 
\\[2mm] 
%\hline
$g_{Y}$ &$ 2\sqrt{M_Z^2-M_W^2}/V$
& full~\hbox{\cite{Sirlin:1980nh,Marciano:1980pb}} & full~[This work]    & --- %full~\cite{Mihaila:2012pz} 
\\[2mm] 
%\hline
$y_t$ &$\sqrt{2} M_t/V$ &   $\Ord(\alpha_3)$~\hbox{\cite{Tarrach:1980up}}& $\Ord(\alpha_3^2,\alpha_3\alpha_{1,2})$~\hbox{\cite{Bezrukov:2012sa}}   & $\Ord(\alpha_3^3)$~\hbox{\cite{Chetyrkin:1999ys,Chetyrkin:1999qi,Melnikov:2000qh}}  \\[-1mm]
 & &   $\Ord(\alpha)$~\hbox{\cite{Hempfling:1994ar}} &   full~[This work] &   \\[2mm]
%\hline
$\lambda$ & $M_h^2/2V^2$& full~\hbox{\cite{SZ}} & for $g_{1,2}=0$~\hbox{\cite{DDEEGIS}}
 &---% \hbox{for $g_{1,2},\lambda \ll y_t $} 
  \\[-1mm]
 & 
 &  & full~[This work]
 &   \\[2mm]
%\hline
$m^2$ & $M_h^2$& full~\hbox{\cite{SZ}} &  full~[This work] & --- \\
 \hline
 %\multicolumn{5}{c}{* We plan to include future results in a revised version of this work.}
\end{tabular}
%REMEMBER: (1)since we insert the input numerical value for $\alpha_3$, $\alpha_{1,2}$ knowing the beta functions tells us also the threshold 
%corrections 
%(2) to know g_1,2 at two loops we can use the relation between G_mu, Delta r, e and theta_W so once we know theta_W and Delta r we know g_1,2;
%theta_W was computed in \cite{Awramik:2006uz} and Delta r \cite{Awramik:2003ee}
\caption{\label{tab:status} \em Present status of higher-order computations included in our code.
With the present paper the calculation of the SM parameters at NNLO precision is complete. Here we have defined $V \equiv (\sqrt{2} G_\mu)^{-1/2}$ and $g_1=\sqrt{5/3} g_Y$.
%the two-loop correction to the quartic and quadratic Higgs terms $\lambda$ and $m^2$ and to 
%the top Yukawa coupling $y_t$ are fully computed 
%($V \equiv (\sqrt{2} G_\mu)^{-1/2}$).
}
\end{center}
\end{table}%

\section{Computing the  $\MS$ parameters with two-loop accuracy}
\label{sec:strategy}
In this section we first outline the general strategy followed to determine the 
 $\MS$ parameters in terms of physical observables at the
two-loop level. Then, in  sects.~\ref{sec2.1} and \ref{sec2.2}, 
we will discuss the results for the quartic and quadratic Higgs couplings, respectively, while
section~\ref{sec2.3} is dedicated to the calculation of two-loop threshold 
corrections to the top Yukawa coupling.

First of all, {\em all $\MS$ parameters have gauge-invariant
  renormalisation group equations~\cite{wil} and are gauge invariant},
as we now prove.\footnote{Gauge invariance of fermion pole masses has
  been proved in
  refs.~\cite{Atkinson:1979ut,Mgaugeinvariance,Mgaugeinvariance1} and here we
  generalise their proof.}  Let us consider a generic $\MS$ coupling
$\theta$ measuring the strength of a gauge invariant term in the
Lagrangian and a generic gauge fixing parametrized by $\xi$ (for
example the $R_\xi$ gauges). Let us first recall the definition of
$\theta$ in terms of the bare coupling $\theta_0$, 
\be 
\mub^{d-4}\theta_0=\sum_{k=0}^{\infty}\frac{c_k(\theta,\xi)}{(d-4)^k}, 
\label{lambda-definition}
\ee
where the $c_k$ are defined to be the residues at the divergence $d=4$. 
The important point is that $c_0 = \theta$, with no dependence on $\xi$.
Since $\theta_0$ is gauge independent, we have
\be 
0= \mub^{d-4}\frac{d \theta_0}{d\xi}=\frac{d \theta}{d\xi}
+\sum_{k=1}^{\infty}\frac{1}{(d-4)^k}\frac{dc_k(\theta,\xi)}{d\xi}.
\ee
Since this equation is valid for any $d$, and $\theta$ has no poles at $d = 4$ by definition, we obtain
$d\theta/d\xi=0$, that is $\theta$ is gauge invariant (as well as all the residues $c_k$).\footnote{Notice
  that this proof does not apply to the Higgs vev $v$, because it is
  not the coefficient of a gauge-invariant term in the Lagrangian.}

\bigskip

To determine the $\MS$ parameters in terms of physical 
observables two strategies can be envisaged.
\begin{itemize}
\item[i)] Perform an  
$\MS$ renormalisation to obtain directly the
  $\MS$ quantity of interest
 in terms of  $\MS$ parameters. Then  express the
$\MS$ parameters in terms of the physical ones via appropriately 
derived two-loop relations. 
\item[ii)] Use a renormalisation scheme in which the
renormalised parameters   are directly expressed in terms of
 physical observables (we call this scheme generically on-shell
(OS) and label quantities in this scheme with an $\os$). Then  relate the 
parameters  as expressed in the OS 
scheme to their $\MS$ counterparts we are looking for. 
\end{itemize}
This last step can be easily done using the relation  
\be
\theta_0 = \theta_\os - \delta \theta_\os = 
\theta(\mub)-{\delta  \theta}_{\MS}
\label{eq:g1} 
\ee
or 
\be
\theta(\mub) = \theta_\os - \delta \theta_\os + 
{\delta  \theta}_{ \MS}\, ,
\label{eq:g2} 
\ee 
where $\theta_0$ is the bare parameter, $\theta (\mub)$ ($\theta_\os$) is
the renormalised $ \MS$ (OS) version and 
$\delta  \theta_{\MS }$ ($\delta  \theta_{\os }$) the
corresponding counterterm. By definition $\delta  \theta_{\MS }$
subtracts only the terms proportional to powers of 
$1/\epsilon$ and $\gamma -\ln (4 \pi)$ in dimensional regularisation, 
with $d= 4- 2\, \epsilon$ being the space-time dimension. Concerning  the structure of the $1/\epsilon$ poles
in the  OS and $ \MS$ counterterms, one  notices that it should 
be   identical once the poles in the OS counterterms are expressed in terms of 
$\MS$ quantities. Then, after this
operation is performed, the desired $\theta(\mub)$ is obtained from
\be
\theta(\mub) = \theta_\os - \left. \delta \theta_\os \right|_{\rm fin} +
\Delta_\theta,
\label{eq:g3} 
\ee
where the subscript `fin' denotes the finite part of the quantity involved 
and $\Delta$ is the two-loop finite contribution that is obtained
when the OS parameters entering the  1/$\epsilon$ pole in the OS counterterm 
are  expressed in terms of $\MS$ quantities, the finite
contribution coming from the ${\cal O}(\epsilon)$ part of the shifts.

\medskip

\begin{table}[t]
\begin{center}
$$\begin{array}{rcllr}
M_W &=& \MWexp\pm\MWerr\,\GeV & \hbox{Pole mass of the $W$ boson} & \hbox{\cite{Wmass}}\\
M_Z &=& 91.1876\pm0.0021\,\GeV & \hbox{Pole mass of the $Z$ boson} & \hbox{\cite{Zmass}}\\
M_h &=& \Mhexperr & \hbox{Pole mass of the higgs} & \hbox{\cite{HiggsMass}}\\
%M_t &=& 173.36\pm0.65\pm0.3\,\GeV & \hbox{Pole mass of the top quark \xxx{OLD}} & \hbox{\cite{topmass}}\\
M_t &=&\Mtexp\pm\Mterr\pm0.3\,\GeV & \hbox{Pole mass of the top quark} & \hbox{\cite{topmass}}\\
V\equiv (\sqrt{2} G_\mu)^{-1/2} &=&246.21971\pm0.00006\GeV& \hbox{Fermi constant for $\mu$ decay} & \hbox{\cite{MuLan}}\\
%G_\mu &=& 1.1663781(6) 10^{-5}/\GeV^2 & \hbox{Fermi constant for $\mu$ decay} & \hbox{\cite{MuLan}}\\
\alpha_3(M_Z) &=& 0.1184\pm 0.0007& \hbox{$\MS$ gauge SU(3)$_c$ coupling (5~flavours)} & \hbox {\cite{alpha3}}
\end{array}$$
\caption{\label{tab:values}\em Input values of the SM observables used to fix the SM fundamental parameters
$\lambda, m, y_t, g_2, g_Y$.  The pole top mass, $M_t$, is a naive average of
TeVatron, CMS, ATLAS measurements, all extracted from difficult MonteCarlo modellings of top decay and production
in hadronic collisions.  Furthermore, $M_t$
is also affected by a non-perturbative theoretical uncertainty of order $\Lambda_{\rm QCD}$,
that we quantify as $\pm0.3\GeV$. Throughout the paper we give explicitly the dependence of all physical quantities on $M_t$, and thus the impact of larger theoretical uncertainties on the top mass is always manifest in our results.}
\end{center}
\end{table}%

In the following we adopt strategy ii). 

The quantities of interests are
$\theta = (m^2, \lambda, v, y_t, g_2, g_1)$, i.e. the quadratic and
quartic couplings in the Higgs potential, the vacuum expectation value (vev),
the top Yukawa coupling, the $\SU(2)_L$ and ${\rm U}(1)_Y$ gauge couplings
$g_2$ and $g_Y$
(with $g_1 = \sqrt{5/3} g_Y$ being the hypercharge coupling rewritten in 
SU(5) normalisation), and are directly 
determined in terms of the  pole masses of the Higgs ($M_h$),
of the top ($M_t$), of the $Z$ ($M_Z$), of the $W$ ($M_W$),  the Fermi
constant $G_\mu$  and
the $\MS$ strong coupling $\alpha_3(M_Z)$.
Their input values are listed in Table~\ref{tab:values}.
Then, using eq.~(\ref{eq:g3}), the $\MS$ quantities are obtained. We notice
that the weak-scale values for the 
$\MS$ gauge couplings at the scale $\mub$ are given in terms
of $G_\mu,\, M_W$ and $M_Z$ 
%(see eqs.~(\ref{eq:g5},\ref{eq:G2},\ref{eq:G1}))
and not in terms of the fine structure constant and the weak mixing angle
at the $M_Z$ scale as usually done.

In order to fix the notation we write the classical Higgs potential as
(the subscript $0$ indicates a bare quantity) 
\be 
V_0=- \frac{m^2_0}2 |H_0|^2+\lambda_0 |H_0|^4 \ . 
\label{higgspot}
\ee
The classical Higgs doublet $H_0$ is defined by
\be 
\label{fermioni} 
\bac H_ 0=\left(\bac \chi \\ (v_0+h+i\, \eta )/\sqrt2 \ea \right)\,
\ea 
\ee 
in terms of the physical Higgs field $h$, and of the neutral
and charged would-be Goldstone bosons $\eta$ and $\chi$.  The
renormalisation of the Higgs potential, eq.~(\ref{higgspot}), was
discussed at the one-loop level in~\cite{SZ} and extended at the
two-loop level in~\cite{DDEEGIS}. We refer to these papers for
details. We recall that in ref.~\cite{DDEEGIS} the renormalised vacuum
is identified with the minimum of the radiatively corrected
potential\footnote{ This condition is enforced choosing the tadpole
  counterterm to cancel completely the tadpole graphs.}  and it is
defined through $G_\mu$. Writing the relation between the Fermi
constant and the bare vacuum as 
\be 
\frac{G_\mu}{\sqrt2}=
\frac{1}{2v_0^2}(1+ \Delta r_\ss0)\, ,
\label{Deltar0} 
\ee
one gets
\be
v_\os^2= \frac1{\sqrt2 G_\mu}, \qquad 
\delta v^2_\os = -\frac{\Delta r_\ss0}{\sqrt2 G_\mu} . 
\label{deltav2}
\ee
The quadratic and quartic couplings in the Higgs potential are defined through
$M_h$ via
\be 
m_\os^2= 2 \lambda_\os v_\os^2\ , \qquad M_h^2=2\lambda_\os v_\os^2\ ,
\label{conditions} 
\ee
 or
\be 
\lambda_{\os}=\frac{G_\mu}{\sqrt2}M_h^2 , \qquad m_\os^2 = M_h^2 .
\ee
Writing the counterterm for the quartic Higgs coupling as
\be
\delta \lambda_\os = \delta^{(1)} \lambda_\os + \delta^{(2)} \lambda_\os\ ,
\ee
where the superscript indicates the loop order, one finds
\bea 
\delta^{(1)} \lambda_\os & = & \frac{G_\mu}{\sqrt2} M_h^2 
\left\{\Delta r_\ss0^{(1)}+ 
\frac{1}{M_h^2} \left[\frac{T^{(1)}}{v_\os}+ \delta^{(1)} M_h^2 
\right]\right\}\, ,
\label{eq:lh1}\\
\delta^{(2)} \lambda_\os &=& \frac{G_\mu}{\sqrt2} M_h^2 \left\{\Delta r_\ss0^{(2)}+ 
\frac{1}{M_h^2} \left[\frac{T^{(2)}}{v_\os}
+\delta^{(2)} M_h^2 \right]+ \right. \nn \\ 
&& \left.-  \Delta r_\ss0^{(1)} 
\left(\Delta r_\ss0^{(1)}+ \frac{1}{M_h^2}\left[\frac{3\, T^{(1)}}{2\,v_\os}
+\delta^{(1)} M_h^2  \right]\right)\right\} \, .
\label{eq:lh2}
\eea 
In eqs.~(\ref{eq:lh1})--(\ref{eq:lh2}) $i T$ represents the sum of the tadpole 
diagrams with external leg extracted, and  $ \delta M_a^2$ labels
the mass counterterm for the particle $a$. 

\medskip

Similarly, one finds for the counterterm of the quadratic Higgs coupling in the
potential
\bea 
\delta^{(1)} m_\os^2 & = & 
 3\frac{T^{(1)}}{v_\os}+ \delta^{(1)} M_h^2  \, ,
\label{eq:mh1}\\
\delta^{(2)} m_\os^2 &=& 
3 \frac{T^{(2)}}{v_\os}
+ \delta^{(2)} M_h^2  - \frac{3\, T^{(1)}}{2\,v_\os}\Delta r_\ss0^{(1)}\, .
\label{eq:mh2}
\eea 
The top Yukawa and gauge  couplings are fixed using $M_t$, $M_W$ and $M_Z$ via
\be
M_t = \frac{y_{t_{\os}}}{\sqrt2} v_\os, \qquad M_W^2 = \frac{g^2_{2_{\os}}}4 v^2_\os, \qquad
 M_Z^2 = \frac{g^2_{2_{\os}} +g^2_{Y_{\os}}}4 v_\os^2    ,
\ee
or 
\be
y_{t_{\os}} = 2 \left( \frac{G_\mu}{\sqrt2} M_t^2 \right)^{1/2}   , ~~
g_{2_{\os}} = 2 \left( \sqrt2\,G_\mu \right)^{1/2}   M_W, ~~
g_{Y_{\os}} = 2 \left( \sqrt2\,G_\mu \right)^{1/2}  \sqrt{M_Z^2 -M^2_W}~.
\label{eq:g5}
\ee
The corresponding counterterms are found to be
\bea 
\delta^{(1)} y_{t_{\os}} &=& 2 \left( \frac{G_\mu}{\sqrt2} M_t^2 \right)^{1/2} 
        \left( \frac{\delta^{(1)} M_t}{ M_t} +    \frac{\Delta r_\ss0^{(1)}}2
 \right) ,
\label{eq:yt1} \\
\delta^{(2)} y_{t_{\os}} &=& 2 \left( \frac{G_\mu}{\sqrt2}  M_t^2 \right)^{1/2}  
        \left( \frac{\delta^{(2)} M_t}{  M_t } +  \frac{\Delta r_\ss0^{(2)}}2 -
 \frac{\Delta r_\ss0^{(1)}}2 \left[ \frac{\delta^{(1)} M_t}{ M_t} + 
   \frac{3\,\Delta r_\ss0^{(1)}}4 \right] \right)  ,
\label{eq:yt2}
\eea
for the top Yukawa coupling, and
\bea
\delta^{(1)} g_{2_{\os}} & = &  \left( \sqrt2\,G_\mu \right)^{1/2} M_W \left(
\frac{\delta^{(1)} M_W^2}{M_W^2} +   \Delta r_\ss0^{(1)} \right),
\label{eq:G2} \\
\delta^{(2)} g_{2_{\os}} & = &  \left( \sqrt2\,G_\mu \right)^{1/2} M_W \left(
\frac{\delta^{(2)} M_W^2}{M_W^2} +   \Delta r_\ss0^{(2)} +\right.\nn\\
&&-\left.\frac{\Delta r_\ss0^{(1)}}{2}\left[\frac{\delta^{(1)} M_W^2}{M_W^2} +   \frac{3\Delta r_\ss0^{(1)}}{2}\right]+\frac{1}{4}\left(\frac{\delta^{(1)} M_W^2}{M_W^2}\right)^{2}\right),
\label{eq:G22} 
\eea
for the $\SU(2)_L$ gauge coupling, and
\bea
\delta^{(1)} g_{Y_{\os}} &=&  \left( \sqrt2\,G_\mu \right)^{1/2}  
\sqrt{M_Z^2 -M^2_W}~ \left( 
\frac{\delta^{(1)} M_Z^2 - \delta^{(1)} M_W^2}{M_Z^2 -M^2_W} +  \Delta r_\ss0^{(1)}
\right),
\label{eq:G1}\\
\delta^{(2)} g_{Y_{\os}} &=&  \left( \sqrt2\,G_\mu \right)^{1/2}  
\sqrt{M_Z^2 -M^2_W}~ \left( 
\frac{\delta^{(2)} M_Z^2 - \delta^{(2)} M_W^2}{M_Z^2 -M^2_W} +  \Delta r_\ss0^{(2)}+\right.\nn\\
&&\left.-\frac{\Delta r_\ss0^{(1)}}{2}\left[\frac{\delta^{(1)} M_Z^2 - \delta^{(1)} M_W^2}{M_Z^2 -M^2_W} +  \frac{3\Delta r_\ss0^{(1)}}{2}\right]+\frac{1}{4}\left(\frac{\delta^{(1)} M_Z^2 - \delta^{(1)} M_W^2}{M_Z^2 -M^2_W} \right)^{2}
\right)~. 
\label{eq:G12}
\eea
for the hypercharge gauge coupling.
%Computations are performed in the linear $R_\xi$ gauge with $\xi=1$.

\subsection{Two-loop correction to the Higgs quartic coupling} 
\label{sec2.1}
The $\MS$ Higgs quartic coupling is given by
\be
\lambda(\mub) = 
\frac{G_\mu}{\sqrt2}M_h^2  + \lambda^{(1)}(\mub) +  \lambda^{(2)}(\mub),
\label{eq:lambda}
\ee
with
\bea
 \lambda^{(1)}(\mub) &=& - \left. \delta^{(1)} \lambda_\os \right|_{\rm fin} 
\label{eq:lmu1} \nn\, ,\\
 \lambda^{(2)}(\mub) &=& -\left. \delta^{(2)} \lambda_\os \right|_{\rm fin} 
+ \Delta_{\lambda}~. 
\label{eq:lmu2}
\eea
The one-loop contribution in eq.~(\ref{eq:lambda}), $\lambda^{(1)}$,
is given by the finite part of eq.~(\ref{eq:lh1}). 
Concerning the two-loop part, $\lambda^{(2)} (\mub)$, the  QCD corrections 
were presented in refs.~\cite{Bezrukov:2012sa,DDEEGIS},
and the  two-loop electroweak (EW) part, $\lambda_{\rm EW}^{(2)}(\mub)$,
 was computed in 
ref.~\cite{DDEEGIS} in the so-called gauge-less limit of the SM, in which the 
electroweak gauge interactions are switched off.
The main advantage of this limit results in a simplified evaluation
of  $ \Delta r_\ss0^{(2)}$. The computation of the two-loop
EW part in the full SM requires instead the
complete evaluation of this quantity and we outline here the 
derivation of $\lambda^{(2)}_{\rm EW}(\mub)$ starting from the term 
$ \Delta r_\ss0^{(2)}$ in $\delta^{(2)} \lambda_\os$.
 
\medskip

We recall that the Fermi constant 
is defined in terms of the muon lifetime $\tau_\mu$ as computed
in the 4-fermion $V-A$ Fermi  theory supplemented by QED interactions. 
We extract $G_\mu$ from  $\tau_\mu$ via
\be 
\label{eq:taumu}
\frac{1}{\tau_\mu} = \frac{G_\mu^2 m_\mu^5}{192\pi^3} F(\frac{m_e^2}{m_\mu^2}) 
(1 + \Delta q)(1+\frac{3m_\mu^2}{5M_W^2}) \ ,
\ee
where $F(\rho)=1-8\rho+8\rho^3-\rho^4-12\rho^2\ln\rho=0.9981295$ 
(for $\rho=m_e^2/m_\mu^2$) is the phase space factor and
$\Delta q=  \Delta q^{(1)} + \Delta q^{(2)}=(-4.234+0.036)\times10^{-3}$
are the QED corrections computed at one~\cite{Kinoshita:1958ru} and two 
loops~\cite{Deltaq}.
From the measurement $\tau_\mu = (2196980.3\pm2.2)\,{\rm ps}$~\cite{MuLan} 
we find
$G_\mu = 1.1663781(6)~10^{-5}/\GeV^2$.
This is $1\sigma$ lower than the value quoted in~\cite{MuLan} because we do not 
follow the convention of including
in the definition of $G_\mu$ itself the last term of\eq{taumu}, which is the
contribution from dimension-8 SM operators. 

\bigskip

The computation
of $ \Delta r_\ss0$ requires  the subtraction of the QED corrections
by matching the result in the SM with that in the Fermi theory. However,
it is well known that the Fermi theory is renormalisable to all order in the
electromagnetic interaction but to lowest order in $G_\mu$ due to a Ward
identity that becomes manifest if the 4-fermion interaction is rewritten 
via a Fierz transformation in the  ``charge retention order''. 
As a consequence, in the
limit of neglecting the fermion masses,  $ \Delta r_\ss0$ as computed in the
Fermi theory vanishes and we are just left with the calculation in the SM\footnote{
We explicitly verified that $ \Delta r_\ss0$ vanishes when computed in the Fermi
theory.}.

Starting from eq.~(\ref{Deltar0}) we write $\Delta r_\ss0$ as a sum of
different terms:
\be 
\Delta r_\ss0= V_\ssW- \frac{A_{\ssW \ssW}}{M^2_{\ssW \ss0}} + 
2\, v_0^2 {\cal B}_\ssW + {\cal E}+{\cal M}\ , 
\label{Deltar-expansion}
\ee
where $M_{\ssW \ss0}$ is the bare W mass; $A_{\ssW \ssW}$ is the $W$
self-energy at zero momentum, $A_{\ssW \ssW}=A_{\ssW \ssW}(0)$;
$V_\ssW$ is the vertex contribution; ${\cal B}_\ssW$ is the box
contribution; ${\cal E}$ is the term due to the renormalisation of the
external legs; ${\cal M}$ is the mixed contribution due to product of
different objects among $V_\ssW$, $A_{\ssW\ssW}$, ${\cal B}_\ssW$ and
${\cal E}$ (see below for an explicit expression at two-loops). All
quantities in eq.~(\ref{Deltar-expansion}) are computed at zero
external momenta. We point out that in the right-hand side of 
eq.~(\ref{Deltar-expansion}) no tadpole
contribution is included because  of our choice of identifying the
renormalised vacuum with the minimum of the radiatively corrected potential. 
As a consequence $\Delta r_\ss0$ is a gauge-dependent quantity.

From eq.~(\ref{Deltar-expansion}) the one-loop  term is  
given by:
\beq 
\Delta r_\ss0^{(1)}=V_\ssW^{(1)} - \frac{A_{\ssW \ssW}^{(1)}}{M^2_{W}}  + 
\frac{\sqrt{2}}{G_\mu} \, {\cal B}_\ssW^{(1)} + {\cal E}^{(1)}  \ ,
\label{dr1}
\eeq
where  we have used that ${\cal M}^{(1)}=0$, while at two-loops
\bea 
\Delta r_\ss0^{(2)} &=& V_\ssW^{(2)}- \frac{A_{\ssW \ssW}^{(2)}}{M^2_{W}} 
+ \sqrt2\, \frac{{\cal B}^{(2)}_\ssW}{G_\mu} + {\cal E}^{(2)}+{\cal M}^{(2)}+\,
 \nn \\
&& -\delta^{(1)}M^2_W \,\frac{A_{\ssW \ssW}^{(1)}}{M^4_{W}} + \frac{\sqrt{2}}{G_\mu}
\, {\cal B}_\ssW^{(1)}\left(V_\ssW^{(1)}- \frac{A_{\ssW \ssW}^{(1)}}{M^2_{W}} 
+ \sqrt2\, \frac{{\cal B}^{(1)}_\ssW}{G_\mu} + {\cal E}^{(1)}\right)\ .
\label{dr2}
\eea
 Here
\be
\delta^{(1)}M^2_W =  \mbox{Re}\, \Pi_{\ssW \ssW}(M_W^2)
\ee 
with $ \Pi_{\ssW \ssW}(M_W^2)$  the $W$ boson  self-energy  evaluated at external 
momentum  equal to $M_W$, and 
 \be
{\cal M}^{(2)} =\frac{\sqrt2}{G_\mu}{\cal E}^{(1)}\,{\cal B}^{(1)}_\ssW
+ \sum_{i< j}{\cal E}^{(1)}_i{\cal E}^{(1)}_j+{\cal E}^{(1)} V^{(1)}-
\left({\cal E}^{(1)}+V^{(1)}\right)\frac{A_{\ssW\ssW}^{(1)}}{M_W^2}~. 
\label{mixedSM}
\ee
The indices $i,j$  in eq.~(\ref{mixedSM}) label
the different species in the muon decay: $\mu$, $e$, $\nu_\mu$ and $\nu_e$
with the sum that runs over $i< j$ because the terms with $i=j$ are included in 
${\cal E}^{(2)}$.

\smallskip

We recall that $\Delta r_\ss0$ is an infrared (IR) safe quantity but not 
ultraviolet (UV) finite. However,  the ${\cal E}$ and ${\cal B}_W$ terms
in eq.~(\ref{dr1}) and (\ref{dr2}) contain IR-divergent contributions from 
photon diagrams. To separate the UV-divergent terms from the IR ones we 
regulated the latter giving a small  mass to the photon. We then explicitly 
verified the cancellation of  all IR divergent contributions.

\bigskip

The other proper two-loop contributions  to $\lambda^{(2)}_{\rm EW} (\mub)$
are the two-loop tadpole diagrams and the two-loop Higgs boson 
mass counterterm. The Higgs mass counterterm, not taking into account 
negligible width effects, is given by  
\be
\delta M_h^2 = \mbox{Re}\, \Pi_{hh}(M_h^2)
\label{eq:mass}
\ee
with $ \Pi_{hh}(M_h^2)$  the Higgs  self-energy  evaluated at external momentum 
equal to $M_h$. The Higgs mass counterterm as defined in eq.~(\ref{eq:mass})
is a gauge-dependent quantity. Yet, 
as proved at the beginning of section~\ref{sec:strategy}, $\lambda(\mub)$ is
a gauge-invariant object.

The diagrams contributing to $\delta^{(2)} \lambda_\os$
were generated using the Mathematica package {\sc Feynarts}~\cite{Feynart}.
The reduction of the two-loop diagrams to scalar integrals was done using 
the code {\sc   Tarcer} \cite{Mertig:1998vk}  that uses the 
Tarasov's algorithm \cite{Tarasov} and it is now part of the {\sc
 Feyncalc}~\cite{Feyncalc} package. In order to extract the 
$V_W$ and ${\cal B}_W$ terms in $\Delta r_\ss0$ from the relevant
diagrams we used the  projector presented in ref.~\cite{Awramik:2002vu}.
The two-loop self-energy diagrams at external momenta different from zero 
were reduced to the set of loop-integral basis functions introduced
in ref.\cite{Martinint}. The evaluation of the basis functions was 
done numerically using the code {\sc TSIL}~\cite{MartinTSIL}.

\medskip

The two loop correction to $\lambda$ is the sum of a QCD term and of an electroweak (EW) term.
The QCD correction $\lambda_{\rm QCD}^{(2)}(\mub)$ is reported as an approximated formula in eq.~(47) of~\cite{DDEEGIS}. 
For simplicity here we present it also in a numerical form:
\be
\lambda_{\rm QCD}^{(2)}(\mub=M_{t}) = 
\frac{g_3^{2}}{(4\pi)^4}\bigg[
- 23.88 +0.12 \left(\frac{M_h}{\rm GeV} - 125\right) - 0.64
\left(\frac{M_t}{\rm GeV} - 173\right)\bigg].
\ee
The result for  $\lambda_{\rm EW}^{(2)}(\mub)$  is  too long to be
displayed explicitly. 
Here we present it in a numerical form valid around the measured values of $M_{h}$ and $M_{t}$. Using the inputs in Table \ref{tab:values} we find
\be
\lambda_{\rm EW}^{(2)}(\mub=M_{t}) = 
\frac{1}{(4\pi)^4}\bigg[
- 9.45 -0.12 \left(\frac{M_h}{\rm GeV} - 125\right) - 0.21
\left(\frac{M_t}{\rm GeV} - 173\right)\bigg].
\label{eq:lh21}
\ee
The numerical expression in eq.~(\ref{eq:lh21}) is accidentally very close to 
the gaugeless limit of the SM presented 
in eq.~(2.45) of~\cite{DDEEGIS}.
Furthermore, as a check of our result, we verified that in the (physically irrelevant) limit $M_h=0$, 
it agrees with an independent computation of $\lambda^{(2)}$  performed
using the known results for the two-loop effective potential in the Landau 
gauge.

\subsection{Two-loop correction to the Higgs mass term} 
\label{sec2.2}
The  result for the mass term in the Higgs 
potential can be easily obtained from that on $\lambda(\mub)$. We write
\be
m^2(\mub) = M_h^2 + \delta^{(1)} m^{2}(\mub) + \delta^{(2)} m^{2}(\mub),
\label{eq:m2}
\ee
with
\bea
\delta^{(1)} m^{2}(\mub) &=&  - \left. \delta^{(1)} m^2_\os \right|_{\rm fin} \nn\ ,\\
\delta^{(2)} m^{2}(\mub) & = & - \left. \delta^{(2)} m^2_\os \right|_{\rm fin} 
+ \Delta_{m^2}~. 
\label{eq:m21}
\eea
The one-loop contribution in eq.~(\ref{eq:m2}), $\delta^{(1)} m^{2}(\mub)$,
is given by the finite part of eq.~(\ref{eq:mh1}). 
The two-loop corrections in eq.~(\ref{eq:m2}), 
$\delta^{(2)} m^{2} (\mub)$, can be divided into a QCD contribution plus an EW contribution.

The QCD contribution,  $\delta^{(2)}_{\rm QCD} m^{2} (\mub)$,
can be obtained evaluating the relevant diagrams via a Taylor series in
$x_{ht} \equiv M_h^2/M_t^2$ up to fourth order
\bea
\delta^{(2)}_{\rm QCD} m^{2}  (\mub) &=& \frac{G_\mu M_t^4}{\sqrt{2}(4 \pi)^4}\, N_c\, C_F
g_3^2 \bigg[  -96 +
(41 -12 \ln^2\frac{M_t^2}{\mub^2} +12 \ln^2\frac{M_t^2}{\mub^2} ) x_{ht}+
\nn\\
& & \qquad\qquad\qquad\qquad
+  \frac{122 }{135} x^2_{ht} +
 \frac{1223 }{3150} x_{ht}^3 +  \frac{43123 }{661500} x_{ht}^4\bigg],\\
 \riga{where $N_c$ and $C_F$ are colour factors ($N_c=3,\, C_F= 4/3$), such that it is numerically approximated as}\\
\delta^{(2)}_{\rm QCD} m^{2}  (\mub=M_t)  &=& \frac{g_{3}^{2}M_{h}^{2}}{(4\pi)^4}\bigg[
-140.50 +2.91 \left(\frac{M_h}{\rm GeV} - 125\right) -3.72
\left(\frac{M_t}{\rm GeV} - 173\right)\bigg].
\eea
The two-loop EW part,  $\delta^{(2)}_{\rm EW} m^{2} (\mub)$,
can be obtained as a byproduct of the calculation of 
$\lambda_{\rm EW}^{(2)}(\mub)$. Also in this case the result is too long to
be displayed and we present an interpolating formula. Using the inputs in 
table~\ref{tab:values} we find
\be
\delta_{\rm EW}^{(2)} m^2 (\mub=M_{t})= \frac{M_{h}^{2}}{(4\pi)^4}\bigg[
-149.47 +2.54 \left(\frac{M_h}{\rm GeV} - 125\right) - 4.69
\left(\frac{M_t}{\rm GeV} - 173\right)\bigg].
\label{eq:m22}
\ee

%We also report the interpolating formula for the QCD contribution to $\delta^{(2)}m^{2}$:
%\be
%\delta_{\rm QCD}^{(2)} m^2 (\mub=M_{t})
%\ee
%These formulae are valid in the interval $M_{h}=125\pm0.2~\GeV$ and $M_{t}=173\pm0.2~\GeV$.
%\xxx{ma che vuol dire valid?  E poi l'intervallo giusto non  quello}

\subsection{Two loop correction to the  top Yukawa coupling}
\label{sec2.3}
The $\MS$ top Yukawa coupling is given by
\be
y_t(\mub) = 
2 \left(\frac{G_\mu}{\sqrt2}  M_t^2\right)^{1/2} + y_t^{(1)}(\mub)
+ y_t^{(2)}(\mub),
\label{eq:topYuk}
\ee
with 
\bea
 y_t^{(1)}(\mub) &=&   - \left. \delta^{(1)} y_{t{_\os}} \right|_{\rm fin} \nn\ ,\\
 y_t^{(2)}(\mub) &=& -  \left. \delta^{(2)} y_{t_{\os}} \right|_{\rm fin} + 
\Delta_{y_t}~. 
\eea
According to eqs.~(\ref{eq:yt1})--(\ref{eq:yt2}) the corrections to the tree-level
value of $y_t$ are given in terms of $ \Delta r_\ss0$ and the top mass
counterterm. Regarding the latter, a general discussion  on the mass 
counterterm for unstable  fermions in parity-nonconserving theories is 
presented in  ref.~\cite{Kniehl:2008cj}. Writing the fermion self-energy as
\bea
\Sigma(p) & =& \Sigma_1 (p) + \Sigma_2(p) \gamma_5, \nn\\
\Sigma_{1,2} (p) &=& \psl B_{1,2} (p^2) + m_0 A_{1,2} (p^2) ,
\label{KS}
\eea 
the fermion propagator is given by
\be 
i S(p) = \frac{i}{\psl - m_0 - \Sigma(p)} = 
\frac{i}{\psl - m_0 - \Sigma_{\rm eff} (p)} 
\left[1 - \frac{\Sigma_2 (p)}{\psl -\Sigma_1(p) + m_0 [1 + 2 A_1(p^2)]} \gamma_5
\right]~,
\label{eq:prop}
\ee
where $m_0$ is the bare fermion mass and
\be
\Sigma_{\rm eff} (p) = \Sigma_1 (p) +
\frac{ \Sigma_2 (p) \left[ \Sigma_2 (p) -2 m_0 A_2 (p^2)\right]}{\psl
          -\Sigma_1(p) + m_0[1 + 2 A_1(p^2)]}~.
\ee
Identifying the position $\psl = \tilde M$ of the complex pole in eq.~(\ref{eq:prop})
by
\be
\tilde M = m_0 + \Sigma_{\rm eff}(\tilde M)
\ee
and parametrizing $\tilde M  = M -i \Gamma/2$ with $M$ the pole mass of
the unstable fermion and $\Gamma$ its width,
the mass counterterm for the unstable fermion is found to be
\be
\delta M   =  \mbox{Re}\, \Sigma_{\rm eff} (\tilde M)~.
\ee
Specialising the above discussion to the top, we find, including up to
two-loop contributions,
\be
\delta M_t = \mbox{Re}\, \left[\Sigma_1 (\widetilde{M}_t) +
\frac{ \Sigma_2 (M_t) \left[ \Sigma_2 (M_t) -2 M_t A_2 (M_t^2)\right]}{2 M_t}
\right]
\label{eq:Mt}
\ee
with $\widetilde{M}_t = M_t -i \Gamma_t/2$.
The mass counterterm defined in
eq.~(\ref{eq:Mt})  is expressed in terms of the self-energy diagrams only,
without including the tadpole  contribution. While this definition follows 
from our choice of identifying the
renormalised vacuum with the minimum of the radiatively corrected potential, it
gives rise to a $\delta M_t$ that is gauge-dependent and, as a consequence,
in this framework, the $\MS$ top mass, $M_t(\mub)$, is a 
gauge-dependent  quantity. However, a $\MS$ mass is not a 
physical quantity nor a Lagrangian parameter
and  therefore the requirement of gauge-invariance is not 
mandatory. A  gauge-invariant definition of $M_t(\mub)$ can be obtained by 
including the tadpole contribution in the mass 
counterterm~\cite{Hempfling:1994ar}.
However, with this choice the relation between the pole and $\MS$
masses of top quark acquires a very large electroweak 
correction~\cite{Jegerlehner:2012kn}.
The top Yukawa coupling computed in this paper is a parameter of the Lagrangian,
and thereby does not suffer of these problems.

Concerning the two-loop contributions in eq.~(\ref{eq:topYuk}), we have computed
the QCD corrections to the one-loop term and the two-loop EW contribution.

These contributions are too long to
be displayed explicitly, and we report them as interpolating formul\ae.
Using the inputs in  table \ref{tab:values} we find
\bea\nonumber
y^{(2)}_{t}(\mub=M_{t}) &=& \frac{1}{(4\pi)^4}\bigg[
6.48 -0.01 \left(\frac{M_h}{\rm GeV} - 125\right) + 0.18 
\left(\frac{M_t}{\rm GeV} - 173\right)\bigg]+\\
 &&+ \frac{g_{3}^{2}}{(4\pi)^4}\bigg[
-7.53 +0.09\left(\frac{M_h}{\rm GeV} - 125\right) - 0.23 
\left(\frac{M_t}{\rm GeV} - 173\right)\bigg]+\nonumber  \\
 &&+ \frac{g_{3}^{4}}{(4\pi)^4}\bigg[
-145.08 - 0.84 
\left(\frac{M_t}{\rm GeV} - 173\right)\bigg],
\eea
where the last term is the well known pure QCD contribution; the second term is the mixed QCD/EW contribution that agrees with~\cite{Bezrukov:2012sa};
 the first term is the pure EW contribution computed in this paper for the first time.

\subsection{Two-loop correction to weak and hypercharge gauge couplings} 
\label{sec2.4}
The $g_{2}$ and $g_{Y}$ gauge couplings are given by 

\bea
g_{2}(\mub) &=&2 (\sqrt{2}G_{\mu})^{1/2} M_{W}+ g_{2}^{(1)}(\mub) + g_{2}^{(2)}(\mub),\nn\\
g_{Y}(\mub) &=&2 (\sqrt{2}G_{\mu})^{1/2} \sqrt{M_{Z}^{2}-M_{W}^{2}}+ g_{Y}^{(1)}(\mub) + g_{Y}^{(2)}(\mub),
\label{eq:g12}
\eea
with
\bea
&& g_2^{(1)}(\mub) =   - \left. \delta^{(1)} g_{2{_\os}} \right|_{\rm fin}\ , ~~~~
 g_2^{(2)}(\mub) = -  \left. \delta^{(2)} g_{2_{\os}} \right|_{\rm fin} + 
\Delta_{g_2}~\nn\ , \\
&& g_Y^{(1)}(\mub) =   - \left. \delta^{(1)} g_{Y{_\os}} \right|_{\rm fin}\ , ~~~~
 g_Y^{(2)}(\mub) = -  \left. \delta^{(2)} g_{Y_{\os}} \right|_{\rm fin} + 
\Delta_{g_y}~.
\label{eq:deltag12}
\eea
The one-loop contributions in eq.~(\ref{eq:deltag12}) are given by the finite part of eqs.~(\ref{eq:G2})~(\ref{eq:G1}). 
Also in this case the results of the two-loop corrections in eq.~(\ref{eq:deltag12}) are too long to be displayed and we present them with interpolating formulas. Using the inputs in 
table~\ref{tab:values} we find
\bea\nn
g^{(2)}_{2} (\mub=M_{t})&=& \frac{1}{(4\pi)^4}\bigg[
2.25 +0.01 \left(\frac{M_h}{\rm GeV} - 125\right) +0.01
\left(\frac{M_t}{\rm GeV} - 173\right)\bigg]+ \\
&&+\frac{g_{3}^{2}}{(4\pi)^4}\bigg[
3.00 +0.01
\left(\frac{M_t}{\rm GeV} - 173\right)\bigg].
\label{eq:g2MS}
\eea
and
\bea\nn
g^{(2)}_{Y} (\mub=M_{t})&=& \frac{1}{(4\pi)^4}\bigg[
-7.55 - 0.01 \left(\frac{M_h}{\rm GeV} - 125\right) - 0.11
\left(\frac{M_t}{\rm GeV} - 173\right)\bigg]+ \\
&&+\frac{g_{3}^{2}}{(4\pi)^4}\bigg[
-14.66 -0.14
\left(\frac{M_t}{\rm GeV} - 173\right)\bigg]
\label{eq:g2MS}
\eea

\section{SM couplings at the electroweak scale}\label{sec:inputs}
In this section we give practical results for the SM  parameters 
$\theta=\{\lambda, m^2, y_t, g_2, g_Y\}$
computed in terms of the observables $M_h,M_t,M_W,M_Z, G_\mu$ and $\alpha_3(M_Z)$, whose measured values are listed in table~\ref{tab:values}.
Each $\MS$ parameter $\theta$ is expanded in loops as
\beq \theta = \theta^{(0)}+\theta^{(1)}+\theta^{(2)}+\cdots\label{eq:exp}
\eeq
where
\begin{enumerate}
\item the tree-level values $\theta^{(0)}$ are listed in table~\ref{tab:status};
\item the one-loop corrections $\theta^{(1)}$ are analytically given in appendix~\ref{1loop};
\item the two-loop corrections $\theta^{(2)}$ are computed in section~\ref{sec:strategy}.
\end{enumerate}
After combining these corrections, we give in the following the numerical values
for the SM parameters renormalised at the top pole mass $M_t$ in the
$\MS$ scheme.

\begin{table}
%$$\begin{array}{c|ccccc}
% \mub=M_t \xxx{OLD}& \lambda  & y_t & g_2 & g_Y & m/\GeV \\ \hline
% \text{LO} & 0.13023 & 0.99571 & 0.65294 & 0.34972 & 125.66 \\
% \text{NLO} & 0.12879 & 0.95096 & 0.64754 & 0.35940 & 132.85 \\
% \text{NNLO} & 0.12710 & 0.93989 &  0.6483&  0.3587& 132.03 \\
%\end{array}
%$$
$$\begin{array}{c|ccccc}
 \mub=M_t & \lambda  & y_t & g_2 & g_Y & m/\GeV \\ \hline
 \text{LO} & 0.12917 & 0.99561 & 0.65294 & 0.34972 & 125.15 \\
 \text{NLO} & 0.12774 & 0.95113 & 0.64754 & 0.35940 & 132.37 \\
 \text{NNLO} & 0.12604 & 0.94018 & 0.64779 & 0.35830 & 131.55 \\
\end{array}
$$
\caption{\em Values of the  fundamental SM parameters computed at tree level, one loop, two loops
in the $\MS$ scheme and renormalised at $\mub=M_t$ for the central values of the measurements listed in table~\ref{tab:values}.
\label{tab:123}}
\end{table}
 
 \subsection{The Higgs quartic coupling}

For the Higgs quartic coupling, defined by writing the SM potential as $V = -\frac{1}{2} m^2 |H|^2+\lambda |H|^4$, 
we find
\beq \lambda(\mub=M_t) = 0.12604+0.00206\left( \frac{M_h}{\GeV}-\Mhexp \right)-0.00004 \left( \frac{M_t}{\GeV}-\Mtexp \right)\pm
{0.00030}_{\rm th}~.  \eeq 
%\xxx{ADD 3 LOOP? The pure 3-loop QCD term is $-0.00028$}
The dependence on $M_t$ is  small because $\lambda$ is renormalised at $M_t$ itself.
Here and below the theoretical uncertainty is estimated from the dependence on $\bar\mu$ 
(varied around $M_t$ by one order of magnitude) of the higher-order unknown 3 loop corrections.
Such dependence is extracted from the known SM RGE at 3 loops (as summarized in appendix~\ref{SM-RGE}).\footnote{Recently the calculation of the three-loop SM 
effective potential at leading order in strong and top Yukawa couplings has 
appeared \cite{Martin:2013gka}. 
The resulting three-loop contributions to the Higgs quartic coupling are within our estimated error.
Combining the three loop effective potential with  1 and 2-loop renormalizations, we
extract the 3-loop pure QCD correction to $\lambda(\bar\mu=M_t)$ in the  limit $M_h,M_W,M_Z\ll M_t$:
$$ \lambda^{(3)}_{\rm QCD}=-\frac{8}{135}
\frac{g_3^4 G_\mu^2 M_t^4}{(4\pi)^6}\bigg[176\pi^4 + 240\pi^2 (3+ 4\ln^2 2 +6\ln 2)+15
(607-64\ln^42 - 1536 \hbox{Li}_4(\frac12) + 576 \zeta(3)
\bigg].$$
}

%\xxx{We can extract the 3-loop QCD correction in the quasi-gauge-less limit 
%from the 3-loop SM potential:
%\beq \lambda^{(3)}_{\rm QCD}=\frac{g_3^4}{(4\pi)^6}\bigg[-326.-
%12 \left(\frac{M_t}{\rm GeV} - 173\right)\bigg]\eeq}

 \subsection{The Higgs mass term}
 For the mass term of the Higgs doublet in the SM Lagrangian (normalised such that $m=M_h$ at tree level)  we find 
 \be \frac{m(\mub=M_t)}{\GeV}=131.55+0.94 \left( \frac{M_h}{\GeV}-\Mhexp \right)+0.17\left( \frac{M_t}{\GeV}-\Mtexp \right) \pm 0.15_{\rm th}.
 \eeq
 
\subsection{The top Yukawa coupling}

For the top Yukawa coupling we get
\bea
y_t(\mub=M_t) &=& 0.93690 +0.00556 \left( \frac{M_t}{\GeV}-\Mtexp \right)+\\
&& -0.00042 \asdiff   \pm{0.00050}_{\rm th}~.  \nonumber
\label{eq:ht_ew}
\eea
The central value differs from the NNLO value in table~\ref{tab:123} because we include here also the NNNLO (3 loop)
pure QCD effect~\cite{Chetyrkin:1999ys,Chetyrkin:1999qi,Melnikov:2000qh}. 
The central value would shift to $0.93446$ if the estimated 4-loop QCD correction of~\cite{Kataev} were added.
The estimated theoretical uncertainty does not take into account the non-perturbative theoretical uncertainty of order $\Lambda_{\rm QCD}$ in the definition of $M_t$.

\subsection{The weak gauge couplings}
For the weak gauge couplings $g_2$ and $g_Y$ computed at NNLO accuracy in terms of $M_W$ and $M_Z$ we find
\begin{eqnarray}
g_2(\bar\mu=M_t) &=& 0.64779 +0.00004 \left( \frac{M_t}{\GeV}-\Mtexp \right)+ 0.00011 \MWdiff
, \\
g_Y(\bar\mu=M_t) &=&0.35830 +0.00011 \left( \frac{M_t}{\GeV}-\Mtexp \right)- 0.00020 \MWdiff,
\end{eqnarray}
where the adopted value for $M_W$ and its experimental error are reported in table~\ref{tab:values}.

%For the moment, we just quote the best-fit values from~\cite{alpha}
%\beq
%\alpha_Y^{-1}(M_Z) =98.35\pm 0.013 ,\qquad 
%\alpha_2^{-1}(M_Z) = 29.587\pm 0.008 .
%\eeq
%Table~\ref{tab:123} reports their central values extrapolated at the renormalisation scale $\mub=M_t$ using the SM two-loop RGE equations.

\subsection{The strong gauge coupling}
Table~\ref{tab:values} contains the value of $\alpha_3(M_Z)$, as extracted from the global fit of~\cite{alpha3} in the effective SM with 5 flavours.
Including RG running from $M_Z$ to $M_t$
at 4 loops in QCD and at 2 loops in the electroweak gauge interactions,
and 3 loop QCD matching at $M_t$ to the full SM with 6 flavours, we get
\beq g_3(\mub=M_t)  =   %1.1645w
1.1666 
+0.00314\asdiff -0.00046 \left( \frac{M_t}{\GeV}-\Mtexp \right).\eeq

The SM parameters can be renormalised to any other desired energy by solving the SM renormalisation group equations summarised in appendix~\ref{SM-RGE}.
For completeness, we  include in the one- and two-loop
RG equations the contributions of the small bottom and tau Yukawa
couplings, as computed from the $\MS$ $b$-quark mass,
$m_b(m_b)=4.2\GeV$, and from $M_\tau=1.777\GeV$.  Within the $\MS$
scheme $\beta$ functions are gauge-independent~\cite{wil}; similarly
the $\MS$ parameters are gauge independent too.

\section{Extrapolation of the SM up to the Planck scale}
\label{sec:Planck}

The most puzzling and intriguing outcome of the Higgs discovery has
been the finding that $M_h$ lies very close to the boundary between
stability and metastability regions. This result is the main
motivation for our refined NNLO calculation of the SM Higgs potential
at large field values. Indeed, the special Higgs mass found by ATLAS
and CMS is so close to criticality that any statement about stability
or metastability of the EW vacuum requires a careful analysis of
theoretical and experimental errors.  The discovered proximity to
criticality also naturally stimulates many theoretical speculations on
its possible hidden significance or on special matching conditions at
very high energy scales. In the rest of the paper, we will explore the
implications of our improved computation of the Higgs quartic coupling
extrapolated to very high scales.

\begin{figure}[t]
$$\includegraphics[width=0.63\textwidth]{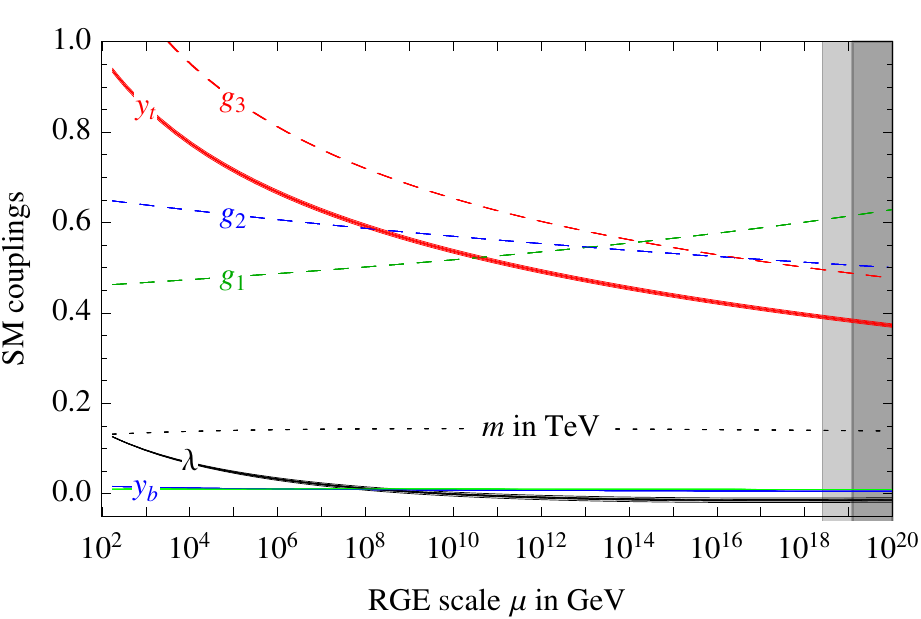}
$$
\caption{\em 
Renormalisation of the SM  gauge couplings 
$g_1=\sqrt{5/3}g_Y, g_2, ~g_3$, of the top, bottom and $\tau$ couplings ($y_t$, $y_b$, $y_\tau$),
 of the Higgs quartic coupling $\lambda$ and of the Higgs mass parameter $m$.
All parameters are defined in the  $\MS$ scheme.
We include two-loop thresholds at the weak scale and three-loop RG equations.
The thickness indicates the $\pm1\sigma$ uncertainties in $M_t,M_h,\alpha_3$.}
\label{fig:run1} 
\end{figure}

\subsection{SM couplings at the Planck scale}\label{sec:scc}

The first issue we want to address concerns the size of the SM
coupling constants. When we try to extract information from the values
of the coupling constants, it is reasonable to analyse their values
not at the weak scale, but at some high-energy scale where we believe
the SM matches onto some extended theory.  So, using our NNLO results, we extrapolate the SM couplings
from their weak-scale values (as determined in section~\ref{sec:inputs}) to higher
energies.  
%  Once we run the SM
%couplings up to a large cut-off scale (as shown in fig.\fig{run1}) we
%discover the following pattern.

The evolution of the SM couplings up to a large cut-off scale is shown in fig.\fig{run1}. 
 At the Planck mass, we find the following values of the SM parameters:
\begin{eqnsystem}{sys:Pl}
g_1(\mpl )&=& 0.6154+0.0003 \Mtdiff -0.0006 \MWdiff \\
g_2(\mpl )&=& 0.5055 \\
g_3(\mpl)&=& 0.4873+ 0.0002 \asdiff \\
y_t(\mpl )&=& 0.3825+ 0.0051\Mtdiff - 0.0021\asdiff  \\
\label{eq:lammp}
\lambda (\mpl ) &=&  -0.0143- 0.0066\Mtdiff +\\   \nonumber
&&+0.0018\asdiff +0.0029\Mhdiff \\
m(\mpl) &=& 129.4\GeV +1.6\GeV\Mhdiff+
\\&&-0.25\GeV\Mtdiff+0.05\GeV\asdiff\nonumber
\end{eqnsystem}
All Yukawa couplings, other than the one of the top quark,
  are very small. This is the well-known flavour problem of the SM,
  which will not be investigated in this paper.

The three gauge couplings and the top Yukawa coupling remain perturbative and 
  are fairly weak at high energy, becoming roughly equal in the
  vicinity of the Planck mass. The near equality of the gauge
  couplings may be viewed as an indicator of an underlying grand
  unification even within the simple SM, once we allow for threshold
  corrections of the order of 10\% around a scale of about
  $10^{16}$~GeV (of course, in the spirit of this paper, we are
  disregarding the acute naturalness problem). It is amusing to note
  that the ordering of the coupling constants at low energy is
  completely overturned at high energy. The (properly normalised)
  hypercharge coupling $g_1$ becomes the largest coupling in the SM
  already at scales of about $10^{14}$~GeV, and the weak coupling
  $g_2$ overcomes the strong coupling at about $10^{16}$~GeV. The top
  Yukawa becomes smaller than any of the gauge couplings at scales
  larger than about $10^{10}$~GeV.

\begin{figure}[p]
$$\includegraphics[width=0.45\textwidth]{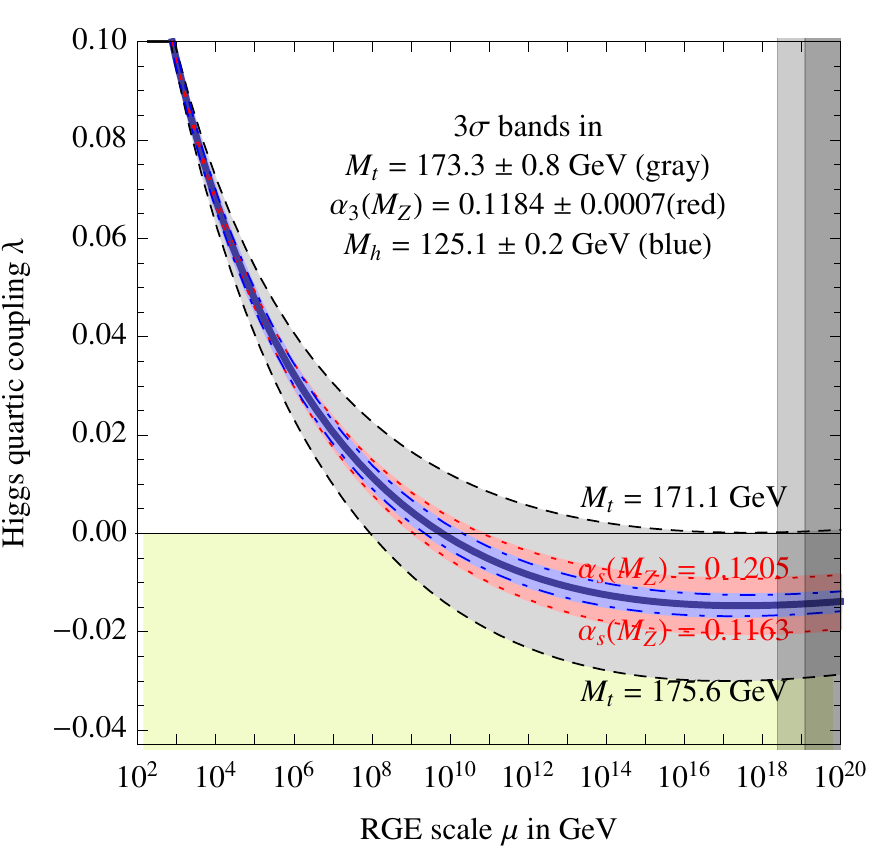}\qquad
\includegraphics[width=0.46\textwidth]{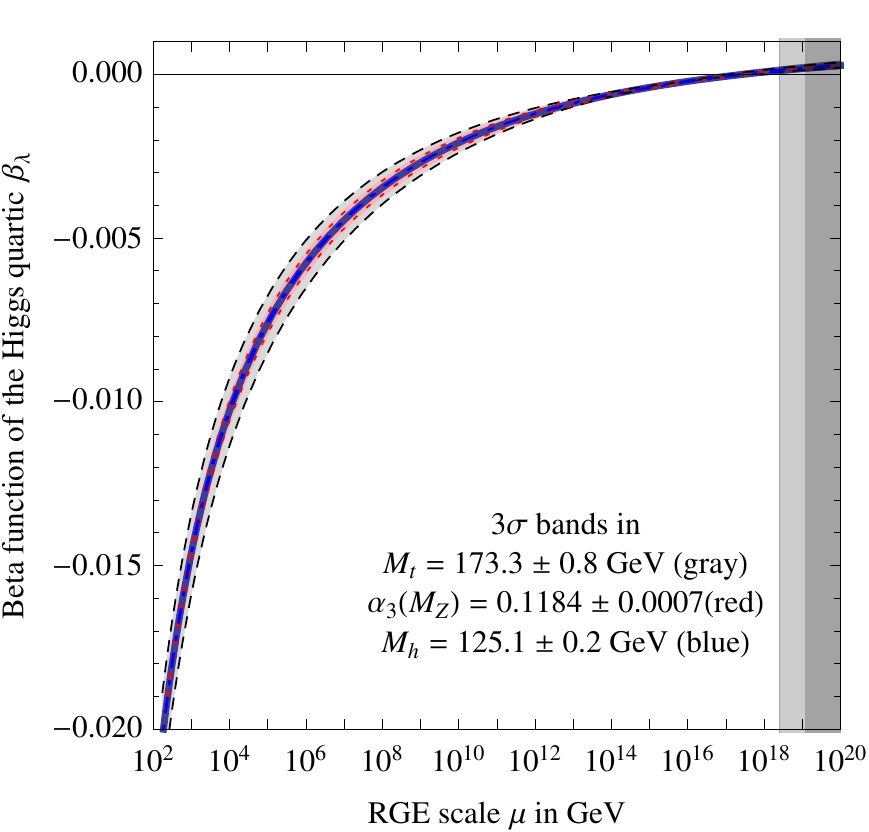}$$
$$\includegraphics[width=0.45\textwidth]{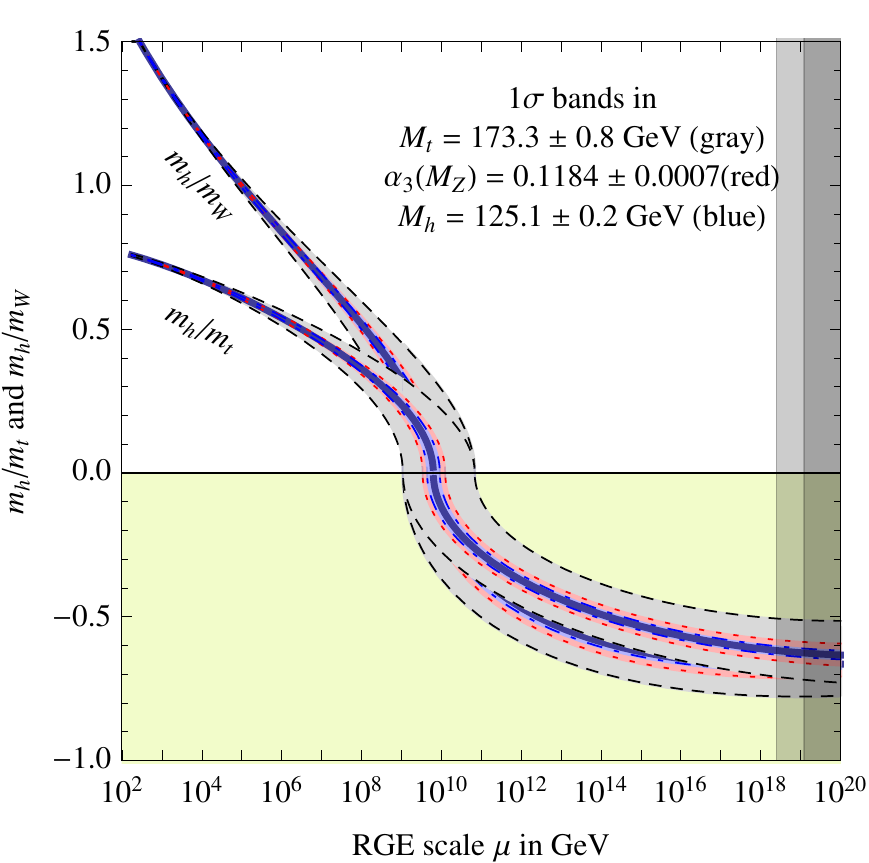}\qquad
\includegraphics[width=0.45\textwidth]{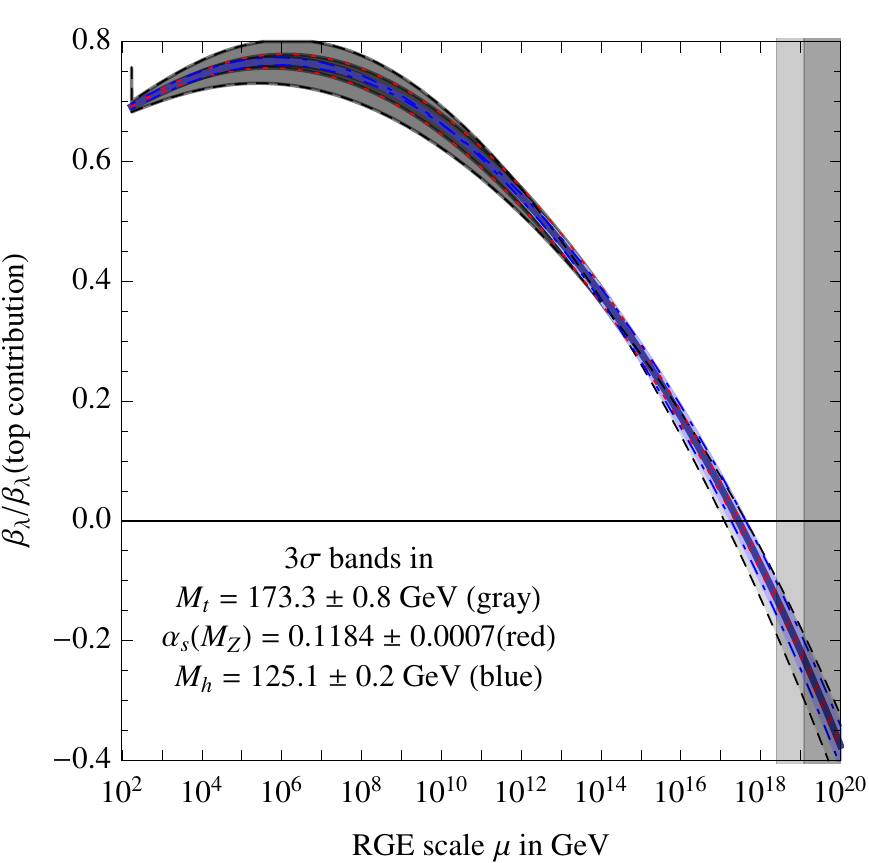}$$
\caption{\em {\bf Upper}: RG evolution of $\lambda$ ({\bf left}) and of $\beta_\lambda$ ({\bf right}) varying 
$M_t$, $\alpha_3(M_Z)$, $M_h$ by $\pm 3\sigma$.
{\bf Lower}:  Same as above, with more ``physical" normalisations. The Higgs quartic coupling is 
compared with the top Yukawa and weak gauge coupling through the ratios ${\rm sign}(\lambda)\sqrt{4|\lambda|}/y_t$ and ${\rm sign}(\lambda)\sqrt{8|\lambda|}/g_2$, which correspond to the ratios of running masses $m_h/m_t$ and $m_h/m_W$, respectively ({\bf left}). The Higgs quartic $\beta$-function is shown in units of its top contribution, $\beta_\lambda$(top contribution) $=-3y_t^4/8\pi^2$ ({\bf right}).
The grey shadings cover values of the RG scale above the Planck mass 
$\mpl \approx 1.2\times 10^{19}\GeV$,
and above the reduced Planck mass $\bar \mpl  = \mpl /\sqrt{8\pi}$.
\label{fig:rattazzi}}
\end{figure}

The Higgs quartic coupling remains weak in the entire energy domain
below $\mpl $. It decreases with energy crossing $\lambda =0$
at a scale of about $10^{10}$~GeV, see fig.\fig{rattazzi}
(upper left). Indeed, $\lambda$ is the only SM coupling that is
allowed to change sign during the RG evolution because it is not
multiplicatively renormalised. For all other SM couplings, the $\beta$
functions are proportional to their respective couplings and crossing
zero is not possible. This corresponds to the fact that $\lambda =0$
is not a point of enhanced symmetry.

\medskip

In fig.\fig{rattazzi} (lower left) we compare the size of $\lambda$
with the top Yukawa coupling $y_t$ and the gauge coupling $g_2$,
choosing a normalisation such that each coupling is equal to the
corresponding particle mass, up to the same proportionality
constant. In other words, we are plotting the ratios 
\beq\hbox{${\rm
  sign}(\lambda)\times\sqrt{4|\lambda|}/y_t\qquad$ and $\qquad{\rm
  sign}(\lambda)\times\sqrt{8|\lambda|}/g_2$} \ ,
  \eeq
equal to the ratios of running
masses $m_h/m_t$ and $m_h/m_W$, respectively.  
Except for the region
in which $\lambda$ vanishes, the Higgs quartic coupling looks fairly
``normal" with respect to the other SM couplings. Nonetheless, the RG
effect reduces significantly the overall size of $\lambda$ in its
evolution from low to high energy. 
Although the central values of Higgs and top masses do not favour a
scenario with vanishing Higgs self coupling at the Planck scale
($\mpl $) --- a possibility originally proposed in
ref.~\cite{Froggatt:1995rt,Froggatt:2001pa} and discussed more recently in
ref.~\cite{BdCE,Isidori:2007vm,Bezrukov:2009db,Shaposhnikov:2009pv,DDEEGIS}
--- the smallness of $\lambda$ around $\mpl $ offers reasons for speculation, 
as we will discuss later. 

% The Higgs quartic coupling is much smaller than gauge and top-Yukawa
% couplings. This is an important lesson we learned from LHC data. Had
% the Higgs mass been found at, say, 150--160~GeV, then $\lambda(\mpl
% )$ would come out of the same size of $g_{1,2,3}(\mpl )$ and
% $y_t(\mpl )$. Of course there is an ambiguity in the normalisation
% of $\lambda$ (definitions differing by a factor of 4 can be found in
% the literature), but the approximate vanishing of $\lambda$ is a
% robust result, independent of the normalisation. Another striking
% feature shown by fig.\fig{run1}b is the smallness of $\beta_\lambda$
% ({\it i.e.} the variation of $\lambda$ with the logarithm of the
% renormalisation scale) at scales larger than about $10^{10}$~GeV. In
% turn, this implies that the smallness of $\lambda$ remains valid
% within a fairly large range of scales in the high-energy domain. It
% also means that tiny variations of the input values for $\lambda$
% lead to wide fluctuations of the instability scale, thus justifying
% our refined calculation. We find

Another important feature of the RG evolution of $\lambda$ is the
slowing down of the running at high energy. As shown in
fig.\fig{rattazzi} (upper right), the corresponding Higgs quartic
$\beta$-function vanishes at a scale of about
$10^{17}$--$10^{18}$~GeV. In order to quantify the degree of
cancellation in the $\beta$-function, we plot in fig.\fig{rattazzi}
(lower right) $\beta_\lambda$ in units of its pure top
contribution. The vanishing of $\beta_\lambda$ looks more like an
accidental cancellation between various large contributions, rather
than an asymptotic approach to zero. Given that the $\beta$-functions
of the other SM couplings are all different than zero, it is not evident to find valid symmetry or dynamical reasons for the vanishing of
$\beta_\lambda$ alone near $\mpl $. However, the smallness of
$\beta_\lambda$ (and $\lambda$) at high energy implies that tiny
variations of the input values of the couplings at $\mpl $ lead to
wide fluctuations of the instability scale, thus justifying our
refined calculation.

\begin{figure}[t]
$$\includegraphics[width=0.45\textwidth]{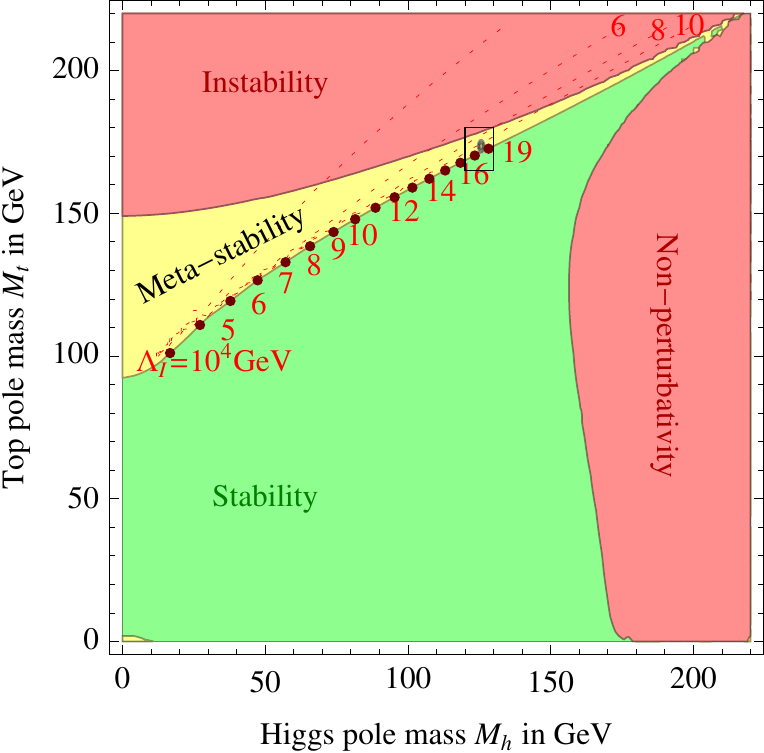}\qquad
  \includegraphics[width=0.46\textwidth]{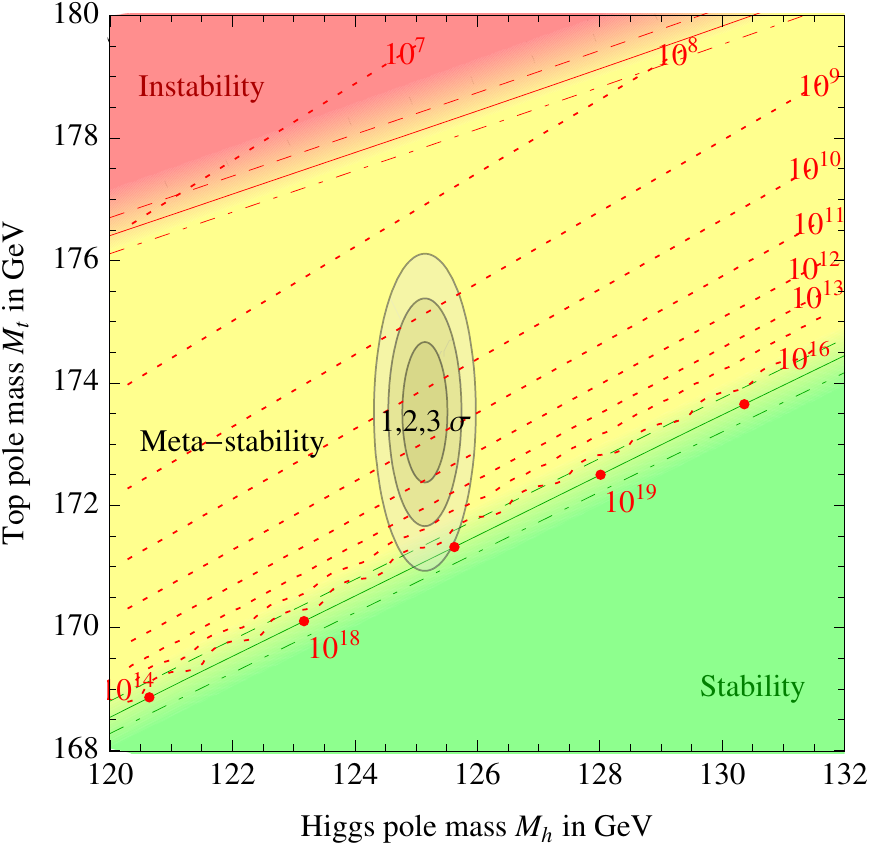}$$
\caption{\em {\bf Left}: SM phase diagram in terms of Higgs and top pole masses.
The plane is divided into regions of absolute stability, meta-stability, 
instability of the SM vacuum, and non-perturbativity of the Higgs quartic 
coupling. The top Yukawa coupling becomes non-perturbative for 
$M_t>230\, \GeV$. The dotted contour-lines show the instability scale 
$\Lambda_I$ in $\GeV$ assuming $\alpha_3(M_Z)=0.1184$.
{\bf Right}: Zoom in the region of the preferred experimental range of $M_h$ 
and $M_t$ (the grey areas denote the allowed region at 1, 2, and 3$\sigma$).
The three  boundary lines correspond to 1-$\sigma$ variations of 
$\alpha_3(M_Z)=0.1184\pm 0.0007$, and the grading of the 
colours indicates the size of the theoretical error.  
\label{fig:regions}}
\end{figure}

\subsection{Derivation of the stability bound}
%\xxx{
%Changing the SM couplings at $M_t$ by $g\to g (1+\delta_g)$ shifts the critical Higgs mass needed for stability
%by
%\beq \delta M_h \approx  (-62\delta_\lambda+364 \delta_t-186\delta_3-43\delta_2 - 15\delta_1)\GeV\eeq}
In order to compute the stability bound on the Higgs
mass one has to study the full effective potential and identify the
critical Higgs field above which the potential becomes smaller than the value at
the EW vacuum. We will refer to such critical energy as the
instability scale $\Lambda_I$.

A first estimate of the instability scale
can be obtained by approximating the effective potential with its
RG-improved tree level expression.  The analysis shows that the instability scale occurs at energies much bigger than
the EW scale. Thus, for our purposes, the approximation of neglecting $v$ with respect to the value of the field $h$ is amply justified. Under this assumption, the effective potential (in the relevant region $h\gg v$) becomes
\beq V_{\rm eff} (h)=\lambda_{\rm eff}(h)\frac{h^4}{4}. \label{eff-potential-high-h}\eeq
The quantity $\lambda_{\rm eff}$ can be extracted from the effective
potential at two loops~\cite{V2} and is explicitly given in appendix~\ref{eff-potential-app}.
% In fig. \ref{fig:regions} we provide, among other things, $\Lambda_I$ for various values of $M_t$ and $M_h$.

\subsection{The SM phase diagram in terms of Higgs and top masses}

The two most important parameters that determine the various EW phases
of the SM are the Higgs and top-quark masses. In fig.\fig{regions} we
update the phase diagram given in ref.~\cite{DDEEGIS} with our
improved calculation of the evolution of the Higgs quartic
coupling. The regions of stability, metastability, and instability of
the EW vacuum are shown both for a broad range of $M_h$ and $M_t$, and
after zooming into the region corresponding to the measured
values. The uncertainty from $\alpha_3$ and from theoretical errors
are indicated by the dashed lines and the colour shading along the
borders. Also shown are contour lines of the instability scale
$\Lambda_I$.

\medskip

As previously noticed in ref.~\cite{DDEEGIS}, the measured values of
$M_h$ and $M_t$ appear to be rather special, in the sense that they
place the SM vacuum in a near-critical condition, at the border
between stability and metastability.  In the neighbourhood of the
measured values of $M_h$ and $M_t$, the stability condition is well
approximated by 
\beq M_h > 129.6 \GeV+ 2.0 (M_t-\Mtexp\GeV) -0.5\GeV\asdiff \pm 0.3 \GeV
%\pm 0.6_{\rm non-pert} 
\ .
\label{eq:stability}
\eeq
The quoted uncertainty comes only from higher order perturbative corrections.
Other non-perturbative uncertainties associated with the relation between the measured value of the top mass
and the actual definition of the top pole mass used here (presumably of the order of $\Lambda_{\rm QCD}$) are buried inside the parameter $M_t$ in \eqg{eq:stability}. For this reason we include a theoretical error in the top pole mass and take
%
%Combining in quadrature
%this theoretical uncertainty with the experimental errors on $\alpha_3$ and $M_t$,
%also plagued by a non-perturbative theoretical uncertainty of order $\Lambda_{\rm QCD}$,
$M_t = (\Mtexp \pm \Mterr_{\rm exp} \pm 0.3_{\rm th})\GeV$. Combining in quadrature
theoretical uncertainties with experimental errors, we find
\beq 
M_h > (129.6 \pm 1.5)\GeV\qquad {\rm (stability~condition).}
\eeq 
From this result we conclude that vacuum stability of the SM up to the
Planck scale is excluded at $2.8\sigma$ (99.8\% C.L.~one-sided). 
Since the main source of uncertainty in \eqg{eq:stability}
comes from $M_t$, any refinement in the measurement of the top mass is
of great importance for the question of EW vacuum stability.

Since the experimental error on the Higgs mass is already fairly small
and will be further reduced by future LHC analyses, it is becoming
more appropriate to express the stability condition in terms of the pole
top mass. We can express the stability condition of \eqg{eq:stability} as 
\beq
M_t < (171.53\pm 0.15\pm 0.23_{\alpha_3} \pm 0.15_{M_h})\GeV=
(171.53\pm 0.42)\GeV .
\eeq
In the latter equation we combined in quadrature the theoretical uncertainty
with the experimental uncertainties on $M_h$ and $\alpha_3$.

\medskip

Notice that the stability bound is scheme and gauge independent.
While intermediate steps of the computation
(threshold corrections, higher-order RG equations, and the effective potential)
are scheme-dependent, the values of the effective potential at its local minima are scheme-independent physical observables,
and thus the stability condition has the same property.

%The line shown in fig.~\ref{} defining the border between stable and metastable regions corresponds to the condition of having a new Higgs vacuum exactly degenerate with the origin ($H=0$ configuration). In formul{\ae}, this is equivalent to $\lambda (M_*)=\beta_\lambda (M_*)=0$, for an arbitrary $M_*$. However, for $M_h>?$~GeV {\bf inserisci numero} the value of the new minimum $M_*$ becomes larger than $\mpl $. Since we do not extrapolate the SM beyond the Planck mass, for $M_*>\mpl $ (or, equivalently,  $M_h>?$~GeV) the stability condition becomes simply $\lambda (\mpl )>0$.

The instability scale $\Lambda_V$ can be defined in a gauge-independent and scheme-independent way
as  $\Lambda_V \equiv (\max_h V_{\rm eff}(h))^{1/4}$,
in terms of the value of the effective SM potential of eq.~(\ref{eff-potential-high-h}) at the maximum of its barrier.
Numerically we find
\beq
\log_{10}\frac{\Lambda_V}{\GeV} = 9.5 + 0.7\Mhdiff-1.0\Mtdiff + 0.3 \asdiff .
\label{eqlambdai}
\eeq
The alternative definition of the instability scale,  as the scale $\Lambda_\lambda$ at which the running coupling $\lambda$ vanishes,
is scheme-dependent.   In the $\MS$ scheme we find $\Lambda_\lambda \approx 2 \Lambda_V$.
The alternative definition of the instability scale,  as the scale $\Lambda_{I}$
at which $\lambda_{\rm eff}$ vanishes, is gauge dependent.
In the Landau gauge we find  $\Lambda_{I} \approx 13 \Lambda_V$
around the observed values of the SM parameters.
%In the Landau gauge we find
%\beq
%\log_{10}\frac{\Lambda_I}{\GeV} = 11.3 + 1.0\Mhdiff-1.2\Mtdiff + 0.4 \asdiff .
%\label{eqlambdai}
%\eeq
%. We find
%$\Lambda_0 \approx 0.15 \Lambda_I$, with the same dependence on the SM parameters as in \eqg{eqlambdai}. 
%Even if $\Lambda_V$ is much smaller than $M_{\rm Pl}$, new physics at the Planck scale can affect the stability condition~\cite{Branchina}.

\begin{figure}[t]
$$\includegraphics[width=0.45\textwidth]{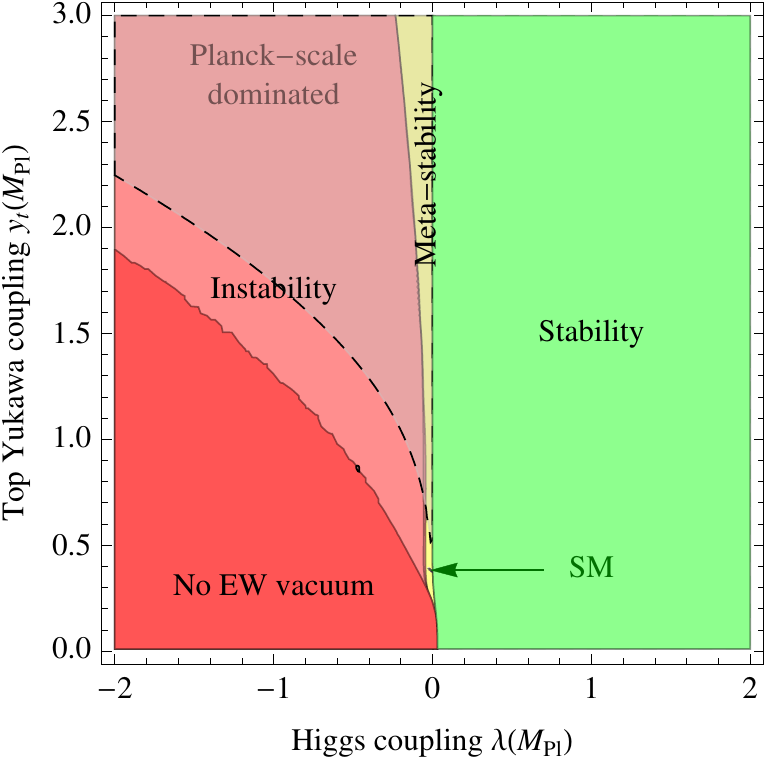}\qquad\includegraphics[width=0.45\textwidth]{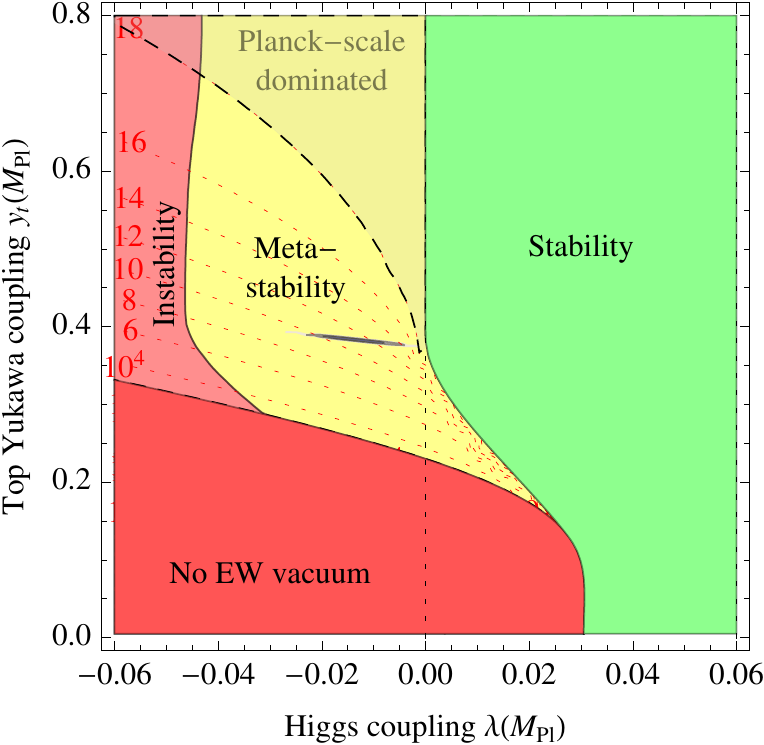}$$
\caption{\em {\bf Left}: SM phase diagram in terms of quartic Higgs coupling $\lambda$ and top Yukawa coupling $y_t$ renormalised at the Planck scale. The region where the instability scale $\Lambda_I$ is larger than $10^{18}\, \GeV$
is indicated as `Planck-scale dominated'.
{\bf Right}: Zoom around the experimentally measured values of the couplings, which correspond to the thin ellipse roughly at the centre of the panel.
The dotted lines show contours of $\Lambda_I$ in $\GeV$.  
\label{fig:regionsPlanck}}
\end{figure}

\subsection{The SM phase diagram in terms of Planck-scale couplings}
The discovery of the SM near-criticality has led to many theoretical
speculations~\cite{Hall,DDEEGIS,Holthausen:2011aa,EliasMiro:2011aa,Chen:2012faa,Lebedev:2012zw,EliasMiro:2012ay,Rodejohann:2012px,Bezrukov:2012sa,Datta:2012db,Alekhin:2012py,Chakrabortty:2012np,Anchordoqui:2012fq,Masina:2012tz,Chun:2012jw,Chung:2012vg,Chao:2012mx,Lebedev:2012sy,Nielsen:2012pu,Kobakhidze:2013tn,Tang:2013bz,Klinkhamer:2013sos,He:2013tla,Chun:2013soa,Jegerlehner:2013cta,Bezrukov:2009db,Shaposhnikov:2009pv}. In
order to address such speculations and to investigate if the measured
value of $M_h$ is really special in the SM, it is more appropriate to
study the phase diagram in terms of the Higgs quartic and the top
Yukawa coupling evaluated at some high-energy scale, rather than at
the weak scale. This is because of our theoretical bias that the SM is
eventually embedded into a new framework at short distances, possibly
as short as the Planck length. Therefore, it is more likely that
information about the underlying theory is directly encoded in the
high-energy coupling constants. For this reason in
fig.~\ref{fig:regionsPlanck} we recast the phase diagram of
fig.\fig{regions} in terms of $\lambda (\mpl )$ and $y_t (\mpl )$. The
diagram is shown in a broad range of couplings allowed by
perturbativity, and also after zooming into the interesting
region. The new area denoted as `no EW vacuum' corresponds to a
situation in which $\lambda$ is negative at the weak scale, and
therefore the usual Higgs vacuum does not exist. In the region denoted
as `Planck-scale dominated' the instability scale $\Lambda_I$ is
larger than $10^{18}\, \GeV$. In this situation we expect that both
the Higgs potential and the tunnelling rate receive large
gravitational corrections and any assessment about vacuum stability
becomes unreliable.
%If Planckian dynamics does not significantly affect the Higgs potential, our estimate of the tunnelling rate in the region $10^{18}\, \GeV < \Lambda_I < \mpl $ could be viewed as a plausible lower bound on the actual value.  

\medskip

From the left panel of fig.~\ref{fig:regionsPlanck} it is evident
that, even when we consider the situation in terms of high-energy
couplings, our universe appears to live under very special
conditions. The interesting theoretical question is to understand if
the apparent peculiarity of $\lambda (\mpl )$ and $y_t (\mpl )$ carry
any important information about phenomena well beyond the reach of any
collider experiment. Of course this result could be just an accidental
coincidence, because in reality the SM potential is significantly
modified by new physics at low or intermediate scales. Indeed, the
Higgs naturalness problem corroborates this possibility. However, both
the reputed violation of naturalness in the cosmological constant and
the present lack of new physics at the LHC cast doubts on the validity
of the naturalness criterion for the Higgs boson. Of course, even
without a natural EW sector, there are good reasons to believe in the
existence of new degrees of freedom at intermediate energies.
Neutrino masses, dark matter, axion, inflation, baryon asymmetry
provide good motivations for the existence of new dynamics below the
Planck mass. However, for each of these problems we can imagine
solutions that either involve physics well above the instability scale
or do not significantly modify the shape of the Higgs potential. As a
typical example, take the see-saw mechanism. As shown in
ref.~\cite{EliasMiro:2011aa}, for neutrino masses smaller than 0.1~eV
(as suggested by neutrino-oscillation data without mass degeneracies),
either neutrino Yukawa couplings are too small to modify the running
of $\lambda$ or the right-handed neutrino masses are larger than the
instability scale. In other words, a see-saw neutrino does not modify
our conclusions about stability of the EW vacuum.  Couplings of
weak-scale dark matter to the Higgs boson are constrained to be small
by WIMP direct searches (although dark-matter particles with weak
interactions would modify the running of the weak gauge couplings,
making the Higgs potential more stable).

\medskip

Thus, it is not inconceivable that the special values of $\lambda
(\mpl )$ and $y_t (\mpl )$ carry a significance and it is worth to
investigate their consequences. In the next section we discuss several
possible classes of solutions that explain the apparent peculiarity of
the SM parameters.

\medskip

Finally, we notice that extrapolating SM parameters above the Planck scale ignoring gravity
(this is a questionable assumption) the hypercharge couplings hits a Landau pole at about $10^{42}\GeV$.
Demanding perturbativity up to such scale (rather than up to the Planck scale),
the  bounds on the top and Higgs masses become stronger by about 10 and 20 GeV stronger respectively,
and their measured values still lie in the region that can be extrapolated up to such high scale.

\section{Interpretations of the high-energy SM couplings}
\label{sec:interpret}

The first possible interpretation, discussed in section~\ref{match}, is the result of new dynamics
occurring at some high-energy scale, the others find their most
natural implementations in the multiverse.

\subsection{Matching conditions}\label{match}

The special value of the Higgs quartic coupling could be the result of
a matching condition with some high-energy theory in the vicinity of
$\mpl $. It is not difficult to imagine theories able to drive
$\lambda(\mpl )$ to zero: high-scale supersymmetry with 
$\tan\beta=1$~\cite{Hall:2009nd,Giudice:2011cg,Cabrera:2011b,Arbey:2011ab,Ibanez:2013gf,Hebecker:2013lha};
partial $N=2$ supersymmetry insuring
$D$-flatness~\cite{Fox:2002bu,Benakli:2012cy}; an approximate
Goldstone or shift symmetry~\cite{Hebecker:2012qp,Redi}; an infrared
fixed-point of some transplanckian physics~\cite{Shaposhnikov:2009pv};
a power-law running in a quasi-conformal theory. Present data suggest
that an exact zero of $\lambda$ is reached at scales of about
$10^{10}$--$10^{12}$~GeV, see eq~(\ref{eqlambdai}), well below the Planck mass. It is not
difficult to imagine theories that give $\lambda(\mpl )$ in agreement
with \eqg{eq:lammp} as a result of a vanishing matching condition
modified by threshold corrections.

 Supersymmetry is probably one of the best candidates
able to explain the vanishing of $\lambda$ as a high-energy boundary
condition, because of the natural appearance of radiatively-stable
flat directions. Such flat directions give a well-grounded
justification for scalar particles with vanishing potentials, and yet
interacting at zero momentum (contrary to the case of Goldstone
bosons).

Note also that the smallness of the Higgs quartic $\beta$-function at high energy is the key ingredient that allows for the possibility of extending the SM up to a matching scale much larger than $\Lambda_I$. If $\lambda$ ran fast above $\Lambda_I$, it would rapidly trigger vacuum instability and the region of metastability would be limited to SM cut-off scales only slightly larger than $\Lambda_I$. This is another peculiarity of the measured values of $M_h$ and $M_t$.  

\subsection{Criticality as an attractor}

Statistical properties of the multiverse offer alternatives to
dynamical determinations of $\lambda(\mpl )$ from matching conditions
with new theories. The first possibility we consider is motivated by
the observation that the measured value of $M_h$ looks special, in the
sense that it corresponds to a near-critical parameter separating two
phases. As remarked in ref.~\cite{Giudice:2006sn}, also Higgs
naturalness can be viewed as a problem of near-criticality between two
phases ({\it i.e.} why is the Higgs bilinear carefully selected just
to place our universe at the edge between the broken and unbroken EW
phases?). This leads to the speculation that, within the multiverse,
critical points are attractors. If this vision is correct, the
probability density in the multiverse is peaked around the boundaries
between different phases, and generic universes are likely to live
near critical lines. Then, near-criticality would be the result of
probability distributions in the multiverse, and would not necessarily
follow from anthropic considerations. In this picture, the Higgs
parameters found in our universe are not at all special. On the
contrary, they correspond to the most likely occurrence in the
multiverse.

There are many natural phenomena in which near-criticality emerges as
an attractor~\cite{SOC}. A typical example is given by the slope angle of
sand dunes. While one could expect to find in a beach sand dunes with
any possible slope angles, in practice the vast majority of dunes have
a slope angle roughly equal to the so-called ``angle of repose". The
angle of repose is the steepest angle of descent, which is achieved
when the material forming the pile is at a critical condition on the
verge of sliding. The angle of repose depends on size and shape of the
material granularity, and for sand is usually about 30--35
degrees. The typicality of finding sand dunes with slope angles near
the critical value is simply understood in terms of the forces that
shape dunes. Wind builds up the dune moving sand up to the top;
gravity makes the pile collapse under its own weight when the dune is
too steep. As a result, near-criticality is the most likely condition,
as a compromise between two competing effects.

\medskip

Something similar could happen with the Higgs parameters in the
multiverse. Suppose that the probability distribution of the Higgs
quartic coupling in the multiverse is not uniform, but is a
monotonically decreasing function of $\lambda$. In other words, there
is a pressure in the multiverse towards the smallest (possibly
negative) $\lambda$. However, in universes where $\lambda$ is
sufficiently negative, the Higgs field is destabilised, forming a
bubble of AdS space with a negative cosmological constant of order
$-\mpl ^4$ in its interior. Such regions of space would rapidly
contract and finally disappear. Therefore, the cosmological evolution
removes regions that correspond to unstable EW vacua, leaving the vast
majority of universes crowded around the critical boundary. It is
amusing to note the strict analogy with the case of sand dunes. The
``wind" of the multiverse pushes the Higgs quartic coupling towards
smaller values until space collapses under the effect of AdS
gravity. As a result, the typical Higgs quartic coupling lies around
the critical value.

We can also imagine alternative scenarios. Suppose that the Higgs
quartic coupling is a function of some new fields $\Phi$ participating
in Planckian dynamics and that their vacuum structure prefers low
values of $\lambda$, as before. Once $\lambda$ becomes smaller than
the critical value, the Higgs potential develops an instability at
large field values. If tunnelling is sufficiently fast, the Higgs
field slides towards Planckian scales. Such large Higgs configurations
will in general affect the scalar potential of the fields $\Phi$,
which will readjust into a different vacuum structure. The new vacua
will give a different probability distribution for the Higgs quartic
coupling $\lambda$ and it is imaginable that now larger values of
$\lambda$ are preferred. In summary: universes in the stable or
metastable phases will experience pressure towards small $\lambda$;
universes in the unstable phase will experience pressure towards large
$\lambda$. As a result, the most probable universes lie around the
critical line separating the two phases.

We stress that these examples do not use anthropic arguments:
near-criticality is achieved by cosmological selection and/or by
probability distributions in the multiverse. Nevertheless, the proximity of
our universe to an inhospitable phase, as shown in
fig.\fig{regionsPlanck}, could be viewed as an indication that the principle of `living
dangerously' is at work, in a way similar to the case of the
cosmological constant~\cite{Weinberg:1987dv}. One can assume, as
before, that the probability distribution function of $\lambda(\mpl )$
in the multiverse is skewed towards the lowest possible values, making
it more likely for our universe to live in the leftmost region of
fig.\fig{regionsPlanck}. The anthropic boundary of EW instability
limits the allowed parameter space, giving a justification of why our
universe is `living dangerously', with conditions for stability barely
satisfied.

\subsection{Double criticality of Higgs and top couplings}

From fig.\fig{regionsPlanck} we can infer more than just criticality
of the Higgs quartic coupling. Indeed, this figure shows that the
measured values of the Higgs and top masses lie in the region
corresponding not only to the lowest possible values of $\lambda (\mpl
)$ allowed by (meta)stability, but also to the smallest possible value
of $y_t(\mpl )$, once $\lambda (\mpl )$ has been selected.
Indeed, for small Higgs quartic ($\lambda (\mpl )<0.02$),
there is a non-vanishing minimum value of $y_t(\mpl )$ required to
avoid instability. 

This special feature is related to the approximate vanishing of $\beta_\lambda$ around the Planck mass.
Indeed, for
fixed $\lambda (\mpl )$, the top Yukawa coupling has the effect to
stabilise the potential, as we evolve from high to low
energies. Without a sizeable contribution from $y_t$, the gauge
couplings tend to push $\lambda$ towards more smaller (and eventually
negative) values, leading to an instability. Therefore, whenever
$\lambda (\mpl )$ is small or negative, a non-zero $y_t(\mpl )$ is
necessary to compensate the destabilising effect of gauge
couplings. These considerations assume that $\lambda (\mpl )$ and
$y_t(\mpl )$ scan widely in the multiverse, while gauge couplings do
not. The case of scanning gauge couplings will be discussed in
section~\ref{sec:gaugescan}.

It is a remarkable coincidence that the measured values of the Higgs
and top masses correspond rather precisely to the simultaneous minima
of both $\lambda (\mpl )$ and $y_t(\mpl )$. In other words, it is
curious that not only do we live in the narrow vertical yellow stripe
of fig.\fig{regionsPlanck} --- the minimum of $\lambda (\mpl )$ ---
but also near the bottom of the funnel -- the minimum of $y_t (\mpl
)$. Near-criticality holds for both the Higgs quartic and the top
Yukawa coupling. Our universe is doubly enjoying a `dangerous life'
with respect to EW stability.

\subsection{Statistics}
 
We can also envisage a different situation within the multiverse
hypothesis, namely that $\lambda(\mpl )$ and $y_t(\mpl )$ are
determined statistically, while neither criticality nor anthropic
arguments play any role.
%One such example was presented in section~\ref{sec:scc}, where we found that $\lambda(\mpl )$ and $y_t(\mpl )$ densely scan in a narrow range of values (of relative size $1/\sqrt{N}$, where $N$ is the number of fields). In that case, we found $\lambda(\mpl )={\cal O}(1/N)$ and $y_t(\mpl )={\cal O}(1/\sqrt{N})$, which is an estimate in agreement with observation. In particular, this example can account for a small, but non-zero, value of $\lambda(\mpl )$.
To illustrate this possibility we argue that some of the features of
the high-energy SM couplings described in section~\ref{sec:scc} can be
explained, at a purely qualitative level, by the existence of a
multiverse in which SM coupling constants scan. We will not try to
address the hierarchies in the Yukawa couplings. These could emerge as
the result of an underlying flavour symmetry, remnant of a sector
external to the SM, although it is not excluded that the pattern of
Yukawa couplings is the result of the statistical properties of the
multiverse~\cite{Donoghue:2005cf,Hall:2007zh,Gibbons:2008su,flamulti}. Here
we keep an agnostic point of view on the issue of flavour and
concentrate only on Higgs quartic, top-Yukawa, and gauge couplings.

In order to describe the multiverse, we introduce some new scalar
fields $\Phi_i$ ($i=1,\dots ,N$), each having $p$ different vacuum
configurations. The total number of possible vacua is $p^N$, which is
huge for large $N$. If this multiverse of vacua is a viable candidate
to solve the cosmological constant problem ({\it i.e.} to explain why
somewhere in the multiverse the cosmological constant could be
$10^{120}$ times smaller than $\mpl ^4$), then it is reasonable that
$p^N$ should be at least $10^{120}$. So we envisage a situation in
which $N$ is at least $\order{10^2}$, which is not inconceivable in a string framework.  

To describe the scanning of the SM couplings within this multiverse, we assume 
that the SM fields are coupled to the fields $\Phi$ in the most general way,
\beq
\Lag=-\frac{Z_G(\Phi)}{4} F_{\mu \nu} F^{\mu \nu} +Z_H(\Phi) \left| D_\mu H\right|^2 +\left( iZ_\psi (\Phi) \bar \psi 
{D\!\!\!\!/\,}\psi +Y_{ab}(\Phi)\bar\psi_a  H \psi_b +{\rm h.c.}\right) -\Lambda(\Phi) |H|^4.
\eeq
Here $F_{\mu \nu}$ and $\psi$ collectively denote the SM gauge and fermion fields, and $H$ is the Higgs doublet. The physical SM coupling constants are given by
\beq
g=Z_G^{-1/2},~~~y_t=Z_{t_L}^{-1/2}Y_tZ_{H}^{-1/2}Z_{t_R}^{-1/2},~~~\lambda = Z_H^{-2}\Lambda ~,
\label{coupmulti}
\eeq
where the functions $Z_{G,\psi,H}$, $Y$, and $\Lambda$ are evaluated
at a vacuum of the fields $\Phi$. Since the fields $\Phi$ have $p^N$
vacua, the SM couplings effectively scan in this multiverse. The
coupling constants in \eqg{coupmulti} are evaluated at the high-energy
scale, here identified with $\mpl $, where the new dynamics is
integrated out.

For simplicity, we consider the toy example of multiverse proposed in
ref.~\cite{ArkaniHamed:2005yv}, in which each field $\Phi_i$ has two
vacua ($p=2$) called $\Phi_i^{(+)}$ and $\Phi_i^{(-)}$. We also assume
that each of the functions $Z_{G,\psi,H}$, $Y$, $\Lambda$ (let us call
them collectively $Z$, to simplify notation) can be split as a sum of
the contributions of the different fields $\Phi_i$,
\beq
Z(\Phi_1,\dots,\Phi_n)=\sum_{i=1}^NZ_i(\Phi_i),~~~~~~~~Z=\{Z_{G}, Z_{\psi},Z_{H}, Y,\Lambda \}.
\eeq
This is a consistent hypothesis, as long as the fields $\Phi$ are
mutually weakly-interacting. In this case, any mixed interaction is
generated only by small loop effects and can be ignored. Under this
hypothesis, the $2^N$ values of $Z$ corresponding to the vacua of
$\Phi$ can written as
\beq
Z =\sum_{i=1}^N\left( Z_i^{(S)}+\eta_iZ_i^{(D)}\right), ~~~Z_i^{(S)}=\frac{Z_i(\Phi_i^{(+)})+Z_i(\Phi_i^{(-)})}{2},~~~Z_i^{(D)}=\frac{Z_i(\Phi_i^{(+)})-Z_i(\Phi_i^{(-)})}{2},
\eeq
where $\eta_i=\pm 1$. Each of the $2^N$ vacua (and each of the $2^N$ values of 
$Z$) is labeled by the vector $\eta =(\eta_i,\dots,\eta_N)$.

The normalised probability distribution of $Z$ within the multiverse of vacua 
is given by
\beq
\rho(Z)=2^{-N} \sum_\eta \delta \left( Z-N\bar Z-\sum_{i=1}^N \eta_iZ_i^{(D)}\right),
\label{sumeta}
\qquad
\bar Z \equiv \frac{1}{N}\sum_{i=1}^NZ_i^{(S)}.
\eeq
Using the central limit theorem, the discrete sum over the $2^N$
configurations of $\eta$ in \eqg{sumeta} can be approximated for large
$N$ with a Gaussian distribution~\cite{ArkaniHamed:2005yv}
\beq
\rho(Z)=\frac{1}{\sqrt{2\pi N\Delta^2}}\exp\left[ -\frac{(Z-N\bar Z)^2}{2N\Delta^2}\right] ,\qquad
%\eeq
%\beq
\Delta^2 =\frac{1}{N}\sum_{i=1}^N {Z_i^{(D)}}^2 .
\eeq
This shows that $Z$ densely scans around $N\bar Z$ with an
approximately flat distribution in the range $|Z-N\bar Z|<\sqrt{N}\Delta$.

For generic couplings, we expect that $\bar Z$ and $\Delta$ are
quantities of order unity, and thus $Z$ is ${\cal O}(N)$ with a
relative uncertainty of order $1/\sqrt{N}$. Plugging this result
(which is valid for $Z=Z_{G,\psi,H}$, $Y$, $\Lambda$) into
\eqg{coupmulti}, we find
\beq
g, y_t \sim \frac{1}{\sqrt{N}},\qquad \lambda \sim \frac{1}{N}.
\label{coupnn}
\eeq
For $N\sim 100$, we obtain that gauge and top-Yukawa couplings are
predicted to be ${\cal O}(10^{-1})$ at around $\mpl $, while the Higgs
quartic coupling is ${\cal O}(10^{-2})$, in good qualitative agreement
with experimental data. Indeed,  adopting a `physical' normalisation of couplings as in fig.\fig{rattazzi} (lower
left), the
SM predicts $g_{1,2,3} (\mpl )/\sqrt{2}\approx y_t (\mpl )\approx
\sqrt{4|\lambda (\mpl )|}\approx 0.3$.   

The different behaviour with $N$ in \eqg{coupnn} arises because
$\lambda$ is a quartic coupling, while $g$ and $y_t$ are cubic
couplings.  Note that this framework suggests a hierarchy between $g$,
$y_t$ on one side, and $\lambda$ on the other side, but does not
predict that $\lambda$ should vanish at $\mpl $, again as indicated by
data. Actually, since $\lambda$ scans by a relative amount ${\cal
  O}(1/\sqrt{N})$, a vanishing value of $\lambda(\mpl )$ turns out to
be fairly improbable in this setup.

\section{More on SM phase diagrams}
\label{sec:more}

\begin{figure}[t]
$$\includegraphics[width=0.45\textwidth]{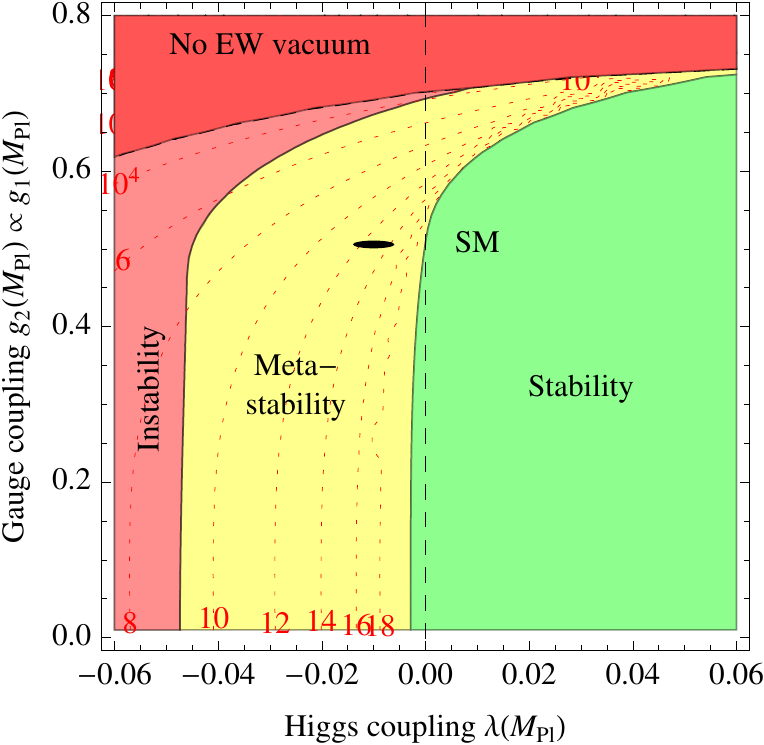}\qquad
\includegraphics[width=0.45\textwidth]{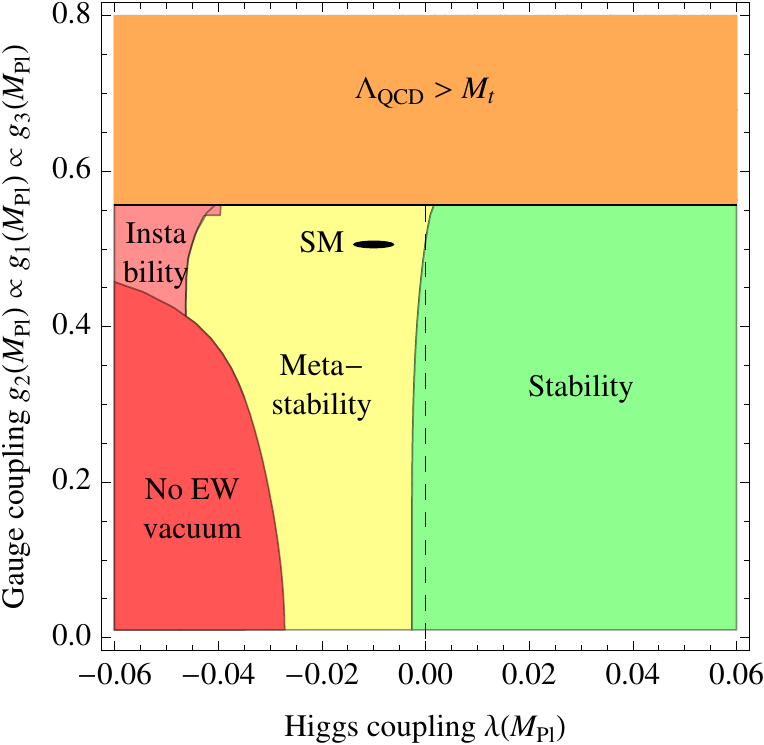}$$
\caption{\em SM phase diagram in terms of the Higgs quartic coupling $\lambda (\mpl )$ and of the gauge coupling $g_2(\mpl )$.
{\bf Left}: A common rescaling factor is applied to the electro-weak gauge couplings $g_1$ and $g_2$, while $g_3$ is kept constant.
{\bf Right}: A common rescaling factor is applied to all SM gauge couplings $g_1,g_2,g_3$, such that a $10\%$ increase in the
strong gauge coupling at the Planck scale makes $\Lambda_{\rm QCD}$ larger than the weak scale.
The measured values of the couplings correspond to the small ellipse marked 
as `SM'.
\label{fig:regionsPlanckg}}
\end{figure}

\subsection{The SM phase diagram in terms of gauge couplings}
\label{sec:gaugescan}

So far we have been studying the phase diagram in terms of Higgs and
top masses or couplings, keeping the other SM parameters fixed. This
is reasonable, since the EW vacuum is mostly influenced by the Higgs
and top quark. However, in the multiverse, other parameters can scan
too and it is interesting to study how they affect our results.

We start by considering the scanning of weak couplings defined at some
high-energy scale, which we identify with $\mpl $. The impact of the
gauge couplings $g_1$ and $g_2$ can be understood from the leading
terms of the RG equation for the Higgs quartic coupling
\beq
(4\pi)^2\, \frac{d\lambda}{d\ln\mub^2} =-3y_t^4+6y_t^2\lambda +12\lambda^2 +\frac{9}{16} \left( g_2^4 +\frac{2}{5} g_2^2g_1^2+\frac{3}{25}g_1^4\right) -\frac{9}{2}\lambda \left( g_2^2+\frac{g_1^2}{5}\right) +\cdots.
\eeq
For small $\lambda(\mpl )$, the weak gauge couplings have the effect
of reducing even further the Higgs quartic coupling in its evolution
towards lower energies, thus contributing to destabilise the
potential. For large $\lambda(\mpl )$, they tend to make $\lambda$
grow at lower energy.

We quantify the situation by plotting in fig.~\ref{fig:regionsPlanckg}
(left) the SM phase diagram in terms of $\lambda(\mpl )$ and $g_2(\mpl)$. 
For simplicity, we scan over the hypercharge coupling $g_1(\mpl )$
by keeping fixed the ratio $g_1(\mpl )/g_2(\mpl )=1.22$ as in the SM,
while $y_t(\mpl )$ and $g_3(\mpl )$ are held to their SM values.  As
in previous cases, also the phase diagram in terms of weak gauge
couplings shows the peculiar characteristic of the SM parameters to
live close to the phase boundary. (Note that the figure is zoomed
around the region of the physical values, so that the proximity to the
boundary is not emphasised.)

Figure~\ref{fig:regionsPlanckg} (left) shows that the weak gauge
couplings in the SM lie near the maximum possible values that do not
lead to a premature decay of the EW vacuum. Were $g_2$ and $g_1$
50\% larger than their actual values, we wouldn't be here speculating on
the peculiarity of the Higgs mass.

%In fig.\fig{} we show the SM phase diagram in terms of $g_2(\mpl )$ and $y_t(\mpl )$, choosing $\lambda (\mpl ) =0$, {\bf (Non sarebbe meglio fissure la mass dell'Higgs? Altrimenti cosa vuol dire l'ellissi sperimentale?)} $g_1(\mpl )=1.22\, g_2(\mpl )$, and fixing $\alpha_3$ to its physical value {\bf (Vero?)}. This figure illustrates how gauge and Yukawa couplings balance each other, as $g_2(\mpl )$ tries to reach its maximum and $y_t(\mpl )$ its minimum.

Next, we discuss the impact of scanning the strong gauge coupling
constant.  In fig.~\ref{fig:regionsPlanckg} (right) we show the phase
diagram in the plane $\lambda (\mpl )$, $g_2(\mpl )$, obtained by
varying all three gauge couplings by a common rescaling factor. The
top Yukawa coupling $y_t(\mpl )$ is held fixed at its SM value and so,
as the other couplings scan, the top mass does not correspond to the
measured value.

The coupling $g_3$ affects $\beta_\lambda$ only at two loops, but it
has a more important role in the RG evolution of the top Yukawa
coupling, whose leading terms are given by
\beq
(4\pi)^2\, \frac{dy_t^2}{d\ln\mub^2} = y_t^2 \left( \frac92 y_t^2-8g_3^2-\frac94 g_2^2-\frac{17}{20} g_1^2\right)+\cdots.
\eeq
When the value of $g_3$ is reduced at fixed $y_t(\mpl )$, the
low-energy top Yukawa coupling becomes smaller. This reduces the
stabilising effect of the top for a given $\lambda (\mpl )$ and
explains the appearance in fig.~\ref{fig:regionsPlanckg} (right) at
small gauge couplings of a `No EW vacuum' region (where $\lambda$ is
negative at the weak scale).

On the other hand, when $g_3$ is increased, the value of $\Lambda_{\rm QCD}$ 
grows rapidly. Whenever
\beq
\alpha_3 (\mpl ) > \frac{6\pi}{21 \ln (\mpl / M_t)},
\eeq
which corresponds to $g_3 (\mpl) > 0.54$, the value of $\Lambda_{\rm QCD}$ 
becomes larger than $M_t$, preventing a perturbative
extrapolation from the Planck to the weak scale. As shown in
fig.~\ref{fig:regionsPlanckg} (right), this region is reached as soon as
the SM gauge couplings are increased by only 11\%. Once again, the SM
gauge couplings live near the top of the range allowed by simple
extrapolations of the minimal theory.

\begin{figure}[t]
$$\includegraphics[width=0.45\textwidth]{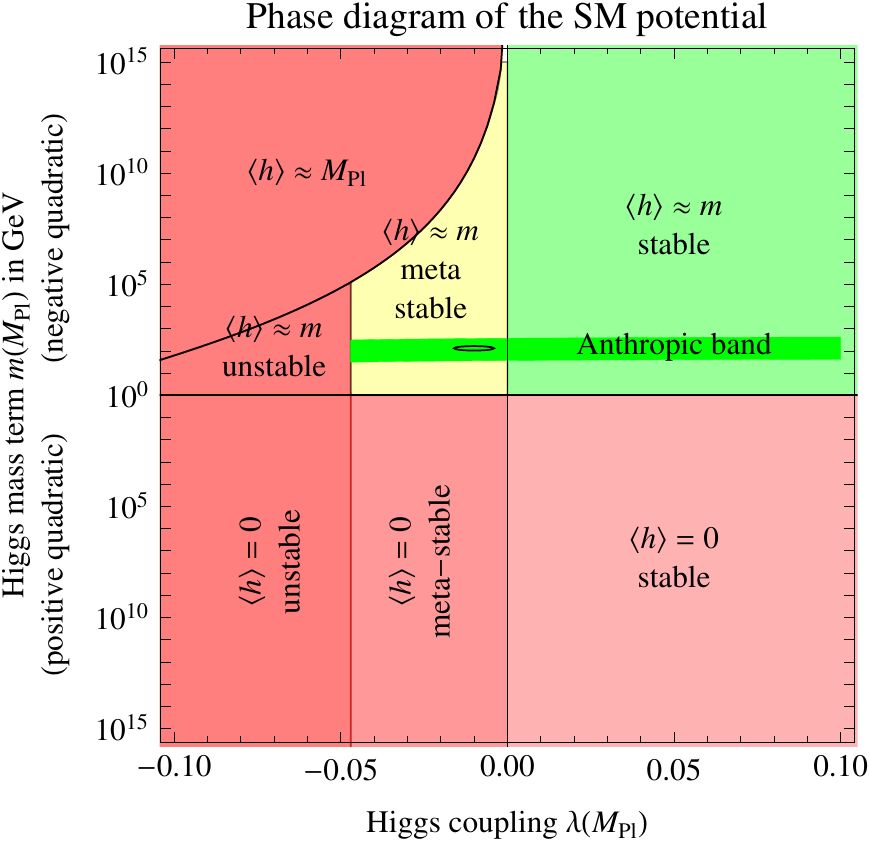}$$
\caption{\em  Phase diagram of the SM in terms of the parameters of the Higgs potential evaluated at the Planck scale. In the metastability region, there is an upper bound on $m$ from the requirement of a Higgs vacuum at a finite field value. 
The green region is simple thanks to the fact that $\beta(\lambda)=0$ at $\mpl $.
On the vertical axis we plot $|m(\mpl )|$, in the case of negative (above) and positive (below) Higgs quadratic term.
\label{fig:VSMphases}}
\end{figure}

\subsection{The SM phase diagram in terms of Higgs potential parameters}

The Higgs mass parameter $m$ in the Higgs potential is the origin of
the well-known naturalness problem. Here we show that the simple
requirement of the existence of a non-trivial EW vacuum sets an upper
bound on $m$, which is completely independent of any naturalness
argument.

Let us start by considering the tree-level Higgs potential in \eqg{higgspot}.  For
$m^2>0$ and $\lambda>0$, the potential has the usual
non-trivial vacuum at $\langle h \rangle
=v=m/\sqrt{2\lambda}$. However, since $v$ is proportional to $m$ and
$\lambda$ is negative above the instability scale $\Lambda_I$, the
Higgs vacuum at finite field value no longer exists when $m^2$ is too
large. The upper bound on $m^2$ can be estimated by considering the
minimisation condition of the potential, including only the
logarithmic running of $\lambda$, but neglecting the evolution of $m$
(which is a good approximation, as shown in fig.\fig{run1}):
\beq
\left[ 2\lambda (v) +\frac{\beta_\lambda (v)}{2}\right]v^2 =m^2.
\label{eq:minbeta}
\eeq
For values of $v$ in the neighbourhood of $\Lambda_I$, we can
approximate\footnote{In this analysis, we can safely neglect the non-logarithmic corrections to the effective potential and so we do not distinguish between $\lambda$ and $\lambda_{\rm eff}$.} $\lambda(v) \approx \beta_\lambda (\Lambda_I ) \ln
v/\Lambda_I$ and $ \beta_\lambda (v) \approx \beta_\lambda (\Lambda_I
)$. Then we see that \eqg{eq:minbeta} has a solution only if
\beq
m^2 <
-\beta_\lambda(\Lambda_I)\, e^{-3/2} \Lambda_I^2.
\label{limstab}
\eeq
Note that $\beta_\lambda(\Lambda_I)$ is negative in the SM.

Figure\fig{VSMphases} shows the SM phase diagram in terms of the
parameters $\lambda(\mpl )$ and $m(\mpl )$. 
The sign of each one of these parameters corresponds to different phases of the theory,
such that  $\lambda(\mpl )=m(\mpl )=0$ is a tri-critical point.

The region denoted by
`$\langle h\rangle \approx \mpl $' corresponds to the case in which \eqg{limstab} is
not satisfied and there is no SM-like vacuum, while the Higgs field
slides to large values. In the region of practical interest, the upper
limit on $m$ is rather far from its actual physical value $m=M_h$,
although it is much stronger than $\mpl $, the ultimate ultraviolet
cutoff of the SM. A much more stringent bound on $m$ can be derived
from anthropic considerations~\cite{Agrawal:1997gf} and the
corresponding band in parameter space is shown in
fig.\fig{VSMphases}. We find it remarkable that the simple request of
the existence of a non-trivial Higgs vacuum, without any reference to
naturalness considerations, gives a bound on the Higgs bilinear
parameter $m$. Unfortunately, for the physical value of $\lambda$, the
actual numerical value of the upper bound is not of great practical
importance.

\begin{figure}[t]
$$\includegraphics[width=0.45\textwidth]{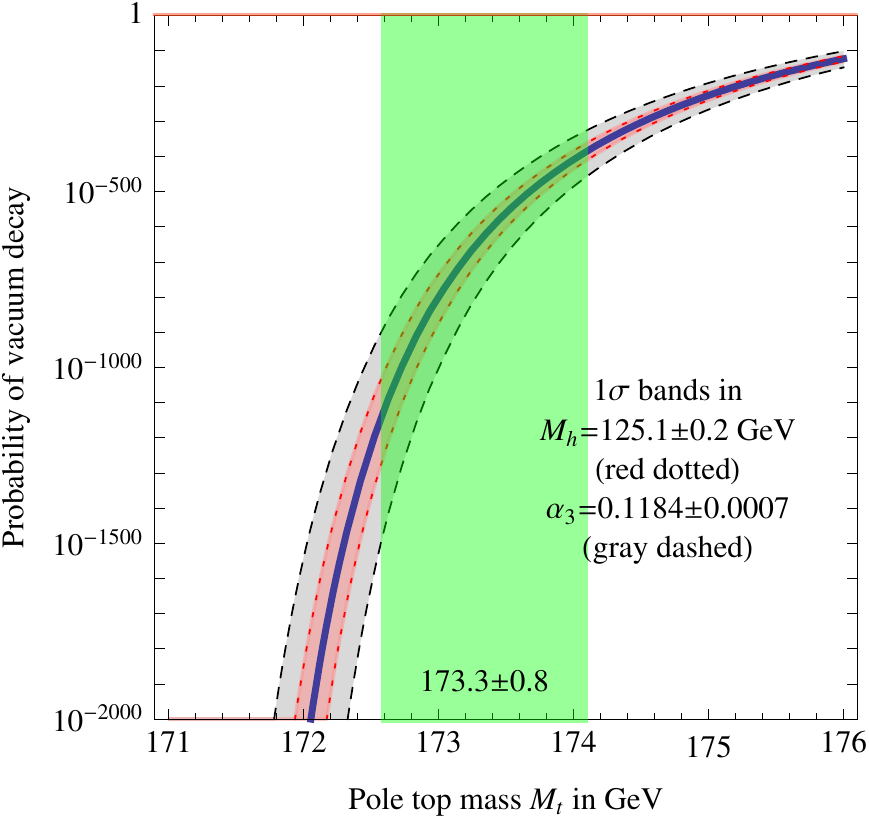}\qquad\includegraphics[width=0.45\textwidth]{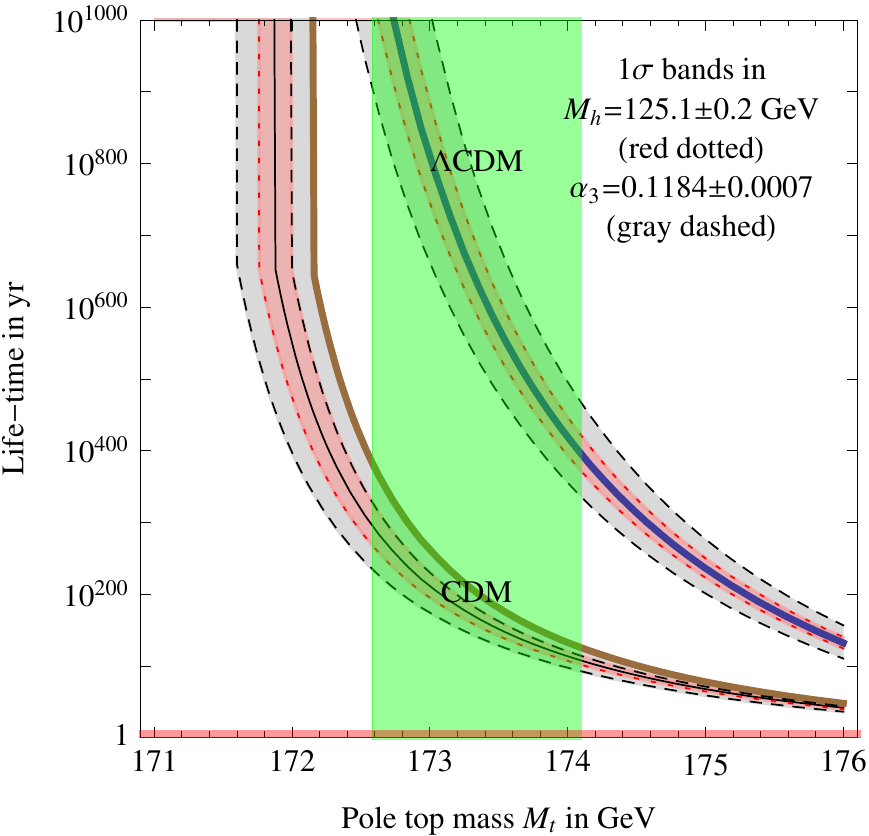}$$
\caption{\em {\bf Left:} The probability that electroweak vacuum decay happened in our past light-cone,
taking into account the expansion of the universe. {\bf Right}:
The life-time of the electroweak vacuum, with two different assumptions for future cosmology: universes dominated by the
cosmological constant ($\Lambda$CDM) or by dark matter (CDM).
\label{fig:lifetime}}
\end{figure}

\subsection{Lifetime of the SM vacuum}

The measured values of $M_h$ and $M_t$ indicate that the SM Higgs
vacuum is not the true vacuum of the theory and that our universe is
potentially unstable. The rate of quantum tunnelling out of the EW
vacuum is given by the probability $d\wp/dV\, dt$ of nucleating a
bubble of true vacuum within a space volume $dV$ and time interval
$dt$~\cite{Kobzarev:1974cp,Coleman:1977py,Callan:1977pt}
\beq
d\wp =dt\,dV ~ \Lambda_B^4\, e^{-S(\Lambda_B)} \, .
\label{eq:probd}
\eeq
In \eqg{eq:probd}, $S(\Lambda_B)$ is the action of the bounce of size 
$R=\Lambda_B^{-1}$, given by
\beq
S(\Lambda_B)=\frac{8\pi^2}{3|\lambda(\Lambda_B)|}.
\eeq

At the classical level, the Higgs theory with only quartic coupling is
scale-invariant and the size of the bounce $\Lambda_B^{-1}$ is
arbitrary. The RG flow breaks scale invariance and the tree level
action gets replaced by the one-loop action, as calculated in
ref.~\cite{IRS}.  
Then, $\Lambda_B$ is determined as the scale at
which $\Lambda_B^4 e^{-S(\Lambda_B)}$ is maximised.  In practice this
roughly amounts to minimising $\lambda(\Lambda_B)$, which corresponds
to the condition $\beta_\lambda (\Lambda_B)=0$.  As long as
$\Lambda_B\ll \mpl $, gravitational effects are irrelevant, since
corrections to the action in minimal Einstein gravity are given by
$\delta S_G = 256\pi^3\Lambda_B^2/45|\lambda|\mpl
^2$~\cite{Isidori:2007vm}. The effect of gravitational corrections is
to slow down the tunnelling rate~\cite{Coleman:1980aw}.  Whenever
$\Lambda_B >\mpl $, one can only obtain a lower bound on the
tunnelling probability by setting $\lambda (\Lambda_B)=\lambda (\mpl)$. 
In some cases, unknown Planckian dynamics can affect the tunnelling rate~\cite{Branchina} .

\bigskip

The total probability $\wp$ for vacuum decay to have occurred during the 
history of the universe can be computed by integrating eq.\eq{probd} over the
space-time volume of our past light-cone,
\beq 
\int dt \,dV = \int_0^{t_0}  dt~\int_{|x|<a(\eta_0-\eta )}d^3x = \frac{4\pi}{3} \int_0^{\eta_0} d\eta\, a^4 (\eta_0-\eta)^3 \approx \frac{0.15}{H_0^4}.
\label{eq:tunrate}
\eeq
Here $a$ is the scale factor, $\eta$ is conformal time ($d\eta /dt =1/a$), $\eta_0\approx 3.4/H_0$ is the present conformal time
and $H_0\approx 67.4\,{\rm km/sec~Mpc}$ is the present Hubble rate.
Equation~(\ref{eq:tunrate}) roughly amounts to saying that the `radius' of the universe is given by $cT_U$, where $T_U\approx 0.96/H_0$ is the present age.
The present value of the vacuum-decay probability $\wp$ is 
\beq
\wp_0 = 0.15~ \frac{\Lambda_B^4}{H_0^4}\, e^{-S(\Lambda_B)} \, ,
\label{eq:probdoggi}
\eeq
and is dominated by late times and this makes our result more robust,
since it is independent of the early cosmological history. In
fig.\fig{lifetime}a we plot, as a function of the top mass, the
probability $\wp_0$ that the EW vacuum had decayed during the past
history of the universe. We find that the probability is spectacularly
small, as a consequence of the proximity of the SM parameters to the
boundary with the region of absolute stability.

\bigskip

The lifetime of the present EW vacuum $\tau_{\rm EW}$ depends on the
future cosmological history. If dark energy shuts off and the future
universe is matter dominated, the space-time volume of the past
light-cone at time $t_0$ is given by
\beq
\int dt \,dV = \frac{4\pi}{3} \int_0^{\eta_0} d\eta\, a^4 (\eta_0-\eta)^3=\frac{16\, \pi}{1485\, H_0^4}.
\eeq
Here $H_0$ is the Hubble parameter at time $t_0$, and we have
performed the integral using the relations
$a^{1/2}=H_0\eta/2=(3H_0t/2)^{1/3}$ and $t_0=2/(3H_0)$, valid in a
matter-dominated flat universe. The lifetime $\tau_{\rm EW}$ is given by the time at which $\wp=1$:
\beq
\tau_{\rm EW} =\left( \frac{55}{3\pi}\right)^{1/4} \frac{e^{S(\Lambda_B)/4}} {\Lambda_B} \approx \frac{T_U} {\wp_0^{1/4}} \qquad
\hbox{\sc (Matter Domination)},
\eeq
where $\wp_0$ is given in \eqg{eq:probdoggi} and shown in fig.\fig{lifetime}a.

If instead the universe keeps being accelerated by the cosmological
constant, entering into a de Sitter phase with Hubble constant $H =
H_0 \sqrt{\Omega_\Lambda}$, at a time $t_0$ in the far future the
volume of the past light-cone will be
\beq 
\int dt~dV = \frac{4\pi}{3} \int_0^{\eta_0} d\eta\, a^4 (\eta_0-\eta)^3= \frac{4\pi }{3H^4} \left[ Ht_0 -\frac{11}{6}+{\cal O}(e^{-Ht_0})\right] .
\eeq
Here we have used the relations $a=(1-H\eta)^{-1}=e^{Ht}$, valid in a 
vacuum-energy dominated universe. The lifetime $\tau_{\rm EW}$ is now equal to
\beq
\tau_{\rm EW} = \frac{3H^3e^{S(\Lambda_B)}} {4\pi \Lambda_B^4} \approx \frac{0.02~T_U}{ \wp_0} \qquad
\hbox{\sc (Vacuum Energy Domination)}.
\eeq
 
The lifetime of the present EW vacuum is plotted in
fig.\fig{lifetime}b in both cases of matter or vacuum-energy
domination. As shown, the SM vacuum is likely to survive for times
that are enormously longer than any significant astrophysical age
({\it e.g.} the sun will exhaust its fuel in about five billion
years).

%\footnote{(INTERNAL NOTE)
%The space-time volume in our past light-cone is
%$$ \int d^4x = \int_0^{t_0} a^3 dt~\int_{x<\eta_0-\eta}d^3x = \frac{4\pi}{3H_0^4} \int_0^{\eta_0} a^4 (\eta_0-\eta)^3 d\eta\approx \frac{0.148}{H_0^4}$$
%where $a$ is the scale factor,
%$d\eta = dt/a=da/a^2H$ is the proper time, $\eta_0\approx 3.4/H_0$ is the present proper time.
%In the far future dominated by the cosmological constant with $H = H_0 \sqrt{\Omega_\Lambda}$,
%$a$ will grow as $e^{Ht}$ and $\eta$ will approach a constant $\eta \simeq \eta_\infty - 1/a$,
%such that the integral will be given by $\ln a + {\cal O}(1) \simeq Ht$, so that 
%$\int d^4 x \simeq 4\pi t/3H^3$.
%}

\section{Summary and conclusions}
\label{sec:conclusions}

The measurement of the Higgs mass $M_h$ has determined the last
unknown parameter of the SM, fixing the Higgs quartic coupling $\lambda$. Now
that the experimental result is in our hands, our task as
theoreticians is to interpret it, investigating whether it contains
any useful information about physics at shorter distances. The first
thing to try is to extrapolate $\lambda$ to high energy in search for
clues. Just as high-energy extrapolations of the gauge coupling
constants gave us hints about a possible grand unification of
fundamental forces, so the extrapolation of $\lambda$ has revealed an
unexpected feature of the SM that opens new avenues for theoretical
speculation. The intriguing result is that, assuming the validity of
the SM up to very high energy scales, the measured value of $M_h$ is
near-critical, in the sense that it places the EW vacuum right at the
border between absolute stability and metastability. Because of the
present experimental uncertainties on the SM parameters (mostly the
top quark mass), we cannot conclusively establish the fate of the EW
vacuum, although metastability is now preferred at 99.3\% CL.

The special coincidence found in the value of $M_h$ warrants a refined
calculation of the high-energy extrapolation of $\lambda$ and this was
the first objective of this paper.
We extracted the fundamental SM parameters $\lambda$ (quartic Higgs coupling), $m$ (Higgs mass term), $y_t$
(top quark Yukawa coupling),
$g_2$ and $g_Y$ (electroweak gauge couplings) 
from the precisely measured values of the Higgs, top, $W$ and $Z$ masses and from
the Fermi constant at full NNLO, by performing dedicated 2-loop computations.
All couplings have been extrapolated to large energies using the RGE equations, now known at NNLO order (3 loops).
We could then compute the effective potential known with 2-loop accuracy.

\bigskip

The second objective of this paper was to investigate the significance
of the measured value of $M_h$, in view of its high-energy
extrapolation. A first observation is that $\lambda$, together with
all other SM coupling constants, remains perturbative in the entire
energy domain between the Fermi and the Planck scales. This gives an
indirect indication that EW-breaking dynamics is probably weakly
interacting. Of course, strongly-interacting dynamics is not excluded,
but there is simply no need for introducing it at any intermediate
energy scale.

The most important observation concerns the stability of the Higgs
potential. The critical condition for stability is defined as the
vanishing of the effective coupling $\lambda_{\rm eff}$, see
\eqg{eff-potential-high-h}, at some energy scale $\Lambda_I$. We find
$\Lambda_I= 10^{10}$--$10^{12}$~GeV, see \eqg{eqlambdai}, suggesting that the instability
is reached well below the Planck mass. The presence of an instability
at an intermediate scale could be interpreted as a sign of a
new-physics threshold around $\Lambda_I$. It is suggestive that
neutrino masses, axion, and inflation give independent indications for
new dynamics at roughly similar energy scales. The hypothetical new
physics could be responsible for a matching condition $\lambda \approx
0$ at a scale near $\Lambda_I$. The vanishing of $\lambda$ could be
the result of special dynamics occurring above $\Lambda_I$, such that
the evolution of the Higgs quartic is power-law suppressed, or the
result of symmetry, as in the case of an approximate Goldstone
boson. One of the most appealing explanations of $\lambda \approx 0$
is offered by supersymmetry, since flat directions provide a valid
justification of vanishing quartic couplings for scalar particles that
have other kinds of interactions at zero momentum. In this way,
supersymmetry convincingly evades the problem, encountered by
Goldstone bosons, of explaining why $\lambda \approx 0$ is compatible
with sizeable gauge and Yukawa couplings of the Higgs boson. The
scheme can be automatically realised in $N=2$ supersymmetry, while a
dynamical vacuum alignment with $\tan\beta \approx 1$ is required in
the case of $N=1$ supersymmetry.

\smallskip

Another peculiarity found in the extrapolation of $\lambda$ is its
slow running at high energy. This is due to a combination of two
factors: the reduction of all SM couplings at high energy and an
accidental zero of $\beta_\lambda$ at a scale of about
$10^{17}$--$10^{18}$~GeV. It is the slow running of $\lambda$ at high
energy that saves the EW vacuum from premature collapse, in a
situation where $\Lambda_I\ll \mpl$. Were $\beta_\lambda$ large and
negative above $\Lambda_I$, we could not live with an instability
scale much smaller than the cutoff scale, without being confronted
with early vacuum decay. Unfortunately, for the moment we have no way
to tell whether this special condition allowing for a prolonged vacuum
lifetime is just a numerical coincidence or an important feature of
the SM.

At any rate, the smallness of $\beta_\lambda$ at high energy makes it
possible to assume that there is no new-physics threshold around
$\Lambda_I$ and that the SM continues to be valid up to the
quantum-gravity scale, since the tunnelling probability remains
small. In this context, the value of $\lambda (\mpl )$ may be regarded
as `normal' for a SM coupling. Indeed, as discussed in
section~\ref{sec:scc}, the ratios $\sqrt{4|\lambda|}/y_t$ and
$\sqrt{8|\lambda|}/g_2$ (which, at low energy, correspond to $M_h/M_t$
and $M_h/M_W$, respectively) are of order unity both at the Fermi and
Planck scales. The vanishing of $\lambda$ at an intermediate scale
could then be purely accidental. After all, the Higgs quartic is the
only SM coupling that can cross zero during its RG evolution, since
$\lambda =0$ is not a point of enhanced symmetry.

\smallskip

In our view, the most interesting aspect of the measured value of
$M_h$ is its near-criticality. In this paper we have thoroughly
studied the condition of near-criticality in terms of the SM
parameters at a high scale, which we identified with the Planck
mass. This procedure is more appropriate than a study in terms of
physical particle masses, since it is more likely that special
features are exhibited by high-energy parameters, just like in the
case of gauge coupling unification.

We have found that near-criticality is manifest also when we explore
the phase diagram as a function of high-energy SM couplings. Moreover,
we found evidence for multiple near-critical conditions. Indeed, the
measured SM parameters roughly correspond to the minimum values of
Higgs quartic coupling $\lambda (\mpl )$ 
and of the top Yukawa
coupling $y_t (\mpl )$ (at fixed gauge couplings) that allow for the existence of a sufficiently
long-lived EW vacuum. Moreover, at fixed top Yukawa coupling, the maximum possible values of the gauge
couplings $g(\mpl )$ are preferred. Incidentally, we have also obtained an upper
bound on the Higgs mass parameter $m$ from the requirement of vacuum
stability, although this bound is too weak to be useful in practice.

\bigskip

We   explored possible interpretations of this multiple
near-criticality. Provided it is not just a fortuitous coincidence, an
explanation of near-criticality almost necessarily requires the
existence of an underlying statistical system. This drives us towards
the multiverse as the most convincing framework in which one can
address the issue. Near-criticality can emerge in the multiverse from
an appropriate probabilistic pressure in the space of coupling
constants, together with the anthropic requirement that selects
universes in which the life-friendly EW vacuum is sufficiently
long-lived. The principle of `living dangerously' populates universes
close to the boundary of a hospitable phase, just as it is conjectured
to happen in the case of the cosmological constant.

In this context, one may wonder whether the LHC measurement of the
Higgs mass corresponds to a point in parameter space that is
sufficiently close to the instability boundary to be justified by the
principle of `living dangerously'. Unfortunately, the answer to this
question depends on the unknown probability distribution of the SM
couplings that scan.
 In the case of the cosmological constant, we have a clear
 understanding of why larger values of $\Lambda_{\rm CC}$ should be
 preferred within the universe: small values of $\Lambda_{\rm CC}$
 require delicate accidental cancellations among the various
 parameters of the theory. On the other hand, here we are dealing with
 dimensionless couplings and it is less clear why there should be any
 probabilistic preference and, especially, in which direction should
 the multiverse pressure act.
 It is plausible that renormalisable couplings,
such as $\lambda$, have a less steep probability distribution than the
cosmological constant and therefore are likely to show a less
pronounced proximity to the critical boundary in a given individual
universe, but of course it is impossible to make definitive statements
at this stage.

It is interesting that near-criticality could find an explanation in
the multiverse, without any reference to anthropic reasoning. In
nature there exist statistical systems in which criticality is an
attractor point of their dynamical evolution. If such a
phenomenon took place in the multiverse, then the majority of
universes would populate regions close to phase transitions. Such a
(non-anthropic) explanation of near-criticality of the Higgs mass
could also provide a link to the naturalness problem, since the
smallness of the mass parameter $m$ in the Higgs potential is
near-critical with respect to the EW symmetry-breaking phase
transition. It is indeed a remarkable experimental fact that both
$\lambda$ and $m$ (the two parameters of the Higgs potential) happen
to lie very close to boundaries between different phases of the
SM. So, according to this interpretation, our universe would not be a
rare occurrence in the multiverse where SM parameters are selected in
such a way that the cosmological evolution of the EW vacuum is
favourable to life. On the contrary, near-criticality of the Higgs
parameters would be a fairly generic property of the multiverse and
our universe would be unexceptional.

\bigskip
In spite of the absence of any signal of new physics, the LHC has
already provided valuable information for theoretical speculations
about physics at very short distances. In that respect, the most
important result has been the near-criticality of the Higgs mass ---
the subject of this paper.

\small

\subsubsection*{Acknowledgments}
This work was supported by the
SF0690030s09 project, by the Research Executive
Agency (REA) of the European Union under the Grant Agreement number PITN-GA-
2010-264564 (LHCPhenoNet); by the EU ITN ``Unification in the LHC Era", contract PITN-GA-2009-237920 (UNILHC) and by MIUR under contract 2006022501; by the Spanish Ministry of Economy and Competitiveness under grant FPA2012-32828, 
Consolider-CPAN (CSD2007-00042), the grant  SEV-2012-0249 of the ``Centro de Excelencia Severo Ochoa'' Programme and the grant  HEPHACOS-S2009/ESP1473 from the C.A. de Madrid.
We thank Simone Alioli, Claudio Bonati, Lawrence Hall, Luis E. Ib\'a${\rm \tilde{n}}$ez, Gino Isidori, and Riccardo Rattazzi for useful discussions.
The work of P.P.G. has been partially funded by the ``Fondazione A.\ Della Riccia''.

\appendix
 \small

\section{Weak scale thresholds at one loop}\label{1loop}
We summarise here the one-loop corrections $\theta^{(1)}$
to the various SM parameters 
\beq 
\theta=\{\lambda, m, y_t, g_2, g_Y\}= 
\theta^{(0)}+\theta^{(1)}+\theta^{(2)}+\cdots.
\eeq
We perform one-loop computations in a generic $\xi$ gauge,
confirming that $\theta^{(1)}$ is gauge-independent, as it should.
Our expressions for $\theta^{(1)}$ are equivalent to the well known
expressions in the literature.  We write $\theta^{(1)}$ in terms of
finite parts of the the Passarino-Veltman functions 
\beq A_0 (M) =
M^2(1-\ln\frac{M^2}{\mub^2})\ ,\qquad B_0(p;M_1,M_2) = -\int_0^1
\ln\frac{xM_1^2+(1-x) M_2^2-x(1-x)p^2}{\mub^2}dx\ .  
\eeq 
The dependence of $\theta^{(1)}$ on the renormalisation scale $\mub$ reproduces
the well known one-loop RGE equations for $\theta$.
Below we report the expressions valid in the limit $M_b=M_\tau=0$; the negligible effect of
light fermions masses is included in our full code.

\subsection{The quartic Higgs coupling}
The one-loop result is obtained from eq.\eq{lh1}:
\bea
\lambda^{(1)} (\mub) &=&
\frac{1}{(4\pi)^2V^4}\Re\bigg[  3M_t^2(M_h^2-4M_t^2) B_0(M_h;M_t,M_t)+3M_h^2 A_0(M_t)+ \nonumber\\
&&+\frac{1}{4}\left(M_h^4-4 M_h^2 M_Z^2+12 M_Z^4\right)B_0(M_h;M_Z,M_Z) +\frac{M_h^2(7M_W^2-4 M_Z^2)}{2(M_Z^2-M_W^2)}  A_0(M_Z)+   \nonumber  \\
&& +\frac{1}{2}(M_h^4-4M_h^2M_W^2+12M_W^4)B_0(M_h;M_W,M_W)-\frac{3M_h^2 M_W^2}{2(M_h^2-M_W^2)} A_0(M_h)+\\
&&+\frac{M_h^2}{2}\left(-11 + \frac{3 M_h^2}{M_h^2-M_W^2} -\frac{3 M_W^2}{M_Z^2 - M_W^2}\right) A_0(M_W) +  \nonumber  \\
&&+\frac{9}{4} M_h^4 B_0(M_h;M_h,M_h) +\frac{1}{4}(M_h^4 +M_h^2(M_Z^2+2M_W^2-6 M_t^2)-8(M_Z^4+2M_W^4))
 \bigg] \ .\nonumber
\label{d1Mh}
\eea
%For the best-fit values of the masses and for $\bar\mu=M_t$ we find $\lambda^{(1)}=-0.22763/(4\pi)^2$.
Each one of the terms in eq.\eq{lh1} is gauge dependent, e.g.\
the one-loop correction to muon decay is
\bea
{\left. \Delta r^{(1)}_0 \right|_{\rm fin}}&=&\frac{1}{(4\pi V)^2}  
\bigg[3M_t^2 -M_W^2-\frac{M_Z^2}{2}-\frac{M_h^2}{2}+\frac{3M_W^2 A_0(M_h)}{M_h^2-M_W^2}+\frac{6M_W^2-3M_Z^2}{M_W^2-M_Z^2}A_0(M_Z)+ \\
&&-6A_0(M_t) + \bigg(9-\frac{3M_h^2}{M_h^2-M_W^2}-\frac{3M_W^2}{M_W^2-M_Z^2}\bigg)A_0(M_W)+ 2A_0(\sqrt{\xi} M_W)+ A_0(\sqrt{\xi} M_Z) \bigg] \nonumber
\eea
%and the correction to the pole Higgs mass
%\bea
%\delta^{(1)} M_h^2 + \frac{T^{(1)}}{v_{\os}} &=&
%\frac{1}{(4\pi V)^2}\Re\bigg[ 6M_t^2(4M_t^2-M_h^2) B_0(M_h;M_t,M_t)+ \nonumber\\
%&& -(M_h^4-4M_h^2M_W^2+12M_W^4)B_0(M_h;M_W,M_W)+2M_h^2 A_0(M_W) +  \nonumber  \\
%&&-\frac{1}{2}\left(M_h^4-4 M_h^2 M_Z^2+12 m_Z^4\right)B_0(M_h;M_Z,M_Z) +M_h^2  A_0(M_Z)+   \nonumber  \\
%&&-\frac{9}{2} M_h^4 B_0(M_h;M_h,M_h) +4(2M_W^4+M_Z^4)- M_h^2 \Xi \bigg] .
%\label{d1Mh}
%\eea
and the gauge dependence cancels out in the sum $\lambda^{(1)}(\mub)$.

\subsection{The Higgs mass term}
The correction is obtained from eq.\eq{mh1}:
 \bea
\delta^{(1)} m^{2} (\mub) &=&
\frac{1}{(4\pi)^2 V^2}\Re \bigg[  6M_t^2(M_h^2-4M_t^2)B_0(M_h;M_t,M_t)+ 24 M_t^2 A_0(M_t) +\nonumber\\
&& +(M_h^4-4M_h^2M_W^2+12M_W^4)B_0(M_h;M_W,M_W)-2(M_h^2+ 6M_W^2) A_0(M_W) +  \nonumber  \\
&&+\frac{1}{2}\left(M_h^4-4 M_h^2 M_Z^2+12 M_Z^4\right)B_0(M_h;M_Z,M_Z) -(M_h^2+ 6M_Z^2) A_0(M_Z) +  \nonumber  \\
&&+\frac{9}{2} M_h^4 B_0(M_h;M_h,M_h)-3M_h^2 A_0(M_h) \bigg] \ .
\eea

\subsection{The top Yukawa coupling}
The gauge-invariant one-loop correction to the top Yukawa coupling is obtained 
from eq.\eq{yt1}
\bea \nonumber
y_t^{(1)} (\mub) &=&\frac{M_t}{\sqrt{2}V^3 (4\pi)^2}
\Re\bigg[
- \left(M_h^2-4 M_t^2\right) B_0\left(M_t;M_h,M_t\right) +
\\ &&+  \nonumber
\frac{M_t^2 \left(80 M_W^2 M_Z^2-64 M_W^4-7
   M_Z^4\right)+40 M_W^2 M_Z^4-32 M_W^4 M_Z^2-17 M_Z^6 }{9  M_t^2
   M_Z^2}  B_0\left(M_t;M_t,M_Z\right)+
  \\ &&+ 
   \frac{\left(M_t^2 M_W^2+M_t^4-2 M_W^4\right)}{ M_t^2} B_0\left(M_t;0,M_W\right)+
   \\ &&+
  \left(\frac{3 M_h^2}{M_h^2-M_W^2}+\frac{2
   M_W^2}{M_t^2}+\frac{3 M_W^2}{M_W^2-M_Z^2}-10\right) A_0\left(M_W\right) +
    \left(\frac{3 M_W^2}{M_W^2-M_h^2}+1\right) A_0\left(M_h\right)+\nonumber
   \\ &&+\nonumber
   \frac{\left(36 M_t^2 M_Z^2-56 M_W^2 M_Z^2+64 M_W^4-17
   M_Z^4\right)}{9  M_t^2 M_Z^2} A_0\left(M_t\right)+
   \\ &&+\nonumber
   \left( \frac{3 M_W^2}{M_Z^2-M_W^2} + \frac{32 M_W^4 - 40 M_W^2 M_Z^2 +17 M_Z^4}{9 M_t^2 M_Z^2} -3\right) A_0\left(M_Z\right) +
   \\ &&+\nonumber
    \frac{M_h^2}{2} - 3 M_t^2 - 9 M_W^2 +\frac{7 M_Z^2}{18}+\frac{64 M_W^4}{9 M_Z^2} \bigg]
+ \frac{M_t}{\sqrt{2}V(4\pi)^2} g_3^2 \left(-\frac{8
   A_0\left(M_t\right)}{M_t^2}-\frac{8}{3}\right) \ .
   \eea 
   
   \subsection{The weak gauge couplings}
 The one-loop correction to the $\SU(2)_L$ gauge coupling is obtained from
eq.\eq{G2}: 
\bea
g_2^{(1)} (\mub) &=& \frac{2M_W}{(4\pi)^2V^3} \Re\bigg[
\left(\frac{M_h^4}{6 M_W^2}-\frac{2 M_h^2}{3}+2 M_W^2\right) B_0\left(M_W,M_h,M_W\right)
+\nonumber
   \\ &&\nonumber
   +\left(-\frac{M_t^4}{M_W^2}-M_t^2+2
   M_W^2\right) B_0\left(M_W,0,M_t\right)+
      \\ &&
      +\frac{1}{6} \left(-\frac{48 M_W^4}{M_Z^2}+\frac{M_Z^4}{M_W^2}-68 M_W^2+16
   M_Z^2\right) B_0\left(M_W,M_W,M_Z\right)+
         \\ &&\nonumber
        + \frac{1}{6}\left(M_h^2
   \left(\frac{9}{M_h^2-M_W^2}+\frac{1}{M_W^2}\right)+\frac{M_Z^2}{M_W^2}+M_W^2
   \left(\frac{9}{M_W^2-M_Z^2}+\frac{48}{M_Z^2}\right)-27\right)  A_0\left(M_W\right) +
      \\ &&\nonumber
      + \left( 2- \frac{M_h^2 \left( M_h^2 + 8 M_W^2\right)}{6 M_W^2 \left(M_h^2 - M_W^2 \right)}\right) A_0\left(M_h\right) + \left(\frac{M_t^2}{M_W^2}+1\right) A_0\left(M_t\right)+
      \\ &&\nonumber
      +\frac{1}{6} \left(\frac{24 M_W^2}{M_Z^2} - \frac{M_Z^2}{M_W^2} + \frac{9 M_W^2}{M_Z^2 - M_W^2} - 17\right) A_0\left(M_Z\right) +
      \\ &&\nonumber
      +\frac{1}{36} \left(-3 M_h^2+18 M_t^2+\frac{288 M_W^4}{M_Z^2}-374 M_W^2-3 M_Z^2\right)\bigg] \ .
\eea

    The one-loop correction to the ${\rm U(1)_Y}$ gauge coupling is obtained 
from eq.\eq{G1}: 
 \bea \nonumber
 g_Y^{(1)} (\mub) &=& \frac{2 \sqrt{M_Z^2-M_W^2}}{(4\pi)^2 V^3}\Re\bigg[
   \left(\frac{88}{9} - \frac{124 M_W^2}{9 M_Z^2} + \frac{M_h^2+34 M_W^2}{6 (M_Z^2-M_W^2)} \right) A_0\left(M_Z\right)+
 \\ &&\nonumber+
    \frac{M_h^2-4 M_W^2}{2 (M_h^2-   M_W^2)}   A_0\left(M_h\right)  +
   \left(-\frac{7}{9} - \frac{M_t^2}{M_Z^2-M_W^2} +\frac{64 M_W^2}{9 M_Z^2} \right) A_0\left(M_t\right)+
    \\ &&\nonumber+
    \frac{M_h^4+ 2M_W^2 (M_W^2-15M_Z^2) +3M_H^2 (2 M_W^2 + 7 M_Z^2) }{6
   \left(M_h^2-M_W^2\right) \left(M_W^2-M_Z^2\right)}A_0\left(M_W\right)
  +  \\ &&\nonumber 
   -\frac{M_t^4+M_W^2 M_t^2-2 M_W^4   }{M_W^2-M_Z^2}B_0\left(M_W,0,M_t\right)
  % + \\ &&\nonumber 
   -\frac{M_h^4-4 M_Z^2 M_h^2+12 M_Z^4 }{6 (M_W^2-M_Z^2)} B_0\left(M_Z,M_h,M_Z\right)+
    \\ &&+
    \frac{M_h^4-4 M_W^2 M_h^2+12 M_W^4 }{6(   M_W^2- M_Z^2)}B_0\left(M_W,M_h,M_W\right)+
    \\ &&\nonumber+
    \frac{M_Z^6-48 M_W^6-68 M_Z^2 M_W^4+16 M_Z^4 M_W^2 }{6 M_Z^2
   \left(M_W^2-M_Z^2\right)}B_0\left(M_W,M_W,M_Z\right)+
    \\ &&\nonumber +
    \frac{1}{9}\left(-23 M_W^2 + 7 M_t^2 +17 M_Z^2 -\frac{64 M_t^2 M_W^2}{M_Z^2} - \frac{9 M_W^2 (M_t^2-M_W^2)}{M_Z^2-M_W^2}\right) B_0\left(M_Z,M_t,M_t\right)+
    \\ &&\nonumber+
    \frac{M_Z^6 -48 M_W^6-68 M_Z^2 M_W^4+16 M_Z^4 M_W^2  }{6M_Z^2 \left(M_Z^2-M_W^2 \right)}B_0\left(M_Z,M_W,M_W\right)+ \\ &&\nonumber +
 \frac{1}{36} \left(\frac{576 M_W^4}{M_Z^2}-242 M_W^2-3 M_h^2 + 257
   M_Z^2 + \frac{36 M_W^2}{M_Z^2-M_W^2}+ M_t^2 \left(82-\frac{256 M_W^2}{M_Z^2}\right)\right)
   \bigg] \ .
   \eea
%  All the $\delta_i^{(1)}$ turn out to be gauge-invariant as they should, 
%their $\mu$ dependence   reproduces the usual RGE for the SM couplings;
%%  $\delta_t^{(1)}$ agrees with the standard expression.
%  Numerically we find, for $\mu=M_t$,
%  \beq \begin{array}{ll}
% \lambda^{(1)}=-0.22716/(4\pi)^2, &
%y_t^{(1)} = y_t^{(0)}[0.4000-\frac{32}{3}g_3^2]/(4\pi)^2,\\
%g_W^{(1)} = g_W^{(0)}[-2.611]/(4\pi)^2, \qquad &
% g_Z^{(1)} = g_Z^{(0)}[-0.0824]/(4\pi)^2,
%  \end{array}\eeq
%\xxx{There are imaginary parts at 1 loop (not sure they are right).
%They describe the width of the particles.  Only the Higgs is immune, as it negligibly decays into light particles.
%I have to think what to do with them, given that imaginary parts in the (1 loop)$^2$ terms affect the real part of the
%(2 loop) full result
%}

\section{SM RGE equations up to three loops}\label{SM-RGE}
We list here the known results for the
renormalisation group equations  up to 3 loop order
for the sizeable SM couplings,
$g_1,g_2,g_3,y_t$ and $\lambda$ in the $\MS$ scheme.
We write numerically those  3-loop coefficients that involve the $\zeta_3$ constant.
Stopping for simplicity at two loops,
we also write RGE equations for the smaller bottom and tau Yukawa coupling and their contributions
to the RGE of the large couplings.
Our numerical code includes full RGE at 3 loops.

\newcommand{\acapo}{\\ &&+}

\subsection{Gauge couplings}
RGE for the hypercharge gauge coupling in GUT normalisation ($g_1^2 = 5 g_Y^2/3$):
\begin{eqnarray}
\frac{dg_1^2}{d\ln\bar\mu^2} &=& \frac{g_1^4}{(4\pi)^2} \bigg[\frac{41 }{10}\bigg]
+ \frac{g_1^4}{(4\pi)^4} \bigg[ \frac{44 g_3^2}{5}+\frac{27 g_2^2}{10}+\frac{199 g_1^2}{50}-\frac{17 y_t^2}{10}
-\frac{y_b^2}{2}-\frac{3 y_{\tau }^2}{2} \bigg]  +
\nonumber \\ && 
+ \frac{g_1^4}{(4\pi)^6} \bigg[ y_t^2\left( \frac{189 y_t^2}{16}-\frac{29 g_3^2}{5}-\frac{471 g_2^2}{32}
-\frac{2827 g_1^2}{800}\right) + \lambda \left( -\frac{9 \lambda}{5} +\frac{9 g_2^2}{10}+ \frac{27 g_1^2}{50}\right) +
\nonumber \\ &&
+\frac{297 g_3^4}{5}+\frac{789 g_2^4}{64}-\frac{388613 g_1^4}{24000}-\frac{3 g_3^2 g_2^2}{5} 
-\frac{137 g_3^2g_1^2}{75}+\frac{123 g_2^2g_1^2}{160} \bigg] \ .
     \end{eqnarray}
%\begin{eqnarray*}
%\frac{dg_1^2}{d\ln\bar\mu^2} &=& \frac{1}{(4\pi)^2} \bigg[\frac{41 g_1^4}{10}\bigg]
%+ \frac{1}{(4\pi)^4} \bigg[g_1^4 \left(-\frac{y_b^2}{2}+\frac{27 g_2^2}{10}+\frac{44 g_3^2}{5}-\frac{17 y_t^2}{10}-\frac{3 y_{\tau }^2}{2}\right)+\frac{199 g_1^6}{50}\bigg]+  \\ 
%   && + \frac{1}{(4\pi)^6} \bigg[g_1^4 \left(\frac{9 g_2^2 \lambda }{10}-(\frac{471 g_2^2}{32}+\frac{29 g_3^2}{5}) y_t^2+\frac{789 g_2^4}{64}-\frac{3}{5} g_3^2 g_2^2+\frac{297 g_3^4}{5}-\frac{9 \lambda ^2}{5}+\frac{2157 y_t^4}{80}\right)+
%  \acapo 
%   g_1^6 \left(\frac{123 g_2^2}{160}-\frac{137 g_3^2}{75}+\frac{27 \lambda }{50}-\frac{2827 y_t^2}{800}\right)-\frac{388613 g_1^8}{24000}\bigg].
%     \end{eqnarray*}
RGE for the SU(2)$_L$ gauge coupling:
\begin{eqnarray}
\frac{dg_2^2}{d\ln\bar\mu^2} &=& \frac{g_2^4}{(4\pi)^2} \bigg[-\frac{19}{6}\bigg]+
\frac{g_2^4}{(4\pi)^4} \bigg[ 12 g_3^2+\frac{35 g_2^2}{6}+\frac{9 g_1^2}{10}-\frac{3 y_t^2}{2}
-\frac{3y_b^2}{2}-\frac{y_{\tau }^2}{2} \bigg]  +
\nonumber \\ && 
+ \frac{g_2^4}{(4\pi)^6} \bigg[ y_t^2\left( \frac{147 y_t^2}{16}-7 g_3^2-\frac{729 g_2^2}{32}
-\frac{593 g_1^2}{160}\right) + \lambda \left( -3 \lambda +\frac{3 g_2^2}{2}+ \frac{3 g_1^2}{10} \right) +
\nonumber \\ &&
+81 g_3^4+\frac{324953 g_2^4}{1728}-\frac{5597 g_1^4}{1600}+39 g_3^2 g_2^2 
-\frac{g_3^2g_1^2}{5}+\frac{873 g_2^2g_1^2}{160} \bigg]  \ .
    \end{eqnarray}
%\begin{eqnarray*}
%\frac{dg_2^2}{d\ln\bar\mu^2} &=& \frac{1}{(4\pi)^2} \bigg[-\frac{19 g_2^4}{6}\bigg] + \frac{1}{(4\pi)^4} \bigg[g_2^4 \left(-\frac{3 y_b^2}{2}+\frac{9 g_1^2}{10}+12 g_3^2-\frac{3 y_t^2}{2}-\frac{y_{\tau }^2}{2}\right)+\frac{35 g_2^6}{6}\bigg]+  \\ 
%   && + \frac{1}{(4\pi)^6} \bigg[g_2^4 \left(\frac{3 g_1^2 \lambda }{10}-(\frac{593 g_1^2}{160}+7 g_3^2) y_t^2-\frac{5597 g_1^4}{1600}-\frac{1}{5} g_3^2 g_1^2+81 g_3^4-3 \lambda ^2+\frac{147 y_t^4}{16}\right)+
%   \acapo
%   g_2^6 \left(\frac{873 g_1^2}{160}+39 g_3^2+\frac{3 \lambda}{2} -\frac{729 y_t^2}{32}\right)+\frac{324953 g_2^8}{1728}\bigg].
%  \end{eqnarray*}
%\subsection{The strong gauge coupling}
RGE for the strong gauge coupling, including also pure QCD terms at 4 loops:
\begin{eqnarray}
\frac{dg_3^2}{d\ln\bar\mu^2} &=& \frac{g_3^4}{(4\pi)^2} \bigg[-7\bigg]
+ \frac{g_3^4}{(4\pi)^4} \bigg[ -26 g_3^2+\frac{9 g_2^2}{2}+\frac{11 g_1^2}{10}-2 y_t^2-2 y_b^2 \bigg]  +
\nonumber \\ && 
+ \frac{g_3^4}{(4\pi)^6} \bigg[ y_t^2\left( 15 y_t^2-40 g_3^2-\frac{93 g_2^2}{8}
-\frac{101 g_1^2}{40}\right) +
\\ &&
+\frac{65 g_3^4}{2}+\frac{109 g_2^4}{8}-\frac{523 g_1^4}{120}+21 g_3^2 g_2^2 
+\frac{77g_3^2g_1^2}{15}-\frac{3 g_2^2g_1^2}{40} \bigg]  + \frac{g_3^{10}}{(4\pi)^8} \bigg[-2472.28 \bigg]\ .
\nonumber     
\end{eqnarray}
%\begin{eqnarray*}
%\frac{dg_3^2}{d\ln\bar\mu^2}&=& \frac{1}{(4\pi)^2} \bigg[-7 g_3^4\bigg]+
% \frac{1}{(4\pi)^4} \bigg[g_3^4 \left(-2 y_b^2+\frac{11 g_1^2}{10}+\frac{9 g_2^2}{2}-2 y_t^2\right)-26 g_3^6\bigg]+  \\ 
%   && + \frac{1}{(4\pi)^6} \bigg[
%g_3^4 \left(- (\frac{101 g_1^2}{40}+\frac{93 g_2^2}{8}) y_t^2-\frac{523 g_1^4}{120}-\frac{3}{40} g_2^2 g_1^2+\frac{109 g_2^4}{8}+15 y_t^4\right)+\acapo
%   g_3^6 \left(\frac{77 g_1^2}{15}+21 g_2^2-40 y_t^2\right)+
%     \frac{65 g_3^8}{2}
% \bigg]
%+ \frac{1}{(4\pi)^8} \bigg[-2472.28 g_3^{10}\bigg].
% \end{eqnarray*}
\subsection{Higgs quartic coupling}
 RGE for the Higgs quartic coupling:
 \begin{eqnarray}
\frac{d\lambda}{d\ln\bar\mu^2}&=& \frac{1}{(4\pi)^2} \bigg[\lambda  \left(12 \lambda +6 y_t^2+6 y_b^2+2 y_{\tau }^2-\frac{9 g_2^2}{2}-\frac{9 g_1^2}{10}\right) -3 y_t^4-3 y_b^4-y_{\tau }^4
+\frac{9 g_2^4}{16}+\frac{27 g_1^4}{400}+\frac{9 g_2^2 g_1^2}{40} \bigg]  +
\nonumber \\ && 
+ \frac{1}{(4\pi)^4} \bigg[\lambda ^2 \left(-156\lambda -72 y_t^2-72 y_b^2 -24 y_{\tau }^2+54 g_2^2+\frac{54 g_1^2}{5}\right)
+\lambda y_t^2 \left( -\frac{3 y_t^2}{2}-21 y_b^2+40 g_3^2+
\right. \nonumber \\ && \left.
+\frac{45 g_2^2}{4}+\frac{17 g_1^2}{4}\right)
+\lambda y_b^2 \left( -\frac{3 y_b^2}{2}+40 g_3^2+\frac{45 g_2^2}{4}+\frac{5 g_1^2}{4}\right)
+\lambda y_\tau^2 \left( -\frac{y_\tau^2}{2}+\frac{15 g_2^2}{4}+\frac{15 g_1^2}{4}\right)
+ \nonumber \\ && \lambda \left( -\frac{73 g_2^4}{16}
+\frac{1887 g_1^4}{400}+\frac{117 g_2^2 g_1^2}{40} \right)
+y_t^4 \left( 15 y_t^2-3y_b^2-16g_3^2-\frac{4 g_1^2}{5}\right)
+
\nonumber \\ && 
+y_t^2 \left( -\frac{9 g_2^4}{8}-\frac{171 g_1^4}{200}+\frac{63 g_2^2g_1^2}{20}\right)+y_b^4 \left( -3 y_t^2+15y_b^2-16g_3^2+\frac{2 g_1^2}{5}\right)+
\nonumber \\ && +y_b^2 \left( -\frac{9 g_2^4}{8}+\frac{9 g_1^4}{40}+\frac{27 g_2^2g_1^2}{20}\right)
+y_\tau^4 \left( 5 y_\tau^2-\frac{6 g_1^2}{5}\right)+y_\tau^2 \left( -\frac{3 g_2^4}{8}-\frac{9 g_1^4}{8}+\frac{33 g_2^2g_1^2}{20}\right)+
\nonumber \\ && 
+\frac{305 g_2^6}{32} -\frac{3411 g_1^6}{4000} -\frac{289 g_2^4 g_1^2}{160} -\frac{1677 g_2^2 g_1^4}{800} 
\bigg]+
 \nonumber \\ && 
 + \frac{1}{(4\pi)^6} \bigg[\lambda ^3 \left(  6011.35 \lambda +873 y_t^2-387.452 g_2^2-77.490 g_1^2\right)   
+\lambda ^2 y_t^2\left( 1768.26 y_t^2+160.77 g_3^2+
\right. \nonumber \\ && \left.
-359.539 g_2^2-63.869 g_1^2\right)   
+\lambda ^2 \left( -790.28 g_2^4-185.532 g_1^4 -316.64 g_2^2 g_1^2\right)
+\lambda  y_t^4\left( -223.382 y_t^2+
\right. \nonumber \\ && \left.
-662.866 g_3^2-5.470 g_2^2-21.015 g_1^2\right)
+\lambda  y_t^2\left(356.968 g_3^4-319.664 g_2^4-74.8599 g_1^4+15.1443 g_3^2 g_2^2+
\right. \nonumber \\ && \left.
+17.454 g_3^2 g_1^2+5.615 g_2^2 g_1^2 \right)
+\lambda  g_2^4\left(-57.144 g_3^2+865.483 g_2^2+79.638  g_1^2\right)
+\lambda  g_1^4\left(-8.381 g_3^2+
\right. \nonumber \\ && \left.
+61.753 g_2^2+28.168 g_1^2 \right)
+  y_t^6\left(-243.149 y_t^2+250.494 g_3^2+74.138 g_2^2+33.930 g_1^2\right)+
\nonumber \\ &&
+  y_t^4\left(-50.201 g_3^4+15.884 g_2^4+15.948 g_1^4+13.349 g_3^2 g_2^2+17.570 g_3^2 g_1^2-70.356 g_2^2 g_1^2 \right)+
\nonumber \\ &&
+  y_t^2g_3^2\left(16.464 g_2^4 +1.016  g_1^4+11.386 g_2^2  g_1^2\right)
+  y_t^2g_2^4\left( 62.500 g_2^2+13.041  g_1^2\right)+
\nonumber \\ &&
+  y_t^2g_1^4\left(10.627 g_2^2 +11.117 g_1^2\right)
+  g_3^2\left( 7.536 g_2^6+0.663  g_1^6+1.507 g_2^4 g_1^2+1.105 g_2^2  g_1^4\right)+
\nonumber \\ &&
-114.091 g_2^8-1.508 g_1^8-37.889 g_2^6 g_1^2+6.500 g_2^4 g_1^4-1.543 g_2^2 g_1^6\bigg] \ .
 \end{eqnarray}

     \subsection{Higgs mass term}

RGE for the Higgs mass term:
\begin{eqnarray}
\frac{dm^2}{d\ln\bar\mu^2}&=& \frac{m^2}{(4\pi)^2} \bigg[ 6 \lambda +3  y_t^2 +3 y_b^2 + y_ {\tau }^2
-\frac{9 g_ 2^2}{4} -\frac{9 g_ 1^2 }{20} \bigg]+
\nonumber \\ &&
 + \frac{m^2}{(4\pi)^4} \bigg[\lambda  \left(-30 \lambda -36 y_t^2 -36 y_b^2 -12  y_ {\tau }^2
 +36 g_ 2^2+\frac{36 g_ 1^2}{5} \right)+
\nonumber \\ &&
 +y_t^2 \left(-\frac{27 y_t^2}{4} -\frac{21 y_b^2}{2}  +20 g_ 3^2
+\frac{45 g_ 2^2}{8}  +\frac{17 g_ 1^2}{8}  \right)
 +y_b^2 \left(-\frac{27 y_b^2}{4}  +20 g_ 3^2
 +\frac{45 g_ 2^2}{8}  +\frac{5 g_ 1^2}{8}  \right)+
\nonumber \\ &&
+y_\tau^2 \left(-\frac{9 y_\tau^2}{4}  +\frac{15 g_ 2^2}{8}  +\frac{15 g_ 1^2}{8}  \right) 
-\frac{145}{32} g_ 2^4 +\frac{1671}{800} g_ 1^4+\frac{9 g_ 2^2 g_ 1^2}{16} \bigg]+
\nonumber \\ &&
 + \frac{m^2}{(4\pi)^6} \bigg[\lambda^2  \left(1026 \lambda +\frac{297 y_t^2}{2}  
 -192.822 g_ 2^2-38.564 g_ 1^2 \right)
 +\lambda y_t^2  \left(347.394  y_t^2+80.385 g_ 3^2 +
 \right. \nonumber \\ && \left.
 -318.591 g_ 2^2-59.699 g_ 1^2\right)
 +\lambda  \left( -64.5145 g_ 2^4-65.8056 g_ 1^4-37.8231 g_ 2^2 g_ 1^2\right)
 + y_t^4  \left( 154.405  y_t^2+
 \right. \nonumber \\ && \left.
 -209.24 g_ 3^2-3.82928 g_ 2^2-7.50769 g_ 1^2 \right)
 + y_t^2  \left( 178.484 g_ 3^4 -102.627 g_ 2^4-27.721 g_ 1^4+
 \right. \nonumber \\ && \left.
 +7.572 g_ 3^2 g_ 2^2
 +8.727 g_ 3^2 g_ 1^2+11.470 g_ 2^2 g_ 1^2 \right)
 +g_ 2^4  \left( -28.572  g_ 3^2+301.724 g_ 2^2+9.931  g_ 1^2 \right)+
 \nonumber \\ &&
 +g_ 1^4  \left( -4.191 g_ 3^2 +9.778 g_ 2^2 + 8.378 g_ 1^2 \right) \bigg] \ .
     \end{eqnarray}

%\begin{eqnarray*}
%\frac{dm^2}{d\ln\bar\mu^2}&=& \frac{m^2}{(4\pi)^2} \bigg[3 y_b^2 -\frac{9}{20} g_ 1^2 -\frac{9}{4} g_ 2^2 +6 \lambda  + y_ {\tau }^2+3  y_t^2\bigg]+  \\ 
%   && + \frac{m^2}{(4\pi)^4} \bigg[\lambda  \left(-36 y_b^2 +\frac{36}{5} g_ 1^2 +36 g_ 2^2 -12  y_ {\tau }^2-36  y_t^2\right)+\frac{5}{8} g_ 1^2 y_b^2 +\frac{45}{8} g_ 2^2 y_b^2 +20 g_ 3^2 y_b^2 +
%   \acapo
%   y_t^2 \left(-\frac{21}{2} y_b^2 +\frac{17}{8} g_ 1^2 +\frac{45}{8} g_ 2^2 +20 g_ 3^2 \right)-\frac{27}{4} y_b^4 +\frac{15}{8} g_ 1^2  y_ {\tau }^2+\frac{15}{8} g_ 2^2  y_ {\tau }^2+\frac{1671}{800} g_ 1^4 +\acapo \frac{9}{16} g_ 2^2 g_ 1^2 -\frac{145}{32} g_ 2^4 -\frac{9}{4}  y_ {\tau }^4-\frac{27}{4}  y_t^4 -30 \lambda ^2\bigg]+  \\ 
%   && + \frac{m^2}{(4\pi)^6} \bigg[\lambda ^2 \left(-38.564 g_ 1^2 -192.822 g_ 2^2 +\frac{297}{2}  y_t^2\right)+\lambda  \left(y_t^2 (-59.699 g_ 1^2 -318.591 g_ 2^2 +
%            \right.\acapo\left.   
%            80.385 g_ 3^2 \right)+
%   -65.8056 g_ 1^4 -37.8231 g_ 2^2 g_ 1^2 -64.5145 g_ 2^4 +347.394  y_t^4)+
%   \acapo
%   y_t^4 \left(-7.50769 g_ 1^2 -3.82928 g_ 2^2 -209.24 g_ 3^2 \right)+154.405  y_t^6+
%         \acapo   
%         y_t^2 \left(-27.721 g_ 1^4 +11.470 g_ 2^2 g_ 1^2 +8.727 g_ 3^2 g_ 1^2 -102.627 g_ 2^4 +178.484 g_ 3^4 +7.572 g_ 2^2 g_ 3^2 \right)+
%         \acapo
%         8.378 g_ 1^6 +9.778 g_ 2^2 g_ 1^4 -4.191 g_ 3^2 g_ 1^4 +9.931 g_ 2^4 g_ 1^2 +301.724 g_ 2^6 -28.572 g_ 2^4 g_ 3^2 +1026 \lambda ^3 \bigg].
%     \end{eqnarray*}

 \subsection{Yukawa couplings}
RGE for the top Yukawa coupling:
\begin{eqnarray}
\frac{dy_t^2}{d\ln\bar\mu^2}&=& \frac{y_t^2}{(4\pi)^2} \bigg[ \frac{9 y_t^2}{2}+\frac{3 y_b^2}{2}+y_{\tau }^2
-8 g_3^2-\frac{9 g_2^2}{4}-\frac{17 g_1^2}{20}\bigg]+
\nonumber \\ &&
+\frac{y_t^2}{(4\pi)^4} \bigg[ y_t^2 \left( -12 y_t^2-\frac{11 y_b^2}{4}-\frac{9 y_\tau^2}{4}-12 \lambda 
+36 g_3^2+\frac{225 g_2^2}{16}+\frac{393 g_1^2}{80}\right)+
\nonumber \\ &&
+y_b^2 \left( -\frac{y_b^2}{4}+\frac{5 y_{\tau }^2}{4} 
+4 g_3^2+\frac{99 g_2^2}{16}  +\frac{7 g_1^2}{80} \right)
+y_\tau^2 \left( -\frac{9 y_\tau^2}{4}+\frac{15}{8} g_2^2+\frac{15}{8} g_1^2\right)+
\nonumber \\ &&
+6\lambda^2 -108 g_3^4 -\frac{23 g_2^4}{4}+ \frac{1187 g_1^4}{600}+9 g_3^2 g_2^2+ 
   \frac{19}{15} g_3^2 g_1^2-\frac{9}{20} g_2^2 g_1^2 \bigg]+
\nonumber \\ &&
+\frac{y_t^2}{(4\pi)^6} \bigg[ 
y_t^4 \left( 58.6028 y_t^2 +198 \lambda -157 g_3^2 -\frac{1593 g_2^2}{16}-\frac{2437 g_1^2}{80}\right)
+\lambda y_t^2 \left( \frac{15 \lambda }{4}+16 g_3^2+
\right. \nonumber \\ && \left.
-\frac{135 g_2^2}{2}-\frac{127 g_1^2}{10}\right)
+y_t^2 \left( 363.764 g_3^4 +16.990 g_2^4-24.422 g_1^4+48.370 g_3^2 g_2^2+18.074 g_3^2 g_1^2+
\right.\nonumber  \\ && \left.
+34.829 g_2^2 g_1^2 \right)
+\lambda^2 \left( -36 \lambda +45 g_2^2 +9 g_1^2\right)
+\lambda \left( -\frac{171 g_2^4}{16}-\frac{1089 g_1^4}{400}+\frac{117 g_2^2 g_1^2}{40} \right)+
 \nonumber \\ &&
-619.35 g_3^6+169.829 g_2^6+16.099 g_1^6+73.654 g_3^4 g_2^2-15.096 g_3^4 g_1^2
-21.072 g_3^2 g_2^4+
\nonumber \\ && 
-22.319 g_3^2 g_1^4-\frac{321}{20} g_3^2 g_2^2 g_1^2-4.743 g_2^4 g_1^2
 -4.442 g_2^2 g_1^4      
\bigg] \ .
 \end{eqnarray}
%\begin{eqnarray*}
% %
%\frac{dy_t^2}{d\ln\bar\mu^2}&=& \frac{1}{(4\pi)^2} \bigg[y_t^2 \left(\frac{3 y_b^2}{2}-\frac{17 g_1^2}{20}-\frac{9 g_2^2}{4}-8 g_3^2+y_{\tau }^2\right)+\frac{9 y_t^4}{2}\bigg]+  \\ 
%   && + \frac{1}{(4\pi)^4} \bigg[y_t^2 \left(\frac{7}{80} g_1^2 y_b^2+\frac{99}{16} g_2^2 y_b^2+4 g_3^2 y_b^2+\frac{5}{4} y_b^2 y_{\tau }^2-\frac{y_b^4}{4}+\frac{15}{8} g_1^2 y_{\tau }^2+\frac{15}{8} g_2^2 y_{\tau }^2+
%    \right.\acapo\left.    
%  \frac{1187 g_1^4}{600}-\frac{9}{20} g_2^2 g_1^2+ 
%   \frac{19}{15} g_3^2 g_1^2-\frac{23 g_2^4}{4}-108 g_3^4+9 g_2^2 g_3^2+6 \lambda ^2-\frac{9 y_{\tau }^4}{4}\right)+
% \acapo  
%   y_t^4 \left(-\frac{11 y_b^2}{4}+\frac{393 g_1^2}{80}+\frac{225 g_2^2}{16}+36 g_3^2-12 \lambda -\frac{9 y_{\tau }^2}{4}\right)-12 y_t^6\bigg]+  \\ 
%   %3loop
%   && + \frac{1}{(4\pi)^6} \bigg[\bigg( (16 g_3^2-\frac{127 g_1^2}{10}-\frac{135 g_2^2}{2}) \lambda -24.422 g_1^4+34.829 g_2^2 g_1^2+18.074 g_3^2 g_1^2+16.990 g_2^4+
%\acapo
%   363.764 g_3^4+48.370 g_2^2 g_3^2+\frac{15 \lambda ^2}{4}\bigg) y_t^4+\bigg(  (9 g_1^2+45 g_2^2) \lambda ^2+(\frac{117}{40} g_2^2 g_1^2-\frac{1089 g_1^4}{400}-\frac{171 g_2^4}{16}) \lambda+
%    \acapo 
%     16.099 g_1^6-4.442 g_2^2 g_1^4-22.319 g_3^2 g_1^4-4.743 g_2^4 g_1^2-15.096 g_3^4 g_1^2-\frac{321}{20} g_2^2 g_3^2 g_1^2+
%\acapo
%      169.829 g_2^6-619.35 g_3^6+73.654 g_2^2 g_3^4-21.072 g_2^4 g_3^2-36 \lambda ^3\bigg) y_t^2+
%      \acapo
%      (-\frac{2437 g_1^2}{80}-\frac{1593 g_2^2}{16}-157 g_3^2+198 \lambda ) y_t^6+58.6028 y_t^8\bigg].
% \end{eqnarray*}
RGE for the bottom Yukawa coupling  (up to two loops):
\begin{eqnarray}
\frac{dy_b^2}{d\ln\bar\mu^2}&=& \frac{y_b^2}{(4\pi)^2} \bigg[ \frac{3 y_t^2}{2}+\frac{9 y_b^2}{2}+y_{\tau }^2
-8 g_3^2-\frac{9 g_2^2}{4}-\frac{g_1^2}{4}\bigg]+
\nonumber \\ &&
+\frac{y_b^2}{(4\pi)^4} \bigg[ y_t^2 \left( -\frac{y_t^2}{4}-\frac{11 y_b^2}{4}+\frac{5 y_\tau^2}{4} 
+4 g_3^2+\frac{99 g_2^2}{16}+\frac{91 g_1^2}{80}\right)+
\nonumber \\ &&
+y_b^2 \left( -12y_b^2-\frac{9 y_{\tau }^2}{4}-12 \lambda 
+36 g_3^2+\frac{225 g_2^2}{16}  +\frac{237 g_1^2}{80} \right)
+y_\tau^2 \left( -\frac{9 y_\tau^2}{4}+\frac{15}{8} g_2^2+\frac{15}{8} g_1^2\right)+
\nonumber \\ &&
+6\lambda^2 -108 g_3^4 -\frac{23 g_2^4}{4}- \frac{127 g_1^4}{600}+9 g_3^2 g_2^2+ 
   \frac{31}{15} g_3^2 g_1^2-\frac{27}{20} g_2^2 g_1^2 \bigg] \ .
 \end{eqnarray}
%\begin{eqnarray*}
%\frac{dy_b^2}{d\ln\bar\mu^2}&=& \frac{1}{(4\pi)^2} \bigg[y_b^2 \left(-\frac{g_1^2}{4}-\frac{9 g_2^2}{4}-8 g_3^2+\frac{3 y_t^2}{2}+y_{\tau }^2\right)+\frac{9 y_b^4}{2}\bigg]+  \\ 
%   && + \frac{1}{(4\pi)^4} \bigg[y_b^2 \left(y_t^2 (\frac{91 g_1^2}{80}+\frac{99 g_2^2}{16}+4 g_3^2+\frac{5 y_{\tau }^2}{4})+\frac{15}{8} g_1^2 y_{\tau }^2+\frac{15}{8} g_2^2 y_{\tau }^2-\frac{127 g_1^4}{600}+
%    \right.\acapo\left.    -\frac{27}{20} g_2^2 g_1^2+
%    \frac{31}{15} g_3^2 g_1^2-\frac{23 g_2^4}{4}-108 g_3^4+9 g_2^2 g_3^2+6 \lambda ^2-\frac{y_t^4}{4}-\frac{9 y_{\tau }^4}{4}\right)+
%    \acapo
%    y_b^4 \left(\frac{237 g_1^2}{80}+\frac{225 g_2^2}{16}+36 g_3^2-12 \lambda -\frac{11 y_t^2}{4}-\frac{9 y_{\tau }^2}{4}\right)-12 y_b^6\bigg].
% \end{eqnarray*}
RGE for the tau Yukawa coupling  (up to two loops):
\begin{eqnarray}
\frac{dy_\tau^2}{d\ln\bar\mu^2}&=& \frac{y_\tau^2}{(4\pi)^2} \bigg[ 3 y_t^2+3 y_b^2+\frac{ 5 y_{\tau }^2}{2}
-\frac{9 g_2^2}{4}-\frac{9 g_1^2}{4}\bigg]+
\frac{y_\tau^2}{(4\pi)^4} \bigg[ +6\lambda^2  -\frac{23 g_2^4}{4}+ \frac{1371 g_1^4}{200}
   +\frac{27}{20} g_2^2 g_1^2+
\nonumber \\ &&
y_t^2 \left( -\frac{27y_t^2}{4}+\frac{3 y_b^2}{2}-\frac{27 y_\tau^2}{4} 
+20 g_3^2+\frac{45 g_2^2}{8}+\frac{17 g_1^2}{8}\right)+
\\ &&
+y_b^2 \left( -\frac{27 y_b^2}{4}-\frac{27 y_{\tau }^2}{4}
+20 g_3^2+\frac{45 g_2^2}{8}  +\frac{5 g_1^2}{8} \right)
+y_\tau^2 \left( -3 y_\tau^2-12\lambda+\frac{165}{16} g_2^2+\frac{537}{80} g_1^2\right) \bigg] \ . \nonumber
 \end{eqnarray}
%\begin{eqnarray*}
%\frac{dy_\tau^2}{d\ln\bar\mu^2}&=& \frac{1}{(4\pi)^2} \bigg[y_{\tau }^2 \left(3 y_b^2-\frac{9 g_1^2}{4}-\frac{9 g_2^2}{4}+3 y_t^2\right)+\frac{5 y_{\tau }^4}{2}\bigg]+ \frac{1}{(4\pi)^4} \bigg[y_{\tau }^2 \left(y_t^2 (\frac{3 y_b^2}{2}+\frac{17 g_1^2}{8}+\frac{45 g_2^2}{8}+20 g_3^2)+
% \right.\acapo\left.  
% \frac{5}{8} g_1^2 y_b^2+\frac{45}{8} g_2^2 y_b^2+20 g_3^2 y_b^2-\frac{27 y_b^4}{4}+  
%   \frac{1371 g_1^4}{200}+\frac{27}{20} g_2^2 g_1^2-\frac{23 g_2^4}{4}+6 \lambda ^2-\frac{27 y_t^4}{4}\right)+
%\acapo
%   y_{\tau }^4 \left(-\frac{27 y_b^2}{4}+\frac{537 g_1^2}{80}+\frac{165 g_2^2}{16}-12 \lambda -\frac{27 y_t^2}{4}\right)-3 y_{\tau }^6\bigg].
%    \end{eqnarray*}
    
    \bigskip

\section{Effective potential at two loops}\label{eff-potential-app}
The effective potential including one-loop and two-loop corrections in Landau gauge
for $h\gg v$ is given by eq.~(\ref{eff-potential-high-h}), where  \cite{Casas:1994qy,DDEEGIS} 
\beq \lambda_{\rm eff}(h) =e^{4\Gamma(h)} \bigg[ \lambda(\bar\mu=h) + \lambda_{\rm eff}^{(1)}(\bar\mu=h) + \lambda_{\rm eff}^{(2)}(\bar\mu=h)\bigg].\eeq
All running couplings are evaluated at $\bar\mu=h$. Here, $\Gamma(h)\equiv \int_{M_t}^h \gamma(\bar\mu)d\ln\bar\mu$, with $\gamma$ the Higgs field anomalous dimension,
\bea \gamma &=& \frac{1}{(4\pi)^2} \bigg[ \frac94 g_2^2 + \frac{9}{20} g_1^2 -3 y_t^2-3y_b^2-y_\tau^2 \bigg] +
\nonumber \\&&+
\frac{1}{(4\pi)^4}\bigg[ 
y_t^2 \left(-\frac{3 y_b^2}{2}-\frac{17 g_1^2}{8}-\frac{45 g_2^2}{8}-20   g_3^2+\frac{27 y_t^2}{4}\right)- y_{\tau }^2
   \left(\frac{15 g_1^2}{8}+\frac{15
   g_2^2}{8}-\frac{9 y_{\tau}^2}{4}\right)   +
      \nonumber\\&& +
  y_b^2        \left(-\frac{5 g_1^2}{8}-\frac{45 g_2^2}{8}-20 g_3^2+\frac{27 y_b^2}{4}\right)
  -\frac{1293 g_1^4}{800}-\frac{27}{80} g_2^2
   g_1^2+\frac{271 g_2^4}{32}-6 \lambda ^2\bigg]+ \nonumber \\ 
&&+
\frac{1}{(4\pi)^6}\bigg[ 
-9 g_1^2 \lambda ^2-45 g_2^2 \lambda ^2+1.07 g_1^4 \lambda +3.57 g_2^2
   g_1^2 \lambda +
   8.92 g_2^4 \lambda +
      \nonumber\\&& +
      14.99 g_1^4 y_t^2+14.13 g_1^2 y_t^4-13.21
   g_2^2 g_1^2 y_t^2-8.73 g_3^2 g_1^2 y_t^2+40.11 g_2^2 y_t^4+
      79.05 g_3^2
   y_t^4+
         \nonumber\\&& +
         23.40 g_2^4 y_t^2-178.48 g_3^4 y_t^2-7.57 g_2^2 g_3^2 y_t^2-5.26
   g_1^6+1.93 g_2^2 g_1^4+
      4.19 g_3^2 g_1^4+
            \nonumber\\&& +
            1.81 g_2^4 g_1^2-158.51 g_2^6+28.57
   g_2^4 g_3^2+36 \lambda ^3+\frac{135}{2} \lambda ^2 y_t^2-45.00 \lambda 
   y_t^4-60.13 y_t^6
\bigg]\ .
 \eea
The one-loop correction is
%
%\beq \lambda_{\rm eff}(h)=e^{4\Gamma(h)} \lambda + \Delta\lambda_{\rm eff},\eeq
%
%
\bea
\lambda_{\rm eff}^{(1)} &=&
\frac{1}{(4\pi)^2}\bigg[  \frac{3g_2^4}{8} ( \ln \frac{g_2^2}{4} - \frac{5}{6}+2\Gamma)
+\frac{3}{16}(g_2^2+g_Y^2)^2 (\ln\frac{g_2^2+g_Y^2}{4} - \frac56  + 2\Gamma)+\nonumber \\
&&
-3 y_t^4 (\ln\frac{y_t^2}{2}-\frac32+2\Gamma)+
3\lambda^2(4 \ln\lambda-6  + 3\ln 3+8\Gamma)\bigg]\ . 
\eea
% A ME PIACE COS" ESPLICITA
%\beq
%\lambda_{\rm eff}^{(1)} =
%\frac{1}{(4\pi)^2}\bigg[ \frac38 g_2^4 ( r_W - \frac{5}{6})
%+\frac{3}{16}(g_2^2+g_Y^2)^2 (r_Z - \frac56  )-3y_t^4 (r_t-\frac32)
%+3\lambda^2 (4r_\lambda -6+3\ln 3)\bigg] ,
%\eeq
The two-loop correction is 
\bea
\lambda_{\rm eff}^{(2)} &=&\frac{1}{(4\pi)^4}\bigg[ 
          8g_3^2y_t^4 \left(3r_t^2-8  r_t+9\right) +\frac{1}{2} y_t^6 \left(-6 r_t r_W-3 r_t^2+48 r_t-6 r_{tW}-69-\pi ^2\right)+
                      \nonumber\\&&+
   \frac{3y_t^2 g_2^4}{16}    \left(8 r_W+4 r_Z-3  r_t^2-6 r_t r_Z-12 r_t+12 r_{tW}+15+2 \pi ^2\right)+
   \nonumber\\&&+
   \frac{y_t^2g_Y^4}{48}
   \left(27 r_t^2-54 r_t  r_Z-68r_t -28 r_Z+189\right)+
   \frac{y_t^2 g_2^2 g_Y^2 }{8} (9 r_t^2-18 r_t r_Z+4r_t+44 r_Z-57)+
      \nonumber\\&&+
      \frac{g_2^6}{192}  (
    36 r_t r_Z\!+\!54 r_t^2\!-\!414 r_W r_Z\!+\!69 r_W^2\!+\!1264 r_W\!+\!156 r_Z^2\!+\!632 r_Z-\!144r_{tW}-2067\!+\!90 \pi ^2)  
      +
      \nonumber\\&&+
     \frac{ g_2^4 g_Y^2}{192} (12 r_t r_Z-6 r_t^2-6 r_W (53 r_Z+50)+213 r_W^2+4 r_Z
   (57 r_Z-91)+817+46 \pi ^2)+
      \nonumber\\&&+
     \frac{ g_2^2 g_Y^4}{576} (132 r_t r_Z-66 r_t^2+306 r_W r_Z-153 r_W^2-36 r_W+924 r_Z^2-4080 r_Z+4359+218 \pi ^2)+\nonumber
      \\&&+
      \frac{g_Y^6}{576} (6
   r_Z (34 r_t+3 r_W-470)-102 r_t^2-9 r_W^2+708 r_Z^2+2883+206 \pi ^2)+
\nonumber       \\&&+
     \frac{   y_t^4 }{6}
\left(4 g_Y^2 (3 r_t^2-8r_t+9)-9 g_2^2 \left(r_t-r_W+1\right)\right)+
   \frac{3}{4} \left(g_2^6-3 g_2^4 y_t^2+4 y_t^6\right)\mathrm{Li}_2\frac{g_2^2}{2y_t^2} +
         \nonumber\\&&+
      \frac{y_t^2}{48}  \xi(\frac{g_2^2+g_Y^2}{2 y_t^2} )
         \left(9 g_2^4
-6 g_2^2 g_Y^2+17 g_Y^4+2y_t^2 \big(7 g_Y^2-73 g_2^2+\frac{64 g_2^4}{g_Y^2+g_2^2}\big) \right)+
   \nonumber\\&&+
    \frac{g_2^2}{64} \xi(\frac{g_2^2+g_Y^2}{g_2^2}) \left(18 g_2^2 g_Y^2+g_Y^4-51 g_2^4-\frac{48
   g_2^6}{g_Y^2+g_2^2}\right)
   \bigg]\ . 
\label{large-h-effeP}\eea
Here we have given $\lambda_{\rm eff}^{(2)}$ in the approximation $\lambda =0$, which is well justified around the instability region. The full expression of $\lambda_{\rm eff}^{(2)}$ can be found in ref.~\cite{DDEEGIS}.
Moreover, we have defined
\be
\xi(z)\equiv \sqrt{z^2-4z}\left[2\ln^2\left(\frac{z-\sqrt{z^2-4z}}{2z}\right)-\ln^2 z-4{\rm Li}_2\left(\frac{z-\sqrt{z^2-4z}}{2z}\right)+\frac{\pi^2}{3}\right],
\ee
where  ${\rm Li}_2%\equiv \sum_{k=1}^{\infty}x^k/k^2
$ is the dilogarithm function, and
\beq r_W = \ln \frac{g_2^2}{4}+2\Gamma \ ,
\qquad r_Z = \ln\frac{g_2^2+g_Y^2}{4}+2\Gamma\ , \qquad
r_t = \ln\frac{y_t^2}{2}+2\Gamma \ ,
%,\qquad
%r_\lambda = \ln\lambda+2\Gamma \ .
\eeq
\be
r_{tW} = (r_t-r_W)\left[ \ln \left(\frac{y_t^2}{2}-\frac{g_2^2}{4}\right)+2\Gamma \right] \ .
\ee

\bigskip

%Moreover, we have introduced the notation and 
%\be
%\xi_{xy}=\xi(x/y)\ ,\;\;r_p\equiv\ln[\kappa_pe^{2\Gamma(h)}]\ ,\;\;
%r_{t/w}\equiv \ln[\kappa_t/\kappa_w]\ ,\;\; 
%r_{(t-w)/t}\equiv \ln[(\kappa_t-\kappa_w)/\kappa_w]\;\; 
%\ee
%and so on, where the index $p$ runs over $p$article species and 
%\beq\begin{array}{c|ccccc}
%p &  t& W & Z & h &\chi\\ \hline
%\kappa_p & y_t^2/2&g^2/4&(g_2^2+g_Y^2)/4&3\lambda&\lambda
%\end{array} \nonumber \eeq

\footnotesize
\begin{multicols}{2}

   \end{multicols}


\begin{thebibliography}{nn}\bibitem{Higgsexp1}
ATLAS Collaboration,
  %``Observation of a new particle in the search for the Standard Model Higgs boson with the ATLAS detector at the LHC,''
  Phys.\ Lett.\ B {716} (2012) 1
 [arXiv:\hhref{1207.7214}].
   
 


\bibitem{Higgsexp2}
  CMS Collaboration,
  %``Observation of a new boson at a mass of 125 GeV with the CMS experiment at the LHC,''
  Phys.\ Lett.\ B {716} (2012) 30
  [arXiv:\hhref{1207.7235}].
  
  


\bibitem{HiggsMass}
CMS Collaboration, \href{http://cds.cern.ch/record/1530524?ln=en}{CMS-PAS-HIG-13-001}.
ATLAS Collaboration, \href{http://cds.cern.ch/record/1523698}{ATLAS-CONF-2013-012}.
CMS Collaboration, \href{http://arxiv.org/abs/arXiv:1312.5353}{arXiv:1312.5353}.
ATLAS Collaboration,   Phys.\ Lett.\ B {726} (2013) 88
  [\href{http://arxiv.org/abs/arXiv:1307.1427}{arXiv:1307.1427}].
M.Bachtis, on behalf of the CMS Collaboration, ``Latest Higgs results from CMS'',  talk at ICHEP 2014.
M. D'Onofrio, on behalf of the ATLAS Collaboration,  \href{http://indico.cern.ch/event/319702/session/1/contribution/22/material/slides/1.pdf}{talk at the 118th LHCC Meeting, 4 June 2014}.
% $M_h$ from $h\to \gamma\gamma$:   
%CMS Collaboration, \href{http://cds.cern.ch/record/1530524?ln=en}{CMS-PAS-HIG-13-001}.
%ATLAS Collaboration, \href{http://cds.cern.ch/record/1523698}{ATLAS-CONF-2013-012}.
%$M_h$ from $h\to ZZ$:
%CMS Collaboration, \href{http://cds.cern.ch/record/1523767?ln=en}{CMS-PAS-HIG-13-002}. 
%ATLAS Collaboration, \href{http://cds.cern.ch/record/1523699}{ATLAS-CONF-2013-013}.
Naive average:
 P.~P.~Giardino, K.~Kannike, I.~Masina, M.~Raidal and A.~Strumia,
  %``The universal Higgs fit,''
  arXiv:\hhref{1303.3570}. 


\bibitem{DDEEGIS}
G.~Degrassi, S.~Di Vita, J.~Elias-Miro, J.~R.~Espinosa, G.~F.~Giudice, G.~Isidori and A.~Strumia,
  %``Higgs mass and vacuum stability in the Standard Model at NNLO,''
  JHEP {1208} (2012) 098
  [arXiv:\hhref{1205.6497}].
    
 


\bibitem{Krive:1976sg}
  I.~V.~Krive and A.~D.~Linde,
  %``On the Vacuum Stability Problem in Gauge Theories,''
  Nucl.\ Phys.\ B {117} (1976) 265.
   
 


\bibitem{Krasnikov:1978pu}
  N.~V.~Krasnikov,
  %``Restriction of the Fermion Mass in Gauge Theories of Weak and Electromagnetic Interactions,''
  Yad.\ Fiz.\  {28} (1978) 549.
   
 


\bibitem{Maiani:1977cg}
  L.~Maiani, G.~Parisi and R.~Petronzio,
  %``Bounds on the Number and Masses of Quarks and Leptons,''
  Nucl.\ Phys.\ B {136} (1978) 115.
     
  


\bibitem{Politzer:1978ic}
  H.~D.~Politzer and S.~Wolfram,
  %``Bounds on Particle Masses in the Weinberg-Salam Model,''
  Phys.\ Lett.\ B {82} (1979) 242
   [Erratum-ibid.\  {83B} (1979) 421].
   
    


\bibitem{Hung:1979dn}
  P.~Q.~Hung,
  %``Vacuum Instability and New Constraints on Fermion Masses,''
  Phys.\ Rev.\ Lett.\  {42} (1979) 873.
   
  


\bibitem{Cabibbo:1979ay}
  N.~Cabibbo, L.~Maiani, G.~Parisi and R.~Petronzio,
  %``Bounds on the Fermions and Higgs Boson Masses in Grand Unified Theories,''
  Nucl.\ Phys.\ B {158} (1979) 295.
    


\bibitem{Linde:1979ny}
  A.~D.~Linde,
  %``Vacuum Instability, Cosmology And Constraints On Particle Masses In The Weinberg-salam Model,''
  Phys.\ Lett.\ B {92} (1980) 119.
   
 


\bibitem{Lindner:1985uk}
  M.~Lindner,
  %``Implications of Triviality for the Standard Model,''
  Z.\ Phys.\ C {31} (1986) 295.
  %%CITATION = ZEPYA,C31,295;%%   


\bibitem{Lindner:1988ww}
  M.~Lindner, M.~Sher and H.~W.~Zaglauer,
  %``Probing Vacuum Stability Bounds at the Fermilab Collider,''
  Phys.\ Lett.\ B {228} (1989) 139.
  


\bibitem{Sher:1988mj}
  M.~Sher,
  %``Electroweak Higgs Potentials and Vacuum Stability,''
  Phys.\ Rept.\  {179} (1989) 273.
  


\bibitem{Arnold:1989cb}
  P.~B.~Arnold,
  %``Can The Electroweak Vacuum Be Unstable?,''
  Phys.\ Rev.\ D {40} (1989) 613.
  


\bibitem{Arnold:1991cv}
  P.~B.~Arnold and S.~Vokos,
  %``Instability of hot electroweak theory: bounds on m(H) and M(t),''
  Phys.\ Rev.\ D {44} (1991) 3620.
  


\bibitem{Sher:1993mf}
  M.~Sher,
  %``Precise vacuum stability bound in the standard model,''
  Phys.\ Lett.\ B {317} (1993) 159
   [Addendum-ibid.\ B {331} (1994) 448]
  [arXiv:\hhref{hep-ph/9307342}].
  


\bibitem{Altarelli:1994rb}
  G.~Altarelli and G.~Isidori,
  %``Lower limit on the Higgs mass in the standard model: An Update,''
  Phys.\ Lett.\ B {337} (1994) 141.
     
   


\bibitem{Casas:1994qy}
  J.~A.~Casas, J.~R.~Espinosa and M.~Quiros,
  %``Improved Higgs mass stability bound in the standard model and implications for supersymmetry,''
  Phys.\ Lett.\ B {342} (1995) 171
  [arXiv:\hhref{hep-ph/9409458}].
   
 


\bibitem{Espinosa:1995se}
  J.~R.~Espinosa and M.~Quiros,
  %``Improved metastability bounds on the standard model Higgs mass,''
  Phys.\ Lett.\ B {353} (1995) 257
  [arXiv:\hhref{hep-ph/9504241}].
 
   


\bibitem{Casas:1996aq}
  J.~A.~Casas, J.~R.~Espinosa and M.~Quiros,
  %``Standard model stability bounds for new physics within LHC reach,''
  Phys.\ Lett.\ B {382} (1996) 374
  [arXiv:\hhref{hep-ph/9603227}].
     
   


\bibitem{Schrempp:1996fb}
  B.~Schrempp and M.~Wimmer,
  %``Top quark and Higgs boson masses: Interplay between infrared and ultraviolet physics,''
  Prog.\ Part.\ Nucl.\ Phys.\  {37} (1996) 1
  [arXiv:\hhref{hep-ph/9606386}].
     
   


\bibitem{Hambye:1996wb}
  T.~Hambye and K.~Riesselmann,
  %``Matching conditions and Higgs mass upper bounds revisited,''
  Phys.\ Rev.\ D {55} (1997) 7255
  [arXiv:\hhref{hep-ph/9610272}].
    
  


\bibitem{IRS}
  G.~Isidori, G.~Ridolfi and A.~Strumia,
  %``On the metastability of the standard model vacuum,''
  Nucl.\ Phys.\ B\ {609} (2001) 387
     [arXiv:\hhref{hep-ph/0104016}].
   
  


\bibitem{Espinosa:2007qp}
  J.~R.~Espinosa, G.~F.~Giudice and A.~Riotto,
  %``Cosmological implications of the Higgs mass measurement,''
  JCAP {0805} (2008) 002
  [arXiv:\hhref{0710.2484}].
    
  


\bibitem{Ellis:2009tp}
  J.~Ellis, J.~R.~Espinosa, G.~F.~Giudice, A.~Hoecker and A.~Riotto,
  %``The Probable Fate of the Standard Model,''
  Phys.\ Lett.\ B {679} (2009) 369
  [arXiv:\hhref{0906.0954}].
  
\bibitem{Hall}
B.~Feldstein, L.~J.~Hall and T.~Watari,
  %``Landscape Prediction for the Higgs Boson and Top Quark Masses,''
  Phys.\ Rev.\ D {74} (2006) 095011
  [hep-ph/0608121].


\bibitem{Holthausen:2011aa}
  M.~Holthausen, K.~S.~Lim and M.~Lindner,
  %``Planck scale Boundary Conditions and the Higgs Mass,''
  JHEP {1202} (2012) 037
  [arXiv:\hhref{1112.2415}].
  


\bibitem{EliasMiro:2011aa}
  J.~Elias-Miro, J.~R.~Espinosa, G.~F.~Giudice, G.~Isidori, A.~Riotto and A.~Strumia,
  %``Higgs mass implications on the stability of the electroweak vacuum,''
  Phys.\ Lett.\ B {709} (2012) 222
  [arXiv:\hhref{1112.3022}].
  


\bibitem{Chen:2012faa}
  C.~-S.~Chen and Y.~Tang,
  %``Vacuum stability, neutrinos, and dark matter,''
  JHEP {1204} (2012) 019
  [arXiv:\hhref{1202.5717}].
  


\bibitem{Lebedev:2012zw}
  O.~Lebedev,
  %``On Stability of the Electroweak Vacuum and the Higgs Portal,''
  Eur.\ Phys.\ J.\ C {72} (2012) 2058
  [arXiv:\hhref{1203.0156}].
  


\bibitem{EliasMiro:2012ay}
  J.~Elias-Miro, J.~R.~Espinosa, G.~F.~Giudice, H.~M.~Lee and A.~Strumia,
  %``Stabilization of the Electroweak Vacuum by a Scalar Threshold Effect,''
  JHEP {1206} (2012) 031
  [arXiv:\hhref{1203.0237}].
  


\bibitem{Rodejohann:2012px}
  W.~Rodejohann and H.~Zhang,
  %``Impact of massive neutrinos on the Higgs self-coupling and electroweak vacuum stability,''
  JHEP {1206} (2012) 022
  [arXiv:\hhref{1203.3825}].
  


\bibitem{Bezrukov:2012sa}
  F.~Bezrukov, M.~Y.~Kalmykov, B.~A.~Kniehl and M.~Shaposhnikov,
  %``Higgs Boson Mass and New Physics,''
  JHEP {1210} (2012) 140
  [arXiv:\hhref{1205.2893}].
  


\bibitem{Datta:2012db}
  A.~Datta and S.~Raychaudhuri,
  %``Vacuum Stability Constraints and LHC Searches for a Model with a Universal Extra Dimension,''
  Phys.\ Rev.\ D {87} (2013) 035018
  [arXiv:\hhref{1207.0476}].
  


\bibitem{Alekhin:2012py}
  S.~Alekhin, A.~Djouadi and S.~Moch,
  %``The top quark and Higgs boson masses and the stability of the electroweak vacuum,''
  Phys.\ Lett.\ B {716} (2012) 214
  [arXiv:\hhref{1207.0980}].
  


\bibitem{Chakrabortty:2012np}
  J.~Chakrabortty, M.~Das and S.~Mohanty,
  %``Constraints on TeV scale Majorana neutrino phenomenology from the Vacuum Stability of the Higgs,''
  arXiv:\hhref{1207.2027}].
  


\bibitem{Anchordoqui:2012fq}
  L.~A.~Anchordoqui, I.~Antoniadis, H.~Goldberg, X.~Huang, D.~Lust, T.~R.~Taylor and B.~Vlcek,
  %``Vacuum Stability of Standard Model$^{++}$,''
  JHEP {1302} (2013) 074
  [arXiv:\hhref{1208.2821}].
  


\bibitem{Masina:2012tz}
  I.~Masina,
  %``The Higgs boson and Top quark masses as tests of Electroweak Vacuum Stability,''
  Phys.\ Rev.\ D {87} (2013) 053001
  [arXiv:\hhref{1209.0393}].
    
  


\bibitem{Chun:2012jw}
  E.~J.~Chun, H.~M.~Lee and P.~Sharma,
  %``Vacuum Stability, Perturbativity, EWPD and Higgs-to-diphoton rate in Type II Seesaw Models,''
  JHEP {1211} (2012) 106
  [arXiv:\hhref{1209.1303}].
    
  


\bibitem{Chung:2012vg}
  D.~J.~H.~Chung, A.~J.~Long and L.~-T.~Wang,
  %``The 125 GeV Higgs and Electroweak Phase Transition Model Classes,''
  Phys.\ Rev.\ D {87} (2013) 023509
  [arXiv:\hhref{1209.1819}].
    
  


\bibitem{Chao:2012mx}
  W.~Chao, M.~Gonderinger and M.~J.~Ramsey-Musolf,
  %``Higgs Vacuum Stability, Neutrino Mass, and Dark Matter,''
  Phys.\ Rev.\ D {86} (2012) 113017
  [arXiv:\hhref{1210.0491}].
P.~S.~Bhupal Dev, D.~K.~Ghosh, N.~Okada and I.~Saha,
  %``125 GeV Higgs Boson and the Type-II Seesaw Model,''
  JHEP {1303} (2013) 150
   [Erratum-ibid.\  {1305} (2013) 049]
  [arXiv:\href{1301.3453}].    
  


\bibitem{Lebedev:2012sy}
  O.~Lebedev and A.~Westphal,
  %``Metastable Electroweak Vacuum: Implications for Inflation,''
  Phys.\ Lett.\ B {719} (2013) 415
  [arXiv:\hhref{1210.6987}].
    


\bibitem{Nielsen:2012pu}
  H.~B.~Nielsen,
  %``PREdicted the Higgs Mass,''
  arXiv:\hhref{1212.5716}.
    
  


\bibitem{Kobakhidze:2013tn}
  A.~Kobakhidze and A.~Spencer-Smith,
  %``Electroweak Vacuum (In)Stability in an Inflationary Universe,''
  Phys.\ Lett.\ B {722} (2013) 130
  [arXiv:\hhref{1301.2846}].
    
  


\bibitem{Tang:2013bz}
  Y.~Tang,
  %``Vacuum Stability in the Standard Model,''
   Mod.\ Phys.\ Lett.\ A {28} (2013) 1330002
  [arXiv:\hhref{1301.5812}].
    
  


\bibitem{Klinkhamer:2013sos}
  F.~R.~Klinkhamer,
  %``Standard Model Higgs field and energy scale of gravity,''
  JETP Letters 97, {297} (2013)
  [arXiv:\hhref{1302.1496}].
    
  


\bibitem{He:2013tla}
  X.~-G.~He, H.~Phoon, Y.~Tang and G.~Valencia,
  %``Unitarity and vacuum stability constraints on the couplings of color octet scalars,''
  JHEP {1305} (2013) 026
  [arXiv:\hhref{1303.4848}].
    
  


\bibitem{Chun:2013soa}
  E.~J.~Chun, S.~Jung and H.~M.~Lee,
  %``Radiative generation of the Higgs potential,''
  arXiv:\hhref{1304.5815}.
    
  


\bibitem{Jegerlehner:2013cta}
  F.~Jegerlehner,
  %``The Standard model as a low-energy effective theory: what is triggering the Higgs mechanism?,''
  arXiv:\hhref{1304.7813}.

    %\cite{Antipin:2013sga}
\bibitem{Antipin:2013sga}
  O.~Antipin, M.~Gillioz, J.~Krog, E.~Mølgaard and F.~Sannino,
  %``Standard Model Vacuum Stability and Weyl Consistency Conditions,''
  JHEP {1308} (2013) 034
  [arXiv:\hhref{1306.3234}].
  %%CITATION = ARXIV:1306.3234;%%


\bibitem{Branchina} V.~Branchina, E.~Messina,
  %``Stability, Higgs Boson Mass and New Physics,''
  arXiv:\hhref{1307.5193}.


   
%%%%%%%%%%%%%%%%%%%%%%%%%%%%%%%%%%%
%%   sect. 2
%%%%%%%%%%%%%%%%%%%%%%%%%%%%%%%%%%%

 


\bibitem{Gross:1973id}
  D.~J.~Gross and F.~Wilczek,
  %``Ultraviolet Behavior of Nonabelian Gauge Theories,''
  Phys.\ Rev.\ Lett.\  {30} (1973) 1343.
   
 


\bibitem{Politzer:1973fx}
  H.~D.~Politzer,
  %``Reliable Perturbative Results for Strong Interactions?,''
  Phys.\ Rev.\ Lett.\  {30} (1973) 1346.
   
 


\bibitem{Caswell:1974gg}
  W.~E.~Caswell,
  %``Asymptotic Behavior of Nonabelian Gauge Theories to Two Loop Order,''
  Phys.\ Rev.\ Lett.\  {33} (1974) 244.
    
  


\bibitem{Jones:1974mm}
  D.~R.~T.~Jones,
  %``Two Loop Diagrams in Yang-Mills Theory,''
  Nucl.\ Phys.\ B {75} (1974) 531.
   
 


\bibitem{Tarasov:1980au}
  O.~V.~Tarasov, A.~A.~Vladimirov and A.~Y.~Zharkov,
  %``The Gell-Mann-Low Function of QCD in the Three Loop Approximation,''
  Phys.\ Lett.\ B {93} (1980) 429.
  


\bibitem{Larin:1993tp}
  S.~A.~Larin and J.~A.~M.~Vermaseren,
  %``The Three loop QCD Beta function and anomalous dimensions,''
  Phys.\ Lett.\ B {303} (1993) 334
   [arXiv:\hhref{hep-ph/9302208}]. 
  


\bibitem{vanRitbergen:1997va}
  T.~van Ritbergen, J.~A.~M.~Vermaseren and S.~A.~Larin,
  %``The Four loop beta function in quantum chromodynamics,''
  Phys.\ Lett.\ B {400} (1997) 379
   [arXiv:\hhref{hep-ph/9701390}].  


\bibitem{Czakon:2004bu}
  M.~Czakon,
  %``The Four-loop QCD beta-function and anomalous dimensions,''
  Nucl.\ Phys.\ B {710} (2005) 485
   [arXiv:\hhref{hep-ph/0411261}].  


\bibitem{Jones:1981we}
  D.~R.~T.~Jones,
  %``The Two Loop beta Function for a G(1) x G(2) Gauge Theory,''
  Phys.\ Rev.\ D {25} (1982) 581.
    
  


\bibitem{Steinhauser:1998cm}
  M.~Steinhauser,
  %``Higgs decay into gluons up to O(alpha**3(s) G(F)m**2(t)),''
  Phys.\ Rev.\ D {59} (1999) 054005
   [arXiv:\hhref{hep-ph/9809507}].  


\bibitem{Machacek:1983tz}
  M.~E.~Machacek and M.~T.~Vaughn,
  %``Two Loop Renormalization Group Equations in a General Quantum Field Theory. 1. Wave Function Renormalization,''
  Nucl.\ Phys.\ B {222} (1983) 83.
  


\bibitem{RGE3-MSS}
L.~N.~Mihaila, J.~Salomon and M.~Steinhauser,
  %``Gauge Coupling Beta Functions in the Standard Model to Three Loops,''
  Phys.\ Rev.\ Lett.\  {108} (2012) 151602
  [arXiv:\hhref{1201.5868}].


\bibitem{Mihaila:2012pz}
  L.~N.~Mihaila, J.~Salomon and M.~Steinhauser,
  %``Renormalization constants and beta functions for the gauge couplings of the Standard Model to three-loop order,''
  Phys.\ Rev.\ D {86} (2012) 096008
  [arXiv:\hhref{1208.3357}].
  


\bibitem{Cheng:1973nv}
  T.~P.~Cheng, E.~Eichten and L.~-F.~Li,
  %``Higgs Phenomena in Asymptotically Free Gauge Theories,''
  Phys.\ Rev.\ D {9} (1974) 2259.
  
 


\bibitem{Fischler:1982du}
  M.~Fischler and J.~Oliensis,
  %``Two Loop Corrections to the Evolution of the Higgs-Yukawa Coupling Constant,''
  Phys.\ Lett.\ B {119} (1982) 385.
   
 


\bibitem{Chetyrkin:2012rz}
  K.~G.~Chetyrkin and M.~F.~Zoller,
  %``Three-loop \beta-functions for top-Yukawa and the Higgs self-interaction in the Standard Model,''
  JHEP {1206} (2012) 033
   [arXiv:\hhref{1205.2892}]. 


\bibitem{Bednyakov:2012en}
  A.~V.~Bednyakov, A.~F.~Pikelner and V.~N.~Velizhanin,
  %``Yukawa coupling beta-functions in the Standard Model at three loops,''
  Phys.\ Lett.\ B {722} (2013) 336
  [arXiv:\hhref{1212.6829}].   



\bibitem{Machacek:1983fi}
  M.~E.~Machacek and M.~T.~Vaughn,
  %``Two Loop Renormalization Group Equations in a General Quantum Field Theory. 2. Yukawa Couplings,''
  Nucl.\ Phys.\ B {236} (1984) 221.
  


\bibitem{Machacek:1984zw}
  M.~E.~Machacek and M.~T.~Vaughn,
  %``Two Loop Renormalization Group Equations in a General Quantum Field Theory. 3. Scalar Quartic Couplings,''
  Nucl.\ Phys.\ B {249} (1985) 70.
  


\bibitem{Luo:2002ey}
  M.~-X.~Luo and Y.~Xiao,
  %``Two loop renormalization group equations in the standard model,''
  Phys.\ Rev.\ Lett.\  {90} (2003) 011601
   [arXiv:\hhref{hep-ph/0207271}].   


\bibitem{Chetyrkin:2013wya}
  K.~G.~Chetyrkin and M.~F.~Zoller,
  %``\beta-function for the Higgs self-interaction in the Standard Model at three-loop level,''
  JHEP {1304} (2013) 091
  [arXiv:\hhref{1303.2890}].
   
 


\bibitem{Bednyakov:2013eba}
  A.~V.~Bednyakov, A.~F.~Pikelner and V.~N.~Velizhanin,
  %``Higgs self-coupling beta-function in the Standard Model at three loops,''
   [arXiv:\hhref{1303.4364}].
  
 


\bibitem{Sirlin:1980nh}
  A.~Sirlin,
  %``Radiative Corrections in the SU(2)-L x U(1) Theory: A Simple 
  % Renormalization Framework,''
  Phys.\ Rev.\ D {22} (1980) 971.
  



\bibitem{Marciano:1980pb}
  W.~J.~Marciano and A.~Sirlin,
  %``Radiative Corrections to Neutrino Induced Neutral Current Phenomena in 
  % the SU(2)-L x U(1) Theory,''
  Phys.\ Rev.\ D {22} (1980) 2695
   [Erratum-ibid.\ D {31} (1985) 213.
 

  


\bibitem{Tarrach:1980up}
  R.~Tarrach,
  %``The Pole Mass in Perturbative QCD,''
  Nucl.\ Phys.\ B {183} (1981) 384.
  
  


\bibitem{Chetyrkin:1999ys}
  K.~G.~Chetyrkin and M.~Steinhauser,
  %``Short distance mass of a heavy quark at order **3(s),''
  Phys.\ Rev.\ Lett.\  {83} (1999) 4001
   [arXiv:\hhref{hep-ph/9907509}].  


\bibitem{Chetyrkin:1999qi}
  K.~G.~Chetyrkin and M.~Steinhauser,
  %``The Relation between the MS-bar and the on-shell quark mass at order alpha(s)**3,''
  Nucl.\ Phys.\ B {573} (2000) 617
  [arXiv:\hhref{hep-ph/9911434}].
    
  


\bibitem{Melnikov:2000qh}
  K.~Melnikov and T.~v.~Ritbergen,
  %``The Three loop relation between the MS-bar and the pole quark masses,''
  Phys.\ Lett.\ B {482} (2000) 99
  [arXiv:\hhref{hep-ph/9912391}].
  
%\cite{Hempfling:1994ar}

\bibitem{Kataev}
  A.L.~Kataev and V.T.~Kim,
  %``Peculiar features of the relations between pole and running heavy quark masses and estimates of the O($\alpha_s^4$) contributions,''
  Phys.\ Part.\ Nucl.\  {41} (2010) 946
  [arXiv:\hhref{1001.4207}].



\bibitem{Hempfling:1994ar}
  R.~Hempfling and B.~A.~Kniehl,
  %``On the relation between the fermion pole mass and MS Yukawa coupling in the standard model,''
  Phys.\ Rev.\ D {51} (1995) 1386
   [arXiv:\hhref{hep-ph/9408313}].  
%\cite{Sirlin:1985ux}


\bibitem{SZ}
  A.~Sirlin and R.~Zucchini,
  %``Dependence Of The Quartic Coupling H(m) On M(h) And The Possible Onset Of New Physics In The Higgs Sector Of The Standard Model,''
  Nucl.\ Phys.\ B {266} (1986) 389.
  


\bibitem{wil}
W.~E.~Caswell and F.~Wilczek,
  %``On the Gauge Dependence of Renormalization Group Parameters,''
  Phys.\ Lett.\ B {49} (1974) 291.
    See also T. Muta, ``Foundations of quantum chromodynamics'', p. 192.


\bibitem{Atkinson:1979ut}
  D.~Atkinson and M.~P.~Fry,
  %``Should One Truncate The Electron Selfenergy?,''
  Nucl.\ Phys.\ B {156} (1979) 301.
    


\bibitem{Mgaugeinvariance}
J.~C.~Breckenridge, M.~J.~Lavelle and T.~G.~Steele,
  %``The Nielsen identities for the two point functions of QED and QCD,''
  Z.\ Phys.\ C {65} (1995) 155
  [arXiv:\hhref{hep-th/9407028}].   


\bibitem{Mgaugeinvariance1}
A.~S.~Kronfeld,
  %``The Perturbative pole mass in QCD,''
  Phys.\ Rev.\ D {58} (1998) 051501
  [arXiv:\hhref{hep-ph/9805215}]. 

  


\bibitem{Wmass}
TeVatron average: FERMILAB-TM-2532-E.
LEP average: CERN-PH-EP/2006-042.  


\bibitem{Zmass}
2012 Particle Data Group average, \href{http://pdg.lbl.gov}{pdg.lbl.gov}.


\bibitem{topmass}
The ATLAS, CDF, CMS, D0 Collaborations, arXiv:\hhref{1403:4427}.
%  Tevatron Electroweak Working Group,
%  %``Combination of CDF and D0 results on the mass of the top quark using up to 5.8~fb-1 of data,''
%  arXiv:\hhref{1107.5255}. 
%  CMS collaboration, \href{https://twiki.cern.ch/twiki/bin/view/CMSPublic/PhysicsResultsTOP}{CMS top physics web site}
%  and talks at the Moriond 2013 conference.
%  ATLAS collaboration, \href{https://twiki.cern.ch/twiki/bin/view/AtlasPublic/TopPublicResults}{ATLAS top physics web site}.

\bibitem{MuLan}
 {{\sc MuLan} Collaboration}.
  arXiv:\hhref{1211.0960}. 

\bibitem{alpha3} 
  S.~Bethke,  
  arXiv:\hhref{1210.0325}.    



\bibitem{Kinoshita:1958ru}
  T.~Kinoshita and A.~Sirlin,
  %``Radiative corrections to Fermi interactions,''
  Phys.\ Rev.\  {113} (1959) 1652.
  
 


\bibitem{Deltaq}    T.~van Ritbergen and R.~G.~Stuart,
  %``On the precise determination of the Fermi coupling constant from the muon lifetime,''
  Nucl.\ Phys.\ B {564} (2000) 343
  [arXiv:\hhref{hep-ph/9904240}]. 
 


\bibitem{Feynart}
  T.~Hahn,
  %``Generating Feynman diagrams and amplitudes with FeynArts 3,''
  Comput.\ Phys.\ Commun.\  {140} (2001) 418
    [arXiv:\hhref{hep-ph/0012260}].  



\bibitem{Mertig:1998vk}
  R.~Mertig and R.~Scharf,
  %``TARCER: A Mathematica program for the reduction of two loop propagator integrals,''
  Comput.\ Phys.\ Commun.\  {111} (1998) 265
    [arXiv:\hhref{hep-ph/9801383}]. 


\bibitem{Tarasov}
  O.~V.~Tarasov,
  %``Generalized recurrence relations for two-loop propagator integrals with arbitrary masses,''
  Nucl.\ Phys.\ B {502} (1197) 455
  [arXiv:\hhref{hep-ph/9703319}]. 
 


\bibitem{Feyncalc}
  R.~Mertig, M.~Bohm and A.~Denner,
  %``FEYN CALC: Computer algebraic calculation of Feynman amplitudes,''
  Comput.\ Phys.\ Commun.\  {64} (1991) 345.
  
  %\cite{Awramik:2002vu}


\bibitem{Awramik:2002vu}
  M.~Awramik, M.~Czakon, A.~Onishchenko and O.~Veretin,
  %``Bosonic corrections to Delta r at the two loop level,''
  Phys.\ Rev.\ D {68} (2003) 053004
   [arXiv:\hhref{hep-ph/0209084}].    


\bibitem{Martinint}
  S.~P.~Martin,
  %``Evaluation of two loop selfenergy basis integrals using differential equations,''
  Phys.\ Rev.\ D {68} (2003) 075002
   [arXiv:\hhref{hep-ph/0307101}].  


\bibitem{MartinTSIL}
  S.~P.~Martin and D.~G.~Robertson
  %``TSIL: a program for the calculation of two-loop self-energy integrals,''
    Comput.\ Phys.\ Commun.\  {174} (2006) 133
   [arXiv:\hhref{hep-ph/0501132}].  
%\cite{Kniehl:2008cj}


\bibitem{Kniehl:2008cj}
  B.~A.~Kniehl and A.~Sirlin,
  %``Pole Mass, Width, and Propagators of Unstable Fermions,''
  Phys.\ Rev.\ D {77} (2008) 116012
   [arXiv:\hhref{0801.0669}].  


\bibitem{Jegerlehner:2012kn}
  F.~Jegerlehner, M.~Y.~Kalmykov and B.~A.~Kniehl,
  %``On the difference between the pole and the MSbar masses of the top quark at the electroweak scale,''
  Phys.\ Lett.\ B {722} (2013) 123
    [arXiv:\hhref{1212.4319}].   
 
 \bibitem{Martin:2013gka}
  S.~P.~Martin,
  %``Three-loop Standard Model effective potential at leading order in strong and top Yukawa couplings,''
  [arXiv:\hhref{1310.7553}].
  %%CITATION = ARXIV:1310.7553;%%



\bibitem{alpha}
Particle Data Group,
  %``Review of particle physics,''
  J.\ Phys.\ G {37} (2010) 075021.
  The LEP Electroweak Working Group, \url{http://lepewwg.web.cern.ch}.
  We thank Jens Erler and Paul Langacker for the latest
  fit we quote, and  Martin Gr\"unewald for useful discussions.
     
  
 


\bibitem{Froggatt:1995rt}
  C.~D.~Froggatt and H.~B.~Nielsen,
  %``Standard model criticality prediction: Top mass 173 +- 5-GeV and Higgs mass 135 +- 9-GeV,''
  Phys.\ Lett.\ B {368} (1996) 96
  [arXiv:\hhref{hep-ph/9511371}].
    
  


\bibitem{Froggatt:2001pa}
  C.~D.~Froggatt, H.~B.~Nielsen and Y.~Takanishi,
  %``Standard model Higgs boson mass from borderline metastability of the vacuum,''
  Phys.\ Rev.\ D {64} (2001) 113014
  [arXiv:\hhref{hep-ph/0104161}].
    


\bibitem{BdCE}
  C.~P.~Burgess, V.~Di Clemente and J.~R.~Espinosa,
  %``Effective operators and vacuum instability as heralds of new physics,''
  JHEP {0201} (2002) 041
  [arXiv:\hhref{hep-ph/0201160}].
  


\bibitem{Isidori:2007vm}
   G.~Isidori, V.~S.~Rychkov, A.~Strumia and N.~Tetradis,
  %``Gravitational corrections to standard model vacuum decay,''
  Phys.\ Rev.\ D\ {77} (2008) 025034
  [arXiv:\hhref{0712.0242}].
  %%CITATION = PHRVA,D77,025034;%%  
  
 


\bibitem{Bezrukov:2009db}
  F.~Bezrukov and M.~Shaposhnikov,
  %``Standard Model Higgs boson mass from inflation: Two loop analysis,''
  JHEP {0907} (2009) 089
  [arXiv:\hhref{0904.1537}].
  %%CITATION = ARXIV:0904.1537;%% 
  
 


\bibitem{Shaposhnikov:2009pv}
  M.~Shaposhnikov and C.~Wetterich,
  %``Asymptotic safety of gravity and the Higgs boson mass,''
  Phys.\ Lett.\ B {683} (2010) 196
  [arXiv:\hhref{0912.0208}].
  %%CITATION = ARXIV:0912.0208;%% 

%

\bibitem{V2}
  C.~Ford, I.~Jack and D.~R.~T.~Jones,
  %``The Standard model effective potential at two loops,''
  Nucl.\ Phys.\ B {387} (1992) 373
   [Erratum-ibid.\ B {504} (1997) 551]
  [arXiv:\hhref{hep-ph/0111190}]. 

\bibitem{Hall:2009nd}
  L.~J.~Hall and Y.~Nomura,
  %``A Finely-Predicted Higgs Boson Mass from A Finely-Tuned Weak Scale,''
  JHEP {1003} (2010) 076
  [arXiv:\hhref{0910.2235]}.
  


\bibitem{Giudice:2011cg}
  G.~F.~Giudice and A.~Strumia,
  %``Probing High-Scale and Split Supersymmetry with Higgs Mass Measurements,''
  Nucl.\ Phys.\ B {858} (2012) 63
  [arXiv:\hhref{1108.6077}].
  


\bibitem{Cabrera:2011b}
M.~E.~Cabrera, J.~A.~Casas and A.~Delgado,
  %``Upper Bounds on Superpartner Masses from Upper Bounds on the Higgs Boson Mass,''
  Phys.\ Rev.\ Lett.\  {108} (2012) 021802
  [arXiv:\hhref{1108.3867}].
   
%%%%%%%%%%%%%%%%%%%%%%%%%%%%%%%%%%%
%%   Appendix
%%%%%%%%%%%%%%%%%%%%%%%%%%%%%%%%%%%


      
  %\cite{Coleman:1973jx}


\bibitem{Arbey:2011ab}
  A.~Arbey, M.~Battaglia, A.~Djouadi, F.~Mahmoudi and J.~Quevillon,
  %``Implications of a 125 GeV Higgs for supersymmetric models,''
  Phys.\ Lett.\ B {708} (2012) 162
  [arXiv:\hhref{1112.3028}].
  


\bibitem{Ibanez:2013gf}
  L.~E.~Ibanez and I.~Valenzuela,
  %``The Higgs Mass as a Signature of Heavy SUSY,''
   JHEP {1305} (2013) 064
  [arXiv:\hhref{1301.5167}].
    
    


\bibitem{Hebecker:2013lha}
  A.~Hebecker, A.~K.~Knochel and T.~Weigand,
  %``The Higgs mass from a String-Theoretic Perspective,''
   Nucl.\ Phys.\ B {874} (2013) 1
  [arXiv:\hhref{1304.2767}].
  


\bibitem{Fox:2002bu}
  P.~J.~Fox, A.~E.~Nelson and N.~Weiner,
  %``Dirac gaugino masses and supersoft supersymmetry breaking,''
  JHEP {0208} (2002) 035
  [arXiv:\hhref{hep-ph/0206096}].
  


\bibitem{Benakli:2012cy}
  K.~Benakli, M.~D.~Goodsell and F.~Staub,
  %``Dirac Gauginos and the 125 GeV Higgs,''
  JHEP {1306} (2013) 073
    [arXiv:\hhref{1211.0552}].   


\bibitem{Hebecker:2012qp}
  A.~Hebecker, A.~K.~Knochel and T.~Weigand,
  %``A Shift Symmetry in the Higgs Sector: Experimental Hints and Stringy Realizations,''
  JHEP {1206} (2012) 093
  [arXiv:\hhref{1204.2551}].
  %%CITATION = ARXIV:1204.2551;%%  
  
 


\bibitem{Redi}
 M.~Redi and A.~Strumia,
  %``Axion-Higgs Unification,''
  JHEP {1211} (2012) 103
  [arXiv:\hhref{1208.6013}]. 
  
  


\bibitem{Giudice:2006sn}
  G.~F.~Giudice and R.~Rattazzi,
  %``Living Dangerously with Low-Energy Supersymmetry,''
  Nucl.\ Phys.\ B {757} (2006) 19
  [arXiv:\hhref{hep-ph/0606105}].
    
  \bibitem{SOC}
   P.~Bak, C.~Tang and K.~Wiesenfeld,  Phys.\ Rev.\ Lett.\  {59} (1987) 381.



\bibitem{Weinberg:1987dv}
  S.~Weinberg,
  %``Anthropic Bound on the Cosmological Constant,''
  Phys.\ Rev.\ Lett.\  {59} (1987) 2607.
    
 %%%%%%%%%%%%%%%%%%%%%%%%%%%%%%%%%%%
%%   sect. 9
%%%%%%%%%%%%%%%%%%%%%%%%%%%%%%%%%%%
    
  


\bibitem{Donoghue:2005cf}
  J.~F.~Donoghue, K.~Dutta and A.~Ross,
  %``Quark and lepton masses and mixing in the landscape,''
  Phys.\ Rev.\ D {73} (2006) 113002
  [arXiv:\hhref{hep-ph/0511219}].
  


\bibitem{Hall:2007zh}
  L.~J.~Hall, M.~P.~Salem and T.~Watari,
  %``Quark and Lepton Masses from Gaussian Landscapes,''
  Phys.\ Rev.\ Lett.\  {100} (2008) 141801
  [arXiv:\hhref{0707.3444}].
    
  


\bibitem{Gibbons:2008su}
  G.~W.~Gibbons, S.~Gielen, C.~N.~Pope and N.~Turok,
  %``Measures on Mixing Angles,''
  Phys.\ Rev.\ D {79} (2009) 013009
  [arXiv:\hhref{0810.4813}].
    
  


\bibitem{flamulti}
   G.~F.~Giudice, G.~Perez, Y.~Soreq,
  %``Flavor Beyond the Standard Universe,''
  arXiv:\hhref{1207.4861}.
    
 


\bibitem{ArkaniHamed:2005yv}
  N.~Arkani-Hamed, S.~Dimopoulos and S.~Kachru,
  %``Predictive landscapes and new physics at a TeV,''
  arXiv:\hhref{hep-th/0501082}.
  %%CITATION = HEP-TH/0501082;%% 
  
  


\bibitem{Agrawal:1997gf}
  V.~Agrawal, S.~M.~Barr, J.~F.~Donoghue and D.~Seckel,
  %``The Anthropic principle and the mass scale of the standard model,''
  Phys.\ Rev.\ D {57} (1998) 5480
  [arXiv:\hhref{hep-ph/9707380}].
    
  


\bibitem{Kobzarev:1974cp}
  I.~Y.~.Kobzarev, L.~B.~Okun and M.~B.~Voloshin,
  %``Bubbles in Metastable Vacuum,''
  Sov.\ J.\ Nucl.\ Phys.\  {20} (1975) 644
   [Yad.\ Fiz.\  {20} (1974) 1229].
    
  


\bibitem{Coleman:1977py}
  S.~R.~Coleman,
  %``The Fate of the False Vacuum. 1. Semiclassical Theory,''
  Phys.\ Rev.\ D {15} (1977) 2929
   [Erratum-ibid.\ D {16} (1977) 1248].
    
  


\bibitem{Callan:1977pt}
  C.~G.~Callan, Jr. and S.~R.~Coleman,
  %``The Fate of the False Vacuum. 2. First Quantum Corrections,''
  Phys.\ Rev.\ D {16} (1977) 1762.
    
    


\bibitem{Coleman:1980aw}
  S.~R.~Coleman and F.~De Luccia,
  %``Gravitational Effects on and of Vacuum Decay,''
  Phys.\ Rev.\ D {21} (1980) 3305.
    
%%%%%%%%%%%%%%%%%%%%%%%%%%%%%%%%%%%
%%   sect. 10
%%%%%%%%%%%%%%%%%%%%%%%%%%%%%%%%%%%
  
  


 
 
 


\end{thebibliography}
\end{document}